\documentstyle[12pt]{article}

\def\beq{\begin{equation}}
\def\eeq{\end{equation}}
\def\beqa{\begin{eqnarray}}
\def\eeqa{\end{eqnarray}}

\newcommand{\cA}{{\cal A}}

\newcommand{\cF}{{\cal F}}
\newcommand{\cL}{{\cal L}}
\newcommand{\cM}{{\cal M}}

\newcommand{\p}{\partial}
\newcommand{\hf}{\frac12}
\newcommand{\bea}{\begin{eqnarray}}
\newcommand{\eea}{\end{eqnarray}}
\newcommand{\be}{\begin{equation}}
\newcommand{\ee}{\end{equation}}
\newcommand{\bt}{\begin{tabular}}
\newcommand{\et}{\end{tabular}}
\newcommand{\ba}{\begin{array}}
\newcommand{\ea}{\end{array}}

\newcommand{\R}{{\rm Re}}

\newcommand{\th}{\theta}
\def\Tr{{\rm Tr}}
\def\C{ {\cal C}}

\newcommand{\resetcounter}{\setcounter{equation}{0}}     

\begin{document}

\begin{titlepage}
\begin{center}
\hfill HUB-EP-97/45\\
\hfill {\tt hep-th/9708009}\\

\vskip .7in

{\bf $N=2$ String-String Duality and Holomorphic Couplings}

\vskip .3in

Gottfried Curio\footnote{email: curio@qft2.physik.hu-berlin.de}
\\
\vskip 1.2cm

{\em Humboldt-Universit\"at zu Berlin,
Institut f\"ur Physik, 
D-10115 Berlin, Germany}

\vskip .1in

\end{center}

\vskip .2in

\begin{center} \end{center}
\begin{quotation}\noindent

We review aspects of N=2 duality between the heterotic and the type IIA string.
After a description of string duality intended for the non-specialist
the computation of the heterotic prepotential and the $F_1$ function for the
ST, STU and STUV model (V a Wilson line) and the matching with the
Calabi-Yau instanton expansions are given in detail. Relations with BPS
spectral sums in various connections are pointed out.

\end{quotation}
\end{titlepage}
\vfill
\eject

\tableofcontents

\newpage

\setcounter{page}{1}

\section{Introduction}

In this section a description of string theory and duality intended for the 
non-specialist is given.\\

The starting point for the following considerations is
represented by the experimentally well-tested standard model
of elementary particle physics with gauge group
$SU(3)\times SU(2) \times U(1)$. \\
Now what is missing in this model ?
First, most of the theoretical insights and also many of the 
phenomenological predictions, which are tested in high precision 
experiments of the Standard Model, are limited to perturbation 
theory. However one would like to include into the picture, especially
in the infrared regime of Quantum Chromodynamics, also solitonic states and
nonperturbative effects at strong coupling. Furthermore there
are 19 free parameters, and so the theory does not look quite fundamental.
Also missing is an understanding of the occurence of exactly these
gauge groups; a problem which is not cured by embedding them in a
grand unified theory (GUT) group like $SU(5)$, $SO(10)$ or even $E_6$. 
Similarly the number of generations has to be explained.
Of course there
is the most fundamental gap of not including gravity at the quantum level. 
Finally one is
confronted with the hierarchy problem of explaining $m_{weak}\ll m_{Pl}$.\\
The last problem can be solved by the introduction of supersymmetry (SUSY),
i.e. by postulating the existence of
superpartners for all particles, which stabilize scalar field masses
by avoiding their quadratic diviergencies in loop orders. Of course,
as these partners were not seen, supersymmetry has to be broken in nature.
To avoid the Goldstino of broken global SUSY and to make a vanishing
cosmological constant possible one is led to local SUSY and thus has
automatically included gravity. 
But as the simplest version of local SUSY, namely $N=1$ supergravity (SUGRA),
is still plagued by infinities one 
is led to introduce higher Supergravities. But -- even those are plagued
by infinities at higher loop-orders; furthermore the maximal group 
occuring is $SO(8)$ -- by far to small to accomodate the data already
known to occur in the atomic world as given by the 
Standard Model gauge group.\\
The superstring \cite{GSW,LT}
is an offer to get rid of all these problems. The ultraviolet
infinities are (probably to all orders)
avoided as point particles are replaced by extended
objects, so there is no 'singular' interaction point in the Feynman
diagrams, which themselves are organised in a series of two-dimensional
Riemann surfaces (coming from the world-sheets) of ascending topological
complexity as measured by the genus. Supergravity is included as 
effective theory in the point particle limit.
The spacetime dimension, which is a free parameter in field theory,
is restricted to be less or equal to 10 for superstring consistency.
More precisely the requirement of 
cancellation of the conformal anomaly 
in the world-sheet approach
fixes the possible space-time dimension $D_{\mbox{crit}}$ 
to be 10 (resp. 26 for the bosonic string). Similarly
cancellation of the chiral anomalies fixes the gauge group.
Note that in theories where spinors are coupled to gauge and 
gravitational fields one has to worry about the issue of
anomalies \cite{gaume,W-anom} and that gravitational anomalies are 
possible in $4k+2$ dimensions like $D=10$.
That the theory is free of anomalies was the clue of the revival
of string theory in the first string revolution of 1984 
as the most promising candidate for a unified theory
of all elementary forces and particles. As said this also constrains 
the tendimensional (10 $D$)
gauge group \cite{green-schwarz}: 
it has to be one of the two rank 16 groups $SO(32)$ 
and $E_8\times E_8$ 
of even self-dual Lie algebra lattice and thus
explains the GUT gauge groups as we will see in a moment.
Furthermore the construction had extremely little freedom -- 
there were only five fundamental consistent theories
\begin{itemize}
\item
the type I string (including) open strings 
of gauge group $SO(32)$ and $N=1$ SUGRA as 10$D$ field theory limit
\item
the heterotic string \cite{GHMR} of gauge group
$SO(32)$ resp. $E_8\times E_8$ and $N=1$ SUGRA as limit
(called heterotic as it consists of a bosonic string 'compactified' on
a rank 16 lattice for the left movers and a superstring for the right movers)
\item
the type II string \cite{GSII}, which does not lead
to non-abelian gauge groups but has $N=2$ SUGRA as limit 
(having for both left- and right- movers a version of the superstring) 
and comes in two variants 
according to the two gravitini having the same (IIB) or different
chirality (IIA)
\end{itemize}

So, to reduce from the critical dimension
$D=10$ to e.g. $D=4$, one has like in the old Kaluza-Klein approaches 
to compactify some dimensions (resp. to take the approach of
going down to $D=4$ not on a space but on a conformal field theory
\cite{ABK,KLT,LLS}). The dynamical evolution of this process 
requires certainly non-perturbative dynamics. 
Now the conformal invariance is maintained
by choosing the six-dimensional compactification space
Ricci-flat (and embedding its spin-connection
into the gauge connection) and the perturbative stability is 
saved by choosing $N=1$ SUSY. This in turn is equivalent 
to the existence of a covariantly constant spinor, whose existence furthermore
affirms that the compactification space is complex K\"ahler. Such a Ricci-flat
K\"ahler space (of complex dimension three) is called a Calabi-Yau space
\cite{CHSW}.\\
Now K\"ahler manifolds of complex dimension $n$ have holonomy group
$U(n)$ instead of the $SO(2n)$ for the general compact manifold. 
Furthermore the
condition of having a Ricci-flat metric is equivalent to have vanishing
first Chern class $c_1$. Then the theorem of Calabi and Yau states that for a 
K\"ahler manifold the further reduction of the holonomy group to $SU(n)$
is equivalent to $c_1=0$ (the hard part is that the existence of a metric 
of $SU(n)$ holonomy is implied by $c_1=0$).\\
Let us now return to our special case $n=3$, i.e.
compactification of the heterotic string on a Calabi-Yau (of strict
$SU(3)$ holonomy) leading to $N=1$ in $4D$. The standard 
embedding of the holonomy
group in the gauge group $E_8\times E_8$ (constituting one of the favoured
possibilities of anomaly cancellation) leaves besides an 'hidden' $E_8$
an $E_6$ GUT group; furthermore the geometry/physics dictionary leads to
a generation number given by (the absolute value of) half the Euler number
of the Calabi-Yau.\\
One is also interested in compactifications 
on spaces of smaller holonomy leading to higher SUSY in $4D$. Now the 
analogue of the CY for $n=3$, which exists in quite diverse topological
forms, is the elliptic curve $T^2$ for $n=1$ and the $K3$ surface for $n=2$.
Compactification of the heterotic string on $K3\times T^2$ of holonomy
$SU(2)$ leads then to $N=2$ SUSY
in $4D$ whereas compactification on the flat space
$T^6$ leads to $N=4$ SUSY in $4D$. If we had started instead with a type II
string, of twice as much SUSY in $D=10$ as the heterotic string, the 
amount of SUSY in $D=4$ would be equally twice as much. So note especially
that the compactifications of the heterotic string 
on $T^4\times T^2$ and the type II string
on $K3\times T^2$ both lead to $N=4$ SUSY in $D=4$ (obviously it is possible
to formulate the observation already in $D=6$) and similarly the heterotic
string on $K3\times T^2$ and the type II string 
on $CY$ both lead to $N=2$ SUSY in $D=4$.\\
Now, if one has direct phenomenelogical wishes, these models with higher
SUSY are certainly not very convincing as extended SUSY has no chiral 
fermions since the representations of the Clifford algebra then split in 
$c+\bar{c}$ for the $D=4$ Dirac algebra. Nevertheless the strategy to
study first a version of the theory with higher symmetry has been here
as in the past extremely succesfull. \\
To put these assertions into perspective
we have to recall now the second string revolution going on since 1994, where
a unification of all string theories as well as an inclusion of the non-
perturbative regimes took place under the headline ``duality". As these 
developments were accompanied and partly proceeded by corresponding
explorations in SUSY field theories let us discuss the latter first.\\
The first basic observation is the connection given by the fact that
electro-magnetic duality of $N=4$ supersymmetric $4D$ Yang-Mills theory
corresponds to the exchange of elementary and solitonic
states (magnetic monopoles) and so also to strong-weak coupling duality
(the non-perturbative solitonic states have $m_{sol}\sim \frac{1}{g^2}$;
in string theory there exist solitons with $m_{sol}\sim \frac{1}{g}$).
At first a corresponding duality was considered impossible to hold also
for $N=2$ supersymmetric gauge theories 
as the gauge bosons and the magnetic monopoles are in different
$N=2$ multiplets (unifying in one multiplet states which
differ in spin up to two steps of spin 1/2): the vector multiplet 
consisting of states of spin 1, 1/2 and 0 resp. the hypermultiplet
consisting of states of spin 1/2, 0 and 1/2 of the other helicity.
Note that because of the scalar $a$ in the vector multiplet one always
has a Coulomb phase with $a$ as Higgs scalar and classical non-abelian
gauge symmetry enhancement, say to $SU(2)$ in the simplest case, for
$a=0$. Now the breaktrough of Seiberg and Witten consisted in the 
determination of the exact non-perturbative low-energy effective 
action of $N=2$ Yang-Mills theory in $D=4$. 
If we call the perturbative expansion regime for a moment
local in the coupling parameter, then they could gain the exact global --
in the qualified sense --  structure of the `characteristic function', 
called prepotential (whose second derivatives give the gauge couplings),
of the theory by combining its holomorphy 
with information about the singularities and the corresponding monodromies.
Namely here the holomorphy corresponds to having $N=2$ SUSY in the
sense that holomorphy unifies analytic dependence on a real and an imaginary
part like SUSY connects bosonic and fermionic states.
Actually one has here - to get the proper conclusions - to consider 
the difference between physical and Wilsonian couplings 
(for a review cf. \cite{KL}), where the latter
are good behaved with respect to the question of the connection between SUSY
and holomorphy.\\
Now, in particular, one has the gauge symmetry enhancement in the Coulomb
phase at $a=0$ only classically, whereas quantum mechanically one finds
a splitting of this one point of enhancement of the massless
spectrum (namely by the non-abelian gauge bosons) 
in two points in the moduli space, where massless monopoles/dyons occur.
The important technical tools to get and to express these results consist in
the translation of this setup: one starts with a variation of data over a 
complex one-dimensional base manifold, where the data degenerate at some 
points leading to monodromies; this is then reformulated in 
a geometrical picture
involving the variation of the Seiberg-Witten torus, where one studies the
periods around 1-cycels of a certain form. Near the points where the 
magnetic solitons occur one has then an effective description of the theory
in variables, for which these states appear as elementary.\\
Now similarly as in supersymmetric field theories 
one has in the $N=4$ heterotic string \cite{FILQ,S+SS} 
a strong-weak coupling duality (S-duality).
Furthermore it was discovered during the last years that there is a duality
of descriptions between perturbative effects in one string theory 
at weak coupling and non-perturbative effects in a dual theory at 
strong coupling, the socalled string-string duality (for a review cf.
\cite{FL}). In this way all string theories are now believed to be
unified just as different perturbative expansions
of one basic underlying theory, often called $M$-theory. 
For instance \cite{Wi} -- to give
an impression -- one can think of the heterotic
string, which is chiral with non-abelian gauge group, as a description of the 
real world at high energies (GUT's) whereas the type II string, non-chiral
and with abelian gauge group only, as a description of the
low energy world. More precisely Hull and Townsend \cite{HT} 
and also Witten \cite{W} found
that the $N=4$ string-string duality holds, 
if one includes in the comparison of heterotic string 
on $T^4$ and type IIA string
on $K3$ also solitonic states, which are given in string theory by higher
dimensional extended objects like membranes and p-branes. The occurence
of non-abelian gauge symmetry on the type II side is exactly explained by
these membranes wrapping certain 2-cycels of $K3$, which become massless as 
the 2-cyclels shrink leading to a $K3$ with singularities as classified
by the Dynkin diagrams $ADE$ of simply laced Lie groups; one can already
see the $E_8\times E_8$ of the heterotic string occuring in the cohomology
of $K3$ where $H^2(K3)=E_8 \oplus E_8 \oplus H \oplus H \oplus H$ ($H$ being 
the hyperbolic plane). This same observation lead also to the detection off
the heterotic string (in $6D$, on $T^4$) as a soliton of the type IIA string
(on $K3$) \cite{HS}.\\
Now in the next step, 
it was demonstrated \cite{KV} that there exists a string-string duality of 
$N=2$ string compactifications to $4D$ given on the one hand by the 
heterotic string on $K3\times T^2$ and on the other hand the 
type IIA string on $CY$. 
This duality becomes clear remembering 
the fact that $K3$ can be represented as fibration of $T^2$ over $P^1$
and so $K3\times T^2$ as fibration of $T^2\times T^2$ over $P^1$;
in this way one gets by this adiabatic extension principle \cite{VW}
that the $4D$ duality can be understood as spreading out the $6D$ duality
over $P^1$. This then leads to the conclusion that one should expect a duality
with type IIA on a $CY$ which is a $K3$ fibration over $P^1$. As we have
just seen that it is possible to match the classical gauge symmetry
enhancements it is of interest to study the mentioned `characteristic
function' (prepotential) and its higher relatives (involving higher
gravitational couplings (socalled $F_1$ function)) in the possible
dual models \cite{FHSV,KV}.\\
More precisely we will study the prepotential in its vector moduli
dependence, i.e. on the Coulomb branch.
On the heterotic side, where the dilaton sits in a vector multiplet, one
can at first study the prepotential in the perturbative regime only. Thanks
to $N=2$ SUSY, perturbative corrections occur only at 1-loop. Actually one
uses then the perturbative quantum symmetries (T-duality generalising
the exchange of momentum and winding modes related to the $R\rightarrow 1/R$
on $S^1$) and the singularity structure corresponding to the gauge symmetry 
enhancement loci to determine the one-loop correction of the gauge couplings, 
which gives then direct information about the second derivatives of the 
prepotential. Here again the crucial role of singularities in the moduli 
space of vacua is demonstrated. So let us make more precise what is really 
meant with such a singularity. If the singularity is physical
(and not  an artefact of perturbation theory and 
smoothed out once quantum corrections are taken into account)
it signals the breakdown of some approximation like 
the appearance of additional massless degrees
of freedom on a subspace of the moduli space. Almost
everywhere on the moduli space these degrees of freedom are heavy and 
thus have been integrated out of the effective theory.
In $4D$ SUSY Yang-Mills below the threshold scale $M$ of the heavy states the 
gauge couplings of the light modes obey ($c$ a model dependent constant)
\begin{eqnarray*} 
g^{-2}_{\rm low} = g^{-2}_{\rm high} + c \log M,\nonumber 
\end{eqnarray*}
where  the mass is a function of the
moduli $M(\phi)$ with a zero somewhere in the moduli space,
where the gauge coupling $g^{-2}_{\rm low}$ becomes
singular due to the inappropriate approximation of integrating out the heavy 
states. As we discussed, quantum mechanically the number and position of the 
singularities and the interpretation of the fields becoming massless at the
singularities can be quite different to the classical picture \cite{SW}.\\
Now by contrast on the type IIA side the vector moduli do not interfere with 
the dilaton, which for type II
lies in a hypermultiplet, and so the space-time quantum field theory is given
by the uncorrected string tree level result.
The technical tool used on the type IIA side to evaluate the prepotential
is mirror symmetry \cite{GP} to the type IIB string on the mirror CY,
where the the vector moduli do not correspond to the $h^{1,1}$ cohomology
like in type IIA but are now related to the $h^{2,1}$ cohomology which is
no longer subjected to corrections by world-sheet instantons (i.e. rational
curves giving 2-cycels).\\

So the subject of this review will be to provide several quantitative
tests for the $N=2$ string-string duality between the heterotic string 
on $K3 \times T^2$ and the type IIA string on a $CY$. We will first pick
specific $CY$'s matching the spectrum, i.e. the number of vector- and
hypermultiplets, of heterotic string models and will then go on to the
heart of the investigation, which consists in making finer tests of the
potentially dual pairs by comparing the couplings.\\

In more detail the content is as follows.\\ 
In {\em section 2} we review the necessary formalism of $N=2$ supergravity,
especially the multiplets the splitting of vector- and hypermultiplet moduli
spaces and the prepotential with its relation to all the other relevant 
quantities in the theory like the Kaehler potential, the gauge couplings
and the $BPS$ masses.\\
In {\em section 3} we start with the perturbative analysis of the $N=2$
heterotic vacua studying the spectrum, enhanced symmetry loci and the 
thereby gained information on the prepotential; then we go on and perform the
corresponding analysis for the first of the higher gravitational couplings,
the socalled $F_1$ function.
Here and in the following we especially focus on three models 
of one resp. two. resp. three vector moduli besides the dilaton
called $ST$ resp. $STU$ resp. $STUV$ model.\\
In {\em section 4} then the corresponding questions on the $CY$ side are 
treated. Especially the instanton expansions for the prepotential and the 
$F_1$ function are given and the geometrical situation in the different
potentially dual $CY$ spaces are discussed.\\
In {\em section 5} we discuss the notion of $BPS$ states, including the
situation in $N=4$ in view of a later application of this to the 
question of the $S-T$ exchange symmetry. Then we go on and study the
notion of $BPS$ spectral sums, i.e. the function given by the sum over the 
$BPS$ spectrum of the theory, in different approaches, especially
in that of Harvey and Moore \cite{HM}, 
where an intimate connection to the prepotential is made.\\
In {\em section 6} at last we match the expressions for prepotential
and $F_1$ function we have got on the heterotic and the type IIA side.\\
In the {\em appendix} we review the necessary background of the 
special functions used.

\resetcounter

\newpage

\section{$N=2$ supergravity}

The $N=2$ supergravity 
multiplets relevant for us to consider\footnote{In addition
there are multiplets containing an antisymmetric tensor:
the vector-tensor multiplet  $VT$ 
features an Abelian gauge field $A_\mu$,
an antisymmetric tensor $b_{\mu\nu}$, two Weyl fermions $\chi^I_\alpha $ 
and one real scalar; it consists of an $N=1$ vector multiplet
and a linear multiplet $L$.
The tensor multiplet $T$ has an antisymmetric tensor $b_{\mu\nu}$,
two Weyl spinors $\chi_{\alpha}^I$, a complex scalar $\phi$ and a real
scalar; it consists of an $N=1$ chiral multiplet
and a linear multiplet.
Finally the double tensor multiplet $\Pi$ contains two antisymmetric
tensors $b_{\mu\nu}$, $b^{\prime}_{\mu\nu}$, 
two Weyl spinors $\chi_{\alpha}^I $,
and two real scalars ; it consists of two linear 
multiplets.
In four space-time dimensions an antisymmetric tensor only contains
one physical degree of freedom and is dual to a real scalar 
$a(x)$ via 
$\epsilon^{\mu\nu\rho\sigma}\partial_\mu b_{\nu\rho}
\sim \partial ^\sigma a$.
This duality can be elevated to a duality between entire
supermultiplets and one finds that 
the vector-tensor multiplet is dual to a vector multiplet
while the tensor multiplet and
the double tensor multiplet are both dual to a hyper multiplet.}
are the following:
\begin{itemize}
\item
the gravitational multiplet containing the graviton $g_{\mu\nu}$,
two gravitini $\psi^{I}_{\mu\alpha}$  and an Abelian
graviphoton $\gamma_\mu$; 
in terms of $N=1$ multiplets it is the sum of the $N=1$
gravitational multiplet and a gravitino multiplet $\Psi$ which contains
a gravitino and an Abelian vector
\item
the vector multiplet $V$ 
with a gauge field $A_\mu$, 
two gauginos $\lambda_{\alpha}^I$ and a complex scalar $\phi$;
it consists of an $N=1$ vector multiplet $V$ and a chiral multiplet $\Phi$
\item
furthermore matter fields arise from 
hypermultiplets $H$ with two Weyl spinors $\chi^{I}_\alpha$ 
and four real scalars $q^{IJ}$; it consists of two chiral multiplets
(chiral plus antichiral)
\end{itemize}

Supersymmetry prohibits gauge neutral interactions 
between vector and
hypermultiplets and therefore the moduli space locally 
has to be a direct product  \cite{dWLvP}

\begin{equation}
{\cal M} = {\cal M}_H \times {\cal M}_V ,
\end{equation}
where ${\cal M}_H$ is the (quaternionic) moduli space parameterized by the
scalars of the hypermultiplets and ${\cal M}_V$ 
is the (complex) moduli space spanned
by the scalars in the vector multiplets.

${\cal M}_V$ has actually the structure of a special Kaehler manifold,
which is  a K\"ahler manifold  whose geometry 
obeys  an additional constraint \cite{CRTvP,dWvP,vP},
which constrains both the  K\"ahler potential and the 
Wilsonian gauge couplings 
of the vector multiplets in an $N=2$ effective Lagrangian to be determined by 
a single holomorphic function  of the scalar fields $\phi$
-- the prepotential $\cF(\phi)$.

One way to express this constraint  is the statement that the 
K\"ahler potential $K$ is not an arbitrary real function 
(as in $N=1$ supergravity) but 
determined in terms of a holomorphic prepotential $F$.
Here $F(X)$ is  a homogeneous function of $X^I$ of degree $2$:
$X^I F_I=2 F$.The $X^I, I=0,\ldots, n_V+1$ are $(n_V+2)$
holomorphic functions of the $n_V+1$ complex scalar fields 
$\phi^I, I=1,\ldots,n_V+1$ which reside 
in the vector multiplets and the dilaton multiplet.
$F_I$ abbreviates the derivative, i.e.
$F_I\equiv \frac{\p F(X)}{\p X^I} $.

The mentioned connection with the Kaehler potential is given by
(cf. also \cite{CdAF,CdAFLL,CRTvP1,Str})
\be\label{Kspecial}
e^{-K}=i\Big( \bar{X}^I (\bar\phi) F_I(X)
-  X^I  (\phi)\bar{F}_I(\bar{X})\Big) \ .
\ee
The above description is slightly redundant. The variable
$X^0$, which represents the graviphoton,
can be eliminated by an appropriate choice of coordinates, 
the socalled 
special coordinates defined by:
\be
\phi^I=\frac{X^I}{X^0} \; ,
\ee
i.e. the graviphoton is set to one.

In these special coordinates 
the K\"ahler potential  (\ref{Kspecial}) reads
\be\label{Kspecial2}
K=-\log\Big(2 (\cF+\bar{\cF})-(\phi^I-\bar{\phi}^I)(\cF_I-\bar{\cF}_I)\Big) ,
\ee
where $\cF(\phi)$ is an arbitrary holomorphic function of 
$\phi^I$ related to $F(X)$ via
$F(X)=-i (X^0)^2 \cF(\phi)$. 

The metric $G_{i \bar{j}}$ on the Kaehler manifold is expressed 
in terms of the K\"ahler potential $K$ by
$G_{i \bar{j}} = \frac{\partial}{\partial \phi^i}
\frac{\partial}{\partial \bar{\phi}^{\bar{j}}}
K\left( \phi, \bar{\phi}\right)$.

Furthermore of central importance in the formalism is 
the period vector
\beqa
\Omega=(X^I,F_I) \; ,
\eeqa
where $X^I$ constitutes the so called electric, $F_I$ the magnetic part.
This terminology is motivated by considering the stringy realisation of
the effective $N=2$ supergravity as a compactification of the type IIB string
on a Calabi-Yau. There the $X^I$ and the $F_J$ arise really as periods of
the holomorphic 3-form $\Omega$

\beqa
X^I=\int _{\Gamma _{\alpha_I}} \Omega \;, \;\;\;\;\;\;
F_J=\int _{\Gamma _{\beta_J}} \Omega \; ,
\eeqa

where the $\Gamma _{\alpha_I}$,$\Gamma _{\beta_J}$,$I,J=1, \cdots ,h^{2,1}+1$
span a symplectic basis of $H_3(CY,{\bf Z})$ with the $\alpha$- and $\beta$-
cycles dual under the intersection product
(spoken in cohomology the "$+1$" corresponding to $h^{3,0}$ corresponds 
to the graviphoton). Furthermore we will discuss later
that one has a $BPS$ mass formula

\beqa
M_{BPS}=M_IX^I+N^JF_J \; ,
\eeqa

where $M_I$ constitutes an electric and $N^J$ a magnetic charge vector. This
by the way makes the easiest contact with the results of Seiberg/Witten
\cite{SW} in the field theory limit by considering the analogy to the situation
involving 1-cycles on the Seiberg/Witten torus \cite{KLMVW}.

This whole structure is acted on by duality transformations
$\Gamma \in Sp(2(n_V+2),{\bf Z})$

\beqa
\Big( \begin{array}{c}
X^I \\ F_I
\end{array} \Big)
\rightarrow 
\Big( \begin{array}{cc}
U & Z  \\ W & V
\end{array} \Big)
\Big( \begin{array}{c}
X^I \\ F_I
\end{array} \Big) \; ,
\eeqa

where $\Big( {\footnotesize \begin{array}{cc}
U & Z  \\ W & V
\end{array}} \Big)$
constitutes a so called monodromy matrix 
(with $U^TV-W^TZ=\underline{1}$, etc.).

The $W_{IKM}=\p_I\p_K\p_M F$ are referred to as
the  Yukawa couplings since for a particular class
of heterotic $N=1$ vacua (vacua which have a
global $(2,2)$ worldsheet supersymmetry)
they correspond to space-time Yukawa couplings 
\cite{DKL}.

$N=2$ supergravity not only constrains the K\"ahler
manifold of the vector multiplets but also relates the gauge couplings 
$N_{IJ}$ to the holomorphic prepotential.
They are essentially \cite{vP}
given by the second derivatives of the prepotential
(which would constitute the exact expression in global $N=2$ SUSY)

\beqa
N_{IJ}=\p_I\p_J F +\cdots
\eeqa

\newpage

\section{$N=2$ Heterotic Vacua}

\resetcounter

So far we reviewed the effective 
$N=2$ supergravity without any reference to a particular 
string theory. The aim of this section is 
to determine the prepotential $\cF$ for 
$N=2$ heterotic vacua (cf. \cite{LF} for a review).

As we already discussed,  for  heterotic vacua 
the dilaton is part of a vector multiplet (actually of the equivalent dual
vector-tensor multiplet) whereas in type IIA (IIB) 
the dilaton sits in a hypermultiplet (actually in the equivalent dual
tensor (double tensor) multiplet). 

From the fact that the dilaton organizes 
the string perturbation theory
together with product structure of the moduli space  one derives a
non-renormalization theorem \cite{dWKLL,S} saying that
the heterotic moduli space of the hypermultiplets 
is determined
at the string tree level and 
receives no further perturbative or non-perturbative corrections, i.e.
heterotically
\be
\cM_H=\cM_H^{(0)}\ .
\ee

On the other hand the PQ-symmetry 
can be used to derive a second
non-renormalization theorem.
The loop corrections of the prepotential $\cF$\footnote{Also often denoted 
in the literature by $h$ and similarly for its components to follow below;
we will follow also this very common notation.} 
are organized in 
an appropriate power series expansion in the dilaton.
The holomorphy of $\cF$
and the PQ-symmetry only allow a very limited number of terms, so
the prepotential $\cF$
only receives contributions at the string tree
level $\cF^{(0)}$ (of order $S$), at one-loop $\cF^{(1)}$ 
(dilaton independent) and non-perturbatively $\cF^{(np)}$
(only constrained by the discrete PQ-symmetry)
, i.e.
\bea\label{Floopexp}
\cF&=&\cF^{(0)} (S,M)\ +\ \cF^{(1)}(M)+\ \cF^{(np)}(e^{-8 \pi^2 S},M)\ .
\eea

\subsection{String Tree Level Prepotential}

The dilaton dependence of the tree level K\"ahler potential 
is constrained by the PQ-symmetry 
and the fact that the dilaton arises in the universal
sector. Therefore  it cannot  mix with any other scalar field
at the tree level and one necessarily has
\be
e^{-K^{(0)}}=(S+\bar{S})e^{-\check{K}(M,\bar{M})} .
\ee

This separation of the dilaton piece together with 
the constraint (\ref{Kspecial2})   uniquely fixes the tree level contribution 
to be \cite{FvP}:
\be\label{Ftreehet0}
\cF^{(0)}=- S M^i \eta_{ij} M^j=-S(TU-\phi^i\phi^j) ,
\ee
where 
\beqa
\eta _{ij}=\left( \begin{array}{ccccc}
0 & 1 &   &      &   \\
1 & 0 &   &      &   \\
  &   & -1&      &   \\
  &   &   &\ddots&   \\
  &   &   &      &-1 
\end{array} \right)
\eeqa
and the $\phi^i$ ($i=1,\cdots ,n_V-3$; we count the dilaton in the $n_V$
for convinience in the later comparison with the type IIA spectrum) are the
vector multiplet moduli of the factor $G^{\prime}$.
Inserted into  (\ref{Kspecial2}) 
the tree level K\"ahler potential is found to be
\bea
K^{(0)}=-\log(S+\bar{S})-\log(\R M^i \eta_{ij}   \R M^j). 
\eea
The metric derived from this K\"ahler potential
is the metric of the coset space 
\be
\cM_V^{(0)}=\frac{SU(1,1)}{U(1)}\times 
\frac{SO(2,n_V-1)}{SO(2)\times SO(n_V-1)}\ ,
\ee

(modulo discrete identifications related to the duality group)
where the first factor of the moduli space is 
spanned by the dilaton and the second factor
by the other vector multiplets. 
Thus one has generically a gauge group $U(1)^{n_V+1}$
(the "$+1$" coming from the graviphoton) 
which enhances at special loci.

There are two Abelian vector
multiplets $T$ and $U$ with moduli scalars
$T=\sqrt{g}+ib,U=g_{22}^{-1}(\sqrt{g}+ig_{12})$
of $T^2$ as well as further  model-dependent
Abelian or possibly  non-Abelian vector multiplets.
The total perturbative gauge group is
\begin{equation}
G= G'\times U(1)_S\times U(1)_T\times U(1)_U\times U(1)_\gamma ,
\end{equation} 
where $G'$ refers to the additional
Abelian or non-Abelian part of the gauge group
and $U(1)_{\gamma}$  corresponds to the graviphoton.\\
Note that one has the $T$-duality group
$Sl(2,{\bf Z})_T \times Sl(2,{\bf Z})_U \times {\bf Z}_2^{T \leftrightarrow U}$
as a symmetry of string perturbation theory, so the effective action 
should be invariant under it.\\
The rank of $G$ is bounded 
by the central charge  $\bar c_{int}=22$.
For $N=2$ vacua one has with the two additional $U(1)$ 
gauge bosons in the universal sector
corresponding to the graviphoton and the 
superpartner of the dilaton that
\be\label{rankhet}
{\rm rank}(G)\ \le\ 22+2\ .
\ee

To get more information about the spectrum 
let us consider now the heterotic string in 6D, i.e. on $K3$.
This supergravity is chiral and thus gauge and gravitational
anomaly cancellation imposes constraints on the allowed spectrum.
The anomaly can be characterized by 
the anomaly eight-form $I_8$, which is exact $I_8=dI_7$, given by  \cite{gaume}
\begin{equation}
I_8\, =\, a\, tr R^4 +\, b\, \left(tr R^2\right)^2 
+\, c\, tr R^2 tr F^2 +\, d\, \left(tr F^2\right)^2 ,
\end{equation}
where $R$ is the curvature two-form,
$F$ is the Yang-Mills two-form
and $a,b,c,d$ are real coefficients
which depend on the spectrum of the theory. 
The anomaly can only be cancelled if 
$a$ vanishes.
One finds \cite{gswest},\cite{schwarzanom}
\begin{equation}\label{eq:Rfourconstraint}
n_H-n_V=244.
\end{equation}

The remaining anomaly eight form has to factorize in order to employ a
Green--Schwarz mechanism. More precisely, one needs
$I_8 \sim  X_4 \wedge \tilde{X}_4$
($i$ labels the factors $G_i$ of the gauge group
$G= \otimes_i G_i$ and  $v_i, \tilde{v}_i$ are constants 
which depend on the
massless spectrum \cite{gswest},\cite{schwarzanom})
\be
X_4 = tr R^2 - \sum_i v_i\left( trF^2\right)_i\ ,\qquad
\tilde{X}_4 = tr R^2 - \sum_i \tilde{v}_i \left( tr F^2\right)_i\ .
\ee

In the Green--Schwarz mechanism one defines a modified field strength $H$
for the antisymmetric tensor
$H = dB +\omega^L - \sum_i v_i\, \omega_i^{YM}$
such that $dH = X_4$
($\omega^L$ is a Lorentz--Chern--Simons term and $\omega_a^{YM}$ 
is the Yang--Mills Chern--Simons term).

Vanishing of $\int_{K3} dH$ gives with $n_i=:\int_{K3}\left( trF^2\right)_i$
\begin{equation}\label{eq:instanton}
\sum_i n_i=\, \int_{K3}trR^2\, =\, 24
\end{equation}
so that one gets for instanton numbers $n_i$ ($i=1,2$) in the two $E_8$
\beqa
n_1+n_2=24.
\eeqa
The dimension (counted quaternionically, i.e. giving 
the number of hypermultiplets) of the moduli space of
instantons on $K3$ in $SU(2)$ of instanton number $n$ is given by 
\begin{equation}
dim {\cal M}_n =2n-3.
\end{equation}
For $n_i\geq 4$ one has the breaking $E_8\times E_8\rightarrow E_7\times E_7$
with $\frac{1}{2}n_i-2$ ${\bf 56}$'s and 62 singlets which consist of
$h^{1,1}=20$ universal hypermultiplets of K3 and the sector 
${\cal M}_{n_1}\times {\cal M}_{n_2}$. Complete Higgsing of the $E_7\times E_7$
is possible in the cases
$(n_1,n_2)=(12,12),(11,13),(10,14)$; one can show that the cases
$(12,12)$ and $(10,14)$ are equivalent \cite{AG,MV}.
This gives $(\frac{1}{2}24-4)\cdot 56-2\cdot 133=182$ further hypermultiplets
giving in total 244 which corresponds in view of the anomaly condition for
the spectrum to the fact that the gauge group is then completely broken.
Alternatively one could count instead the moduli of instantons in the group
$E_8$ (of dual Coxeter number $h=30$) from the beginning giving directly
the $30\cdot 24-2\cdot 248=224$ moduli.\\

We will study the following three cases

\begin{itemize}
\item
$ST$-model: $\;\;\;\;\;\;\;$ at $T=U$ with $(10,10;4)$ instanton embedding
\item
$STU$-model:$\;\;\;\;\;$ generic $T^2$ with $(10,14)$ resp. $(12,12)$ embedding
\item
$STUV$-model: generic $T^2$ with $(10,14)$ embedding and breaking only  
$~~~~~~~~~~~~~~~~~~~~~~ E_7^{(2)}\rightarrow SU(2)_V$
\end{itemize}

\subsubsection{STU model}

To consider this in grater detail start
with a compactification of the heterotic $E_8^{(1)} \times 
E_8^{(2)}$ string
on $K3$ with $SU(2)$ bundles with instanton numbers $(d_1,d_2)=(12-n,12+n)$
($n\ge 0$).
For $0\leq n\leq 8$, 
the gauge group is $E_7^{(1)}\times E_7^{(2)}$, and the spectrum of
massless hypermultiplets 
follows from the index theorem \cite{gswest,KV} as
\beqa
{1\over 2} (8-n) ({\bf 56},{\bf 1})+{1\over 2}(8+n)({\bf 1},{\bf 56})+
62({\bf 1},{\bf 1}).\label{hmspectrum}
\eeqa
For the standard embedding, $n=12$, 
the gauge group is $E_8^{(1)}\times E_7^{(2)}$
with massless hypermultiplets
\beqa
10({\bf 1},{\bf 56})+65({\bf 1},{\bf 1}).\label{hmspectrum1}
\eeqa
These gauge groups can be further broken by giving vev's to the charged
hypermultiplets. Specifically, $E_7^{(2)}$ can be completely
broken through the chain
\beqa
E_7\rightarrow E_6\rightarrow SO(10)\rightarrow SU(5)\rightarrow
SU(4)\rightarrow SU(3)\rightarrow SU(2)\rightarrow SU(1),\label{chain}
\eeqa
where $SU(1)$ denotes the trivial group consisting of the identity only.
In the following, we will concentrate on the cases where we 
break $E_7^{(2)}$ either completely or down to $SU(2)$.
On the other hand, $E_7^{(1)}$ can be perturbatively 
broken only to some terminal
group $G_0^{(1)}$ that depends on $n$
(see \cite{CF} for details); e.g. for $n=4$ this group is given by $G_0^{(1)}=
SO(8)$.
For $n=8$ it is $G_0^{(1)}=E_8$. It is only for $n=0,1,2$ that $E_7^{(1)}$ 
can be completely 
broken.
Finally, when 
compactifying to four dimensions on $T_2$, three additional vector
fields arise, namely the fields $S$, $T$ and $U$.

The enhancements in the $T,U$ moduli space arise 
along the critical line $T=U$ (mod $SL(2,{\bf Z})$)
where two additional massless gauge fields appear and 
the $U(1)_T\times U(1)_U$ is
enhanced to $SU(2)\times U(1)$.
Further enhancement appears at $T=U=1$, 
which is the intersection of the two critical 
lines $T=U$ and $T=1/U$. In this case one has $4$ 
extra gauge bosons and an
enhanced gauge group $SU(2)\times SU(2)$. 
The intersection at the critical point $T=U=\rho=e^{2\pi i/12}$ gives 
rise to $6$ massless gauge bosons 
corresponding to the gauge group $SU(3)$.
Altogether we have:
\begin{center}
\bt{ll}
$T=U$:& $\!\!\! U(1)_T\times U(1)_U \to SU(2)\times U(1)$\\
$T=U=1$:& $\!\!\! U(1)_T\times U(1)_U \to SU(2)\times SU(2)$\\
$T=U=\rho$:& $\!\!\! U(1)_T\times U(1)_U \to SU(3)$\\
\et
\end{center}

\subsubsection{ST model}

A variation of the theme is going in the moduli space of the $T^2$
(on which the sixdimensional theory on $K3$ is further compactified)
to the locus $T=U$ leading to an extra $SU(2)$ gauge symmetry which opens 
the possibility of a threefold distribution of the required 24 among the
gauge group factors in the form $(10,10;4)$ say, leading to
$n_H=2[(3\cdot 56-133)+(2\cdot 10 -3)]+(2\cdot 4 -3)+20=129$. 
Note that this is not
a simple truncation of the $STU$ model as the $SU(3)$ enhancement at
$T=U=\rho$ is lost and here only the possible further $SU(2)$ enhancement at
$T=U=i$ remains.

\subsubsection{STUV model}

We will be also interested in the case where the first $E_7$ is not completely
broken but a $SU(2)_V$ survives; here we let $V$ denote the fourth vector 
field besides the three fields $S,T,U$, i.e. the Wilson line in the 
Cartan subalgebra of $SU(2)^{(2)}$; so this model is connected to the pure 
$STU$-model by a Higgs transition. The commutant of $SU(2)^{(2)}$
in $E_7^{(2)}$ is $SO(12)^{(2)}$.
Then, it follows from the index 
theorem  that the charged spectrum consists of ${1\over 2}(8-n)$
${\bf 56}$ of $E_7^{(1)}$,  as well as of ${1\over 2}(8+n)$ ${\bf 32}$ of
$SO(12)^{(2)}$ plus
62 gauge neutral moduli.

As for the $STU$ models, it is only possible to perturbatively
higgs the $E_7^{(1)} \times SO(12)^{(2)}$ completely for $n=0,1,2$.
Thus, these heterotic models 
will have a massless spectrum comprising $n_V=4$
vector multiplets, $S$, $T$, $U$, $V$ (plus the graviphoton), as well as
\beqa
n_H = {1\over 2}(8+n) 32 - 66 + 
{1\over 2}(8-n) 56 -133 + 62 = 12 d_1 + 71 = 215 - 12n\label{numberhyper}
\eeqa 
neutral hypermultiplets. Note that, unlike in the $STU$ models with 
$n_H=244$, the number of hypermultiplets now depends on $n$. Furthermore,
as we will discuss,  for the four-parameter models also the
vector multiplet couplings are sensitive to $n$ 
already at the perturbative level.

Now consider the mentioned Higgs transition.
At the transition point $V=0$, the $U(1)$ associated with the 
Wilson line modulus $V$ becomes enhanced to an $SU(2)$.
Let $n_V'=2 $ and $n_H'$ denote the number of additional
vector- and hypermultiplets becoming massless at this transition point.  Then
\beqa
\frac{1}{2} \left( n_H' - n_V' \right) = 6n + 15 \;\;.
\label{nvnh}
\eeqa
This will prove to be a useful relation later on.
It follows from the fact that the Euler number of the Calabi--Yau
space $\chi(X_n)$ and of the $STU$ models ($\chi = -480$) differ by
$ 2(  n_H' - n_V' ) =\chi(X_n) + 480$.

Let us consider the question of gauge symmetry enhancement loci for the
$STUV$ model in more detail.
In addition to the $V=0$ locus of gauge symmetry enhancement, there
are also the enhancement loci (such as $T=U$), associated with the 
toroidal moduli $T$ and $U$, already known from the $STU$ model.
All these loci correspond to surfaces/lines of gauge symmetry enhancement in 
the heterotic perturbative moduli space 
${\cal H}_2 = \frac{SO(3,2)}{SO(3)\times SO(2)}$
and have a common description as follows.

Consider the Narain lattice $\Gamma=\Lambda\oplus U(-1)$ of signature $(3,2)$, 
where $U(-1)$ denotes the hyperbolic plane 
${\tiny
\left(\begin{array}{cc}  0&-1\\-1&0\end{array}\right)}$, and where
$\Lambda=U(-1)\oplus<2>={\tiny
\left(\begin{array}{ccc}0&-1&0\\-1&0&0\\0&0&2\end{array}\right)}$
in a basis that we will denote by $(f_2,f_{-2},f_3)$; we will use the 
coordinate $z=iTf_{-2}+iUf_2-iVf_3$ in $\Lambda\otimes {\bf C}$. Note here 
that the perturbative moduli space $\frac{SO(3,2)}{SO(3)\times SO(2)}$,
which is a Hermitian symmetric space, has a representation as a bounded domain
of type IV, that is, as a connected component of 
$ {\cal D}=\{[\omega]\in {\bf P}(\Gamma \otimes {\bf C})|\omega ^2=0,\;
\omega\cdot\bar{\omega}>0\}=\Lambda\otimes {\bf R}+iC(\Lambda)
\subset \Lambda\otimes {\bf C}$, where 
$C(\Lambda)=\{x\in \Lambda\otimes{\bf R}|x^2<0\}$
(this last condition ensures again that 
$2 {\rm Re}T\,{\rm Re} U-2({\rm Re}V)^2>0$; the connected component can then 
be realized as ${\cal D}^+=\Lambda\otimes {\bf R}+iC^+(\Lambda)$,
where $C^+(\Lambda)$ denotes the future light cone component of $C(\Lambda)$).

Now in the basis  
$\varepsilon_1=f_{-2}-f_2,\varepsilon_2=f_3,\varepsilon_3=f_2-f_3$, 
$\Lambda$ is equivalent to the intersection matrix 
$A_{1,0}={\tiny
\left( \begin{array}{ccc} 2&0&-1\\0&2&-2\\-1&-2&2\end{array}\right)}$
associated to the Siegel modular form ${\cal C}_{35}$ of \cite{GN2}.
To each element $\varepsilon_i$, which squares to $2$, is associated the 
Weyl reflection $s_i:x\rightarrow x-(x\cdot\varepsilon_i)\varepsilon_i$. The 
fixed loci of these Weyl reflections give the enhancement loci \cite{CLM2}.
As these reflection planes are given by planes orthogonal to the elements 
$\varepsilon_i$, this gives rise to the following loci: 
the orthogonality conditions $(a\varepsilon_1+b\varepsilon_2+c\varepsilon_3)
\varepsilon_i=0$ yield $c=2a$, $b=c$ and $a=2(c-b)$.  Since $a$, $b$ and
$c$ are related to $T$, $U$ and $V$ by 
$a=iT$, $b=iT+iU-iV$ and $c=iT+iU $, as can be seen by comparing 
$a\varepsilon_1+b\varepsilon_2+c\varepsilon_3=a(f_{-2}-f_2)+bf_3+
c(f_2-f_3)=af_{-2}+(c-a)f_2+(b-c)f_3$ 
with $z=iTf_{-2}+iUf_2-iVf_3$, the above orthogonality
conditions result in the enhancement loci $T=U$, $V=0$ and $T-2V=0$.
Note that these are the conditions for enhancement loci related to 
${\cal C}_{35}={\cal C}_{30}\cdot{\cal C}_{5}$ (cf. Appendix A2). 
Also note that the locus $T - 2V=0$ goes over into the locus $T-U=0$ 
under the target space duality transformation \cite{CLM1} 
$ T \rightarrow T + U + 2 V, U \rightarrow U, V \rightarrow V + U$.
Thus, the enhancement lines of the $STU$ 
model have become the Humbert surfaces $H_4$ and $H_1$
(see the discussion about 
rational quadratic divisors given in ch. 5 of \cite{HM} ($s=1$)
as well as in 
\cite{GN2}).\\

\subsection{Perturbative Corrections to the Prepotential}

We use the perturbative quantum symmetries 
and the singularity structure 
corresponding to the gauge symmetry enhancement loci
to determine the one-loop
correction of the gauge couplings.

\subsubsection{STU model}

Now $h^{(1)}(T,U)$ does have singularities inside
the fundamental domain. As mentioned there are additional (gauge neutral)
massless abelian vector multiplets on the subspace $T=U$,
which induce a logarithmic singularity
in the $U(1)$ gauge couplings, which are related to 
$h^{(1)}$ via its second derivatives.
So the 1-loop prepotential $h^{(1)}$ exhibits logarithmic singularities
exactly at the lines (points) of the classically enhanced gauge symmetries
and is therefore not a single valued function when transporting the moduli
fields around the singular lines \cite{AFGNT,CLM1,dWKLL}.
The ratio of the number of additional massless states 
is $1:2:3$, which leads in connection with the 
modular properties and the 
corresponding vanishing properties of the j-function at the special points 
to the occurence of ($r\in {\bf Z}_{\neq 0}$) a term
$\log (j(T)-j(U))^r)$ (cf. \cite{CLM1}).\\

$h^{(1)}$ had to be a modular form of weight $-2$ if it 
were nowhere  singular.
In the presence of singularities one has to allow
for integer ambiguities of the $\theta$-angles which results in
\be\label{htrans}
h^{(1)}(T,U)\to \frac{h^{(1)}(T,U)+\Xi(T,U)}{(i c T+d)^2}
\ee
for an $SL(2,{\bf Z})_T$ transformation.
For $SL(2,{\bf Z})_U$  $T$ and $U$
are interchanged in eq.~(\ref{htrans}).
$\Xi$, which is related to the monodromy around the enhancement loci,
is an arbitrary quadratic polynomial in the 
variables $(1,i T, i U, TU)$  and parameterizes
the most general allowed ambiguities in the
$\theta$-angles; so $\Xi$ obeys
$\p_T^3 \Xi =\p_U^3 \Xi=0$.
As said, for $\Xi =0$ 
$h^{(1)}$ is a modular  form
of weight $(-2,-2)$ with respect to 
$SL(2,{\bf Z})_T\times SL(2,{\bf Z})_U$
but for non-zero  $\Xi$ it has no good modular
properties; instead 
$\p_T^3 h^{(1)}$ is a single valued modular form of 
weight $(4,-2)$  and similarly 
$\p_U^3 h^{(1)}$ has weight $(-2,4)$.

The singularity of the prepotential near $T=U$ can be determined
by purely field theoretic considerations \cite{Seib}.
For the case at hand one finds \cite{dWKLL,Seib,SW}
\bea\label{hsing}
h^{(1)} (T\sim U)=
\frac{1}{16\pi^2}(T-U)^2\log(T-U)^2+{\rm finite}
\eea
where the coefficient $\frac{1}{16\pi^2}$ is set by the 
$SU(2)$ $\beta$-function. 
More precisely one finds from the number of additional massless states
the following behaviour of $2\pi {\cal F}^1$ at the various enhancement loci

\beqa
(T-U)^2\log(T-U)^2 \;\;\;\;\;\;\;\;\;\;\;\; T=U\;\;\;\;\;\;\; \\
(T-1)^2\log(T-1)^4 \;\;\;\;\;\;\;\;\;\;\;\; T=U=1 \;\\
(T-\rho)^2\log(T-\rho)^6 \;\;\;\;\;\;\;\;\;\;\;\; T=U=\rho .
\eeqa

This can be summarised in a covarient way as 

\beqa
{\cal F}_{TU}^1=-\frac{1}{4\pi ^2}\log(j(iT)-j(iU)) +\mbox{finite}
\eeqa

which furthermore can be thought of as the sum over massive states 
in the internal line for the gauge coupling ${\cal F}_{TU}^1=\log \det M=
\sum \log {\cal M}_{TU}$ (cf. section 5).
 
Since the moduli dependence of the gauge coupling
is related to the second derivative of $h$ 
and
in the decompactification limit ($T,U\to \infty$)
the gauge coupling cannot grow faster than a single
power of $T$ or $U$
one sees that 
$\p_T\p_U h^{(1)}$, $\p^2_T h^{(1)}$
and $\p^2_U h^{(1)}$ cannot grow faster than 
$T$ or $U$ in the decompactification limit.
Hence one has the asymptotic behaviour
$\p_T^3 h^{(1)}\to {\rm const.}$
for $T\to \infty$ (analogously with $T$ and $U$ interchanged).
The properties of  $h^{(1)}$ which we have
assembled so far can be
combined in the Ansatz 
\be\label{ansatz}
\p_T^3 h^{(1)}=\frac{X_{4,-2}(T,U)}{j(i T)-j(i U)} .
\ee
If one furthermore uses a factorisation ansatz for the numerator 
then its factors
cannot have any pole inside the modular domain 
while for large $T,U$ they have - divided by the $j$ function - to go 
to a constant, which 
properties uniquely determine the factors to be 
$E_4(iT)$ and $\frac{E_4 E_6}{\eta^{24}}(iU)$, so
\be
\p_T^3 h^{(1)}\ =\ \frac{1}{2\pi}
\frac{E_4(i T)\, E_4 (i U)\, E_6 (i U)}{[j(i T)-j(i U)]\, \eta^{24}(i U)}\ ,
\ee
where the coefficient is determined by eq.~(\ref{hsing}) or rather the 
$SU(2)$ $\beta$-function.
The same analysis holds for $T$ and $U$ interchanged.

In addition to the transformation law of $h$ (eq.~(\ref{htrans}))
also the $N=2$ dilaton is no longer invariant at the quantum level 
\cite{dWKLL}.
Instead, under an  $SL(2,{\bf Z})_T$ transformation one finds
\be\label{Strans}
S\to S+\hf\p_T\p_U \Xi- 
\frac{ic\p_U (h^{(1)}+\Xi)}{2(i c T+d)} 
 + {\rm const.}
\ee
This result can be understood from the fact that
in perturbative string theory the relation between the dilaton and the
vector-tensor multiplet is fixed. 
However, the duality relation between the vector-tensor multiplet
and its dual vector multiplet containing
$S$ is not fixed but suffers from perturbative corrections 
in both field theory and string theory. 
Nevertheless,
it is possible to define an invariant dilaton
\be\label{Sinv}
S^{inv}=S-\hf\p_T\p_U h^{(1)}
-\frac{1}{8\pi^2}\log[j(iT) - j(iU)]\ .
\ee
The last term is added such that $S^{inv}$ is finite
so that altogether
$S^{inv}$ is modular invariant and finite.
However,  $S^{inv}$
it is no longer an $N=2$ special coordinate.

The analysis just performed only determines the third derivatives
of $h^{(1)}$ because these are modular forms.
$h^{(1)}$ itself has been calculated in 
\cite{HM} by explicitly calculating
the appropriate string loop diagram.

It was shown in \cite{AGN,DKL} that threshold corrections in $N=2$
heterotic string compactifications can be written in terms 
of the supersymmetric index
\beqa
\frac{1}{\eta^2}\Tr_R F(-1)^F q^{L_0-c/24}\bar{q}^{\tilde{L}_0-\tilde{c}/24}
\;\;.
\label{susyindex}
\eeqa
This quantity is, as shown in \cite{HM}, also related to the computation of 
the perturbative heterotic $N=2$ prepotential. 
For the model with instanton number embedding
$(d_1,d_2)=(0,24)$, the supersymmetric index (\ref{susyindex})
was calculated in \cite{HM} and found to be equal to 
\begin{eqnarray}
\frac{1}{\eta^2}\Tr_R F(-1)^F q^{L_0-c/24}\bar{q}^{\tilde{L}_0-\tilde{c}/24}
=-2iZ_{2,2}\frac{E_4E_6}{\Delta} \;\;,
\label{sindhm}
\end{eqnarray}
where $Z_{2,2}$ denotes the sum over the Narain lattice $\Gamma^{2,2}$,
$Z_{2,2}=\sum_{p\in \Gamma^{2,2}}q^\frac{p_L^2}{2}\bar{q}^\frac{p_R^2}{2}$,
and where
$\frac{E_4E_6}{\Delta} = \sum_{n\ge -1}\tilde c_{STU}(n)q^n$. 
Here the subscript on 
the trace indicates the Ramond sector as right-moving boundary condition;
$F$ denotes the right-moving fermion number, $F=F_R$.\\
Let us recall how this expression came about. First, one can reduce 
(\ref{susyindex}) to 
$\frac{1}{\eta^2}\Tr_R (-1)^F q^{L_0-c/24}\bar{q}^{\tilde{L}_0-\tilde{c}/24}$,
where the contributions are weighted with $\pm 2\pi i$ depending on
whether a BPS hyper- or vector multiplet contributes.
The expression resulting from the evaluation of the trace consists of 
the product of three terms, namely of 
$Z_{2,2}/\eta^4$, of the partition function for the first 
$E_8^{(1)}$ in the bosonic 
formulation (leading to the contribution $E_4/\eta^8$),
and of the elliptic genus for the second $E_8^{(2)}$ 
containing the gauge connection on $K3$.\\
This last quantity now decomposes additively (taking into account the 
appropriate weightings) into contributions from the
following sectors, namely: 1) the 
$(NS,R)$ sector, which we will also denote by $(NS^+,R)$, 
2) the ``twisted" sector
$(NS^-,R)$, where a factor $(-1)^{F_L}$ is inserted in the trace 
(this contribution is weighted with $(-1)$) 
and 3) the $(R,R)$ sector, which we will also denote by $(R^+,R)$.
Since we are using the fermionic representation for $E_8^{(2)}$, we decompose
the fermionic $D_8^{(2)}\subset E_8^{(2)}$, so that
each of these contributions splits again multiplicatively into a 
free $D_6^{(2)}$ part and a $D_2$ part, to be called $D_2^{(2)}K3$, 
containing the gauge connection $A_1$ which describes the corresponding
gauge bundle on $K3$.  The corresponding contributions are 
summarized in the following table, where we also indicate the connection
to the generic elliptic genus 

$Z(\tau,z)=
\Tr_{R,R} y^{F_L}(-1)^{F_L+F_R}q^{L_o-c/24}\bar{q}^{\tilde{L}_o-\tilde{c}/24}=
6\frac{\th_2^2\th_4^2}{\eta^4}\frac{\th_3^2(\tau,z)}{\eta^2}
-2\frac{\th_4^4-\th_2^4}{\eta^4}\frac{\th_1^2(\tau,z)}{\eta^2}$,
where
$y={\bf e}[z]={\rm exp} 2\pi i z$  (see \cite{EOTY,KYY}).

\hspace {0.5cm}

\begin{center}
\begin{tabular}{|c||c|c|}\hline
Tr&$D_6$&$K3D_2$\\\hline\hline
$(NS^+,R)$&$\frac{\th_3^6}{\eta^6}$
           &$-2\frac{\th_4^4-\th_2^4}{\eta^4}\frac{\th_3^2}{\eta^2}
           =q^{\frac{1}{4}}Z(\tau,\frac{\tau +1}{2})$\\
$(NS^-,R)$&$\frac{\th_4^6}{\eta^6}$
           &$2\frac{\th_2^4+\th_3^4}{\eta^4}\frac{\th_4^2}{\eta^2}
           =q^{\frac{1}{4}}Z(\tau,\frac{\tau}{2})$\\
$(R^+,R)$ &$\frac{\th_2^6}{\eta^6}$           
           &$2\frac{\th_3^4+\th_4^4}{\eta^4}\frac{\th_2^2}{\eta^2}
           =Z(\tau,\frac{1}{2})$\\
$(R^-,R)$ &$\frac{\th_1^6}{\eta^6}=0$
           &$6\frac{\th_2^2\th_3^2\th_4^2}{\eta^4\cdot\eta^2}=24
           =Z(\tau,0)$\\ \hline
\end{tabular}
\end{center}

\hspace {0.5cm}

Now recall
that $E_4$ and $E_6$ have the following $\theta$-function decomposition:
\beqa
2 E_4
&=&\th_2^6\cdot\th_2^2 +\th_3^6\cdot\th_3^2 +
\th_4^6\cdot\th_4^2  \nonumber\\
2 E_6&=&-\th_2^6(\th_3^4+\th_4^4)\cdot\th_2^2+
\th_3^6(\th_4^4-\th_2^4)\cdot\th_3^2+
\th_4^6(\th_2^4+\th_3^4)\cdot\th_4^2  \;\; ;
\label{e4e6}
\eeqa
the $\theta_i^2$ 
contributions ($i=2,3,4$) are due to the $SO(4)$ piece in the 
fermionic decomposition of $E_8 \supset SO(12) \times SO(4)$.
Hence the sum of the three non-vanishing terms in the table 
precisely leads to (\ref{sindhm}).

On the other hand, in the case of a general $(d_1,d_2)$ embedding 
(using now a fermionic representation for both $E_8$'s),
one first has to decompose the $D_2^{(1)}K3D_2^{(2)}$ part
 into 
$D_2^{(1)}K3\times D_2^{{(2)},free}+D_2^{{(1)},free}\times K3D_2^{(2)}$, 
where the factors in each summand
are now in different, and hence commuting, $E_8$'s.  Furthermore, since
the rudimentary $K3$ gauge bundles are structurally completely the same
as before, the amount of contribution realized by them can --- by 
comparison with the ``complete" $K3$ bundle considered above --- be read off 
from the $R^-$ sector.  Note that $Z(\tau,0)$ is the Witten
index, which gives the Euler number of $K3$ resp. the second Chern class of
the relevant vector bundle. \\
This results in a contribution proportional to 
\begin{eqnarray}
\frac{1}{\Delta}(\frac{d_1}{24}E_6\cdot E_4+E_4\cdot \frac{d_2}{24}E_6)
=\frac{1}{\Delta}E_4E_6 
=\sum_{n\ge -1} c_{STU}(n)q^n=:f_{STU}(q)\;\;,
\end{eqnarray}
which is independent of the particular instanton embedding.

Via an explicit string 1-loop computation involving the quantity

\beqa
I\sim \int _{{\cal F}}\frac{d^2\tau}{\tau _2}
[\sum _{p \in \Gamma^{2,2}}
q^{\frac{p_L^2}{2}}\bar{q}^{\frac{p_R^2}{2}}]f_{STU}(q)
\eeqa

the semiclassical
prepotential can then be written in the following explicit form \cite{HM}
($y=(T,U)$, cf. Appendix A1)

\beqa
F &=& - STU + 
\frac{1}{384 \pi^2} \tilde{d}^{2,2}_{ABC} y^A y^B y^C\nonumber\\
  & &- \frac{1}{(2 \pi)^4} \sum_{k,l\geq 0} c_1(kl)
Li_3(e^{-2 \pi (k T + l U)})-{1\over (2\pi )^4}Li_3(e^{-2\pi (T-U)}),
\label{fhet}
\eeqa

$F$ has a branch locus at $T=U$.  $F$
given in (\ref{fhet}) is defined
in the fundamental Weyl chamber $T>U$ (meaning 
that the real part of $T$ is larger than the real part of $U$).
The cubic coefficients $\tilde{d}^{2,2}_{ABC}$ will be  
determined below.  We have ignored a possible constant term as well as
a possible additional quadratic polynomial in $T$ and $U$.
The cubic terms cannot be uniquely fixed, since the prepotential
contains an ambiguity \cite{AFGNT,dWKLL}
which is a quadratic polynomial in the
period vector $(1,T,U,TU)$. Hence, the ambiguity is at most quartic in the
moduli and at most quadratic in $T$ and in $U$. It follows
that the third derivative in $T$ or in $U$ is unique; 
${\partial^2 h^{(1)}\over\partial T\partial U}$, however, is still ambiguous.
Specifically, in the chamber $T>U$, the cubic terms have the following
general form \cite{HM}
\beqa
\tilde{d}^{2,2}_{ABC} y^A y^B y^C
=  -32\pi \left( 3(1+\beta)T^2U + 3\alpha T U^2 + U^3 \right).
\label{cubichet}
\eeqa
The cubic term in $U$ is unique, whereas the  parameters $\alpha$ and $\beta$
correspond to the change induced by adding a quadratic polynomial
in $(1,T,U,TU)$. So with our later favoured choice $\alpha =0,\beta =-1$
the cubic term will come out as $-\frac{1}{12\pi}U^3$.
As discussed 
it is convenient to introduce a dilaton field $S^{inv}$,
which is invariant under the perturbative $T$-duality transformations at the
one-loop level. It is defined as follows
\beqa
S^{inv}&=&S-{1\over 2}{\partial h^{(1)}(T,U)\over \partial T\partial U}
-{1\over 8\pi^2}\log(j(T)-j(U))\nonumber\\
&=&S+{1\over 4\pi}(1+\beta )T+{\alpha\over 4\pi}U
+ \frac{1}{8\pi^2} \sum_{k,l\geq 0} klc_1(kl)
Li_1(e^{-2 \pi (k T + l U)})\nonumber\\
& &-\frac{1}{8\pi^2} 
Li_1(e^{-2 \pi (T -  U)})
-{1\over 8\pi^2}\log(j(iT)-j(iU)).
\label{sinv}
\eeqa
In the decompactification limit 
to $D=5$ \cite{AFT}, obtained by sending $T,U\rightarrow\infty$ ($T>U$),
the invariant dilaton $S^{inv}$ has
a particularly simple dependence on $T$ and $U$.  Namely, by 
using $\log j(T)\rightarrow 2\pi T$, one obtains that 
\beqa
S^{inv}\rightarrow S^{inv}_{\infty}=S+{\beta\over 4\pi}T+{\alpha\over 4\pi}U ,
\label{largetsinv}
\eeqa
which gives for the mentioned later favoured choice $\alpha =0,\beta =-1$
\beqa
S^{inv}_{\infty}=S-\frac{1}{4\pi}T .
\eeqa

Substituting $S^{inv}_{\infty}$ back into the heterotic prepotential 
(\ref{fhet}) yields that 
\beqa
F &=&  - S^{inv}_{\infty} TU -{1\over 12\pi}U^3 - {1\over 4\pi}T^2 U 
  - \frac{1}{(2 \pi)^4} \sum_{k,l \geq 0} c_1(kl)
Li_3(e^{-2 \pi (k T + l U)}) \nonumber\\
& &- \frac{1}{(2 \pi)^4} 
Li_3(e^{-2 \pi (T - U)}).
\label{fheta} 
\eeqa
Note that the ambiguity in $\alpha$ and $\beta$
is hidden away in 
$S^{inv}_{\infty}$.

Let us make some remarks on a specific symmetry occuring here.
The string--string duality\cite{DK} in $d=6$ which 
is closely related to the duality of the heterotic string on $T^4$
and the type IIA on $K3$ manifests itself also
in $K3$ compactifications of the heterotic string.

For $n_1=n_2=12$ 
there is no strong coupling singularity
so that this class of heterotic vacua is well
defined for all values of the dilaton.
It has been shown 
that there is a selfduality among the
$n_1=n_2=12$ heterotic vacua \cite{AG,DuMinWit,MV}.
More precisely, 
the theory is invariant under 
\bea
D &\to & - D \ , \qquad
g_{\mu\nu}\to  e^{-D}g_{\mu\nu} \nonumber\\
H & \to & e^{-D} * H\ , \qquad
X_4 \leftrightarrow \tilde{X}_4
\eea
if in addition perturbative gauge fields
are replaced
by non-perturbative gauge fields
and the hypermultiplet moduli spaces are mapped
non-trivially onto each other;
note that for $n_1=n_2=12$ 
there is no strong coupling singularity
so that this class of heterotic vacua is well
defined for all values of the dilaton\footnote{This duality 
requires the existence of 
non-perturbative gauge fields with 
properties reminding one of 
$SO(32)$ heterotic strings.
This posed a slight puzzle since
for $E_8\times E_8$ vacua the singularity
of small instantons is caused by a non-critical tensionless string
rather than additional
massless gauge bosons. However, by mapping the
$n_1=n_2=12$ heterotic vacuum 
to a particular type I vacuum \cite{gimon}
which indeed does have non-perturbative gauge
fields this issue has been resolved and 
the gauge fields are  shown to exist
also for this class of heterotic vacua \cite{bele}.}

\subsubsection{ST model} 

Now one has (let $\check{S}=4\pi S$, so $q_{i\check{S}}=e^{-8\pi ^2 S}$)

\beqa
{\cal F}=\frac{1}{2}ST^2+h(T)+h^{np}(e^{-8\pi ^2 S},T) .
\eeqa

The T duality group $Sl_2({\bf Z})_T$ acts as $iT\rightarrow 
\frac{aiT+b}{ciT+d}$ with

\beqa
h\rightarrow \frac{h(T)+\Xi (T)}{(ciT+d)^4} ,
\eeqa

where the quartic polynomial $\Xi$ arises from the multivaluedness of $h(T)$,
i.e. in the absence of the logarithmic singularities $h$ would be a 
modular form of weight $-4$. $\partial ^5_T h$ is then a true modular form
of weight $6$ and the $U(1)_T$ factor of the 
gauge group $U(1)_T\times U(1)_S\times U(1)_{\gamma}$ is at $iT=i$
enhanced: $U(1)_T\rightarrow SU(2)_T$, so 

\beqa
\partial ^2_T h=-\frac{b_{SU(2)}}{8\pi ^2}\log (iT-i)+ \mbox{finite}
=\frac{2}{4\pi ^2}\log (iT-i) + \mbox{finite} ,
\eeqa

which gives\footnote{One can also
go on in an analogous way as for the $STU$ model now \cite{K2}.
But as, in contrast to the 
$STUV$ model with its very special function theory, the $ST$ and $STU$ models
have a lot of the formalism in common we will take in discussing both of these
models two different attitudes: whereas in the $ST$ model we will read of the 
mirror map from the instanton expansion in closed form - as $1/j$ - and then 
go on with closed expressions in the business of comparing couplings between 
the heterotic and the type II side, by contrast - to change methods - 
in the case of the $STU$ model
we will go on the heterotic side to the fully (and not only up to integrations
to be made) solved prepotential with its coefficients (in front of the 
trilogarithm terms) of the specifying modular form - the supersymmetric
index - being then later compared to the explicit instanton numbers on the
type IIA $CY$.}
by modularity ($j(iT)-j(i)$ vanishes to second order)

\beqa
\partial ^2_T h=\frac{1}{4\pi ^2}\log (j(iT)-j(i)) + \mbox{finite} .
\eeqa

Also one defines again an invariant dilaton

\beqa
S^{\rm inv}-S=\sigma=\frac{1}{3}(\partial ^2_T h-\frac{1}{4\pi ^2}
\log (j(iT)-j(i)) .
\eeqa

\subsubsection{STUV model}

In the presence of a Wilson line, which we will take to lay in the
second $E_8^{(2)}$, 
the symmetry between the two $E_8$'s is broken and thus,
contrary to the three-parameter case,  the prepotential
will already depend perturbatively on the type $(d_1,d_2)$ of the 
instanton embedding (we take $d_2\ge d_1$).\\
The supersymmetric index (\ref{susyindex}) will now have the form 
\beqa
\frac{1}{\eta^2}\Tr_R F(-1)^F q^{L_0-c/24}\bar{q}^{\tilde{L}_0-\tilde{c}/24}
=-2iZ_{3,2}(\tau, {\bar \tau}) \star F(\tau) \;\;,
\label{indexstuv}
\eeqa
where \cite{K1,NEU}
\beqa
Z_{3,2}(\tau, {\bar \tau}) \star F(\tau) = 
\left( \sum_{b \; even} q^\frac{p_L^2}{2}\bar{q}^\frac{p_R^2}{2}\right)
F_0 (\tau)
+ \left( \sum_{b \; odd} q^\frac{p_L^2}{2}\bar{q}^\frac{p_R^2}{2} \right) 
F_1(\tau) \;\;.
\eeqa
Here, $F(\tau) = F_0 (\tau) + F_1 (\tau)$, and
$Z_{3,2}$ denotes the sum over the Narain lattice $\Gamma_{3,2}$,
$Z_{3,2}=\sum_{p\in \Gamma^{3,2}}q^\frac{p_L^2}{2}\bar{q}^\frac{p_R^2}{2}
= \left( \sum_{b \; even} + \sum_{b \; odd} \right)
q^\frac{p_L^2}{2}\bar{q}^\frac{p_R^2}{2}$.

The presence of the Wilson line in $E_8^{(2)}$
has the following effect on the $\theta_i^2$ pieces
appearing in the decomposition (\ref{e4e6}) of $E_4$ and $E_6$
\beqa
2 \widehat{E_{4,1}(\tau,z)}
&=&\th_2^6\cdot \widehat{\th_2^2(\tau,z)} +\th_3^6\cdot 
\widehat{\th_3^2(\tau,z)}+
\th_4^6\cdot \widehat{\th_4^2(\tau,z)}  \;\;,  \\
2 \widehat{E_{6,1}(\tau,z)
}&=&-\th_2^6(\th_3^4+\th_4^4)\cdot\widehat{\th_2^2(\tau,z)}
+ \th_3^6(\th_4^4-\th_2^4)\cdot \widehat{\th_3^2(\tau,z)}+
\th_4^6(\th_2^4+\th_3^4)\cdot  \widehat{\th_4^2(\tau,z)} \;\;,\nonumber
\eeqa
where
\begin{eqnarray}
\widehat{\th_1^2(\tau,z)}&=&\th_2(2\tau)-\th_3(2\tau)\;\;, \nonumber\\
\widehat{\th_2^2(\tau,z)}&=&\th_2(2\tau)+\th_3(2\tau)\;\;,\\
\widehat{\th_3^2(\tau,z)}&=&\th_3(2\tau)+\th_2(2\tau)\;\;,\nonumber\\
\widehat{\th_4^2(\tau,z)}&=&\th_3(2\tau)-\th_2(2\tau) \nonumber
\end{eqnarray}
are the two $SU(2)$ characters of the surviving $A_1$ when written in
the boundary condition picture instead of the usual conjugacy class picture.
We refer to appendices \ref{thefjaf} and A.5 for a 
description and interpretation
of the hatting procedure.

The replacement $E_4  \rightarrow \widehat{E_{4,1}}$, in particular,
amounts to replacing the $E_8$ partition function 
$P_{E_8}=P_{E_7^{(0)}}\cdot P_{A_1^{(0)}}+P_{E_7^{(1)}}\cdot P_{A_1^{(1)}}$
with $P_{E_7^{(0)}}+P_{E_7^{(1)}}$.  This precisely describes the breaking of
the $E_8^{(2)}$ to $E_7^{(2)}\times U(1)$ when turning on a Wilson line.

Thus, the effect of turning on a Wilson line can be described as follows.
Introducing 
\begin{eqnarray}
F(\tau) \equiv 
A_n(\tau)=\frac{1}{\Delta} \left(\frac{d_1}{24}E_6\cdot\widehat{E_{4,1}}+
E_4\cdot\frac{d_2}{24}\widehat{E_{6,1}}\right) 
= \sum_{N \in  {\bf Z}, {\bf Z} + \frac{3}{4}}
c_n(4N) q^N 
\label{An}
\end{eqnarray}
as well as 
\beqa
F_0(\tau) &=& \sum_{N \in  {\bf Z}}
c_n(4N) q^N 
 \;\;, \nonumber\\
F_1(\tau) &=& \sum_{N \in  {\bf Z} + \frac{3}{4}}
c_n(4N) q^N  \;\;,
\eeqa
it follows that turning on a Wilson line results in the replacement
\beqa
Z_{2,2} 
\frac{1}{\Delta} E_4 E_6
 &\rightarrow & Z_{3,2} \star F = 
\left( \sum_{b \; even} q^\frac{p_L^2}{2}\bar{q}^\frac{p_R^2}{2}\right)
F_0 
+ \left( \sum_{b \; odd} q^\frac{p_L^2}{2}\bar{q}^\frac{p_R^2}{2} \right) 
F_1\;\;.
\eeqa
The product
$\tau_2 Z_{3,2} \star F$ 
is invariant under modular transformations \cite{K1,NEU}.

As discussed in the previous section, the supersymmetric index
is given in terms of 
\beqa
F(\tau) = A_n = \sum_{N \in  {\bf Z}, {\bf Z} + \frac{3}{4}}
c_n(4N) q^N \;\;.
\eeqa
As explained in appendix A3, the modular function 
$A_n(\tau)$ is in one-to-one correspondence with the index-one 
Jacobi form with the same expansion coefficients $c_n(k,b)=c_n(4k-b^2)$:
$A_n(\tau)=\widehat{A_n(\tau,z)}$, $A_n(\tau,z)={1\over \Delta(\tau)}
\left(\frac{d_1}{24}E_6(\tau)\cdot E_{4,1}(\tau,z)+
E_4(\tau)\cdot\frac{d_2}{24}E_{6,1}(\tau,z)\right)=
\sum_{k,b}c_n(4k-b^2)q^kr^b$. 
(The first few expansion coefficients of $A_0$, $A_1$, $A_2$ and $A_{12}$
are listed in the second table in appendix A7)

The expansion coefficients $c_n(4N)$ of $F(\tau)$  govern
the perturbative, i.e. one-loop,
corrections to the heterotic prepotential $F_0^{\rm het}$
\cite{HM}.   
For the class of $STUV$ models considered
here, the perturbative
heterotic prepotential is given by\footnote{It is shown in \cite{CCL} Appendix 
for the computation how the worldsheet expansion coefficients
$c_n(4N)$ turn into the target-space coefficients $c_n(4kl-b^2)$ appearing
in the prepotential cf. \cite{CCL} Appendix B} 
\beqa
F_0^{\rm het} &=& - S (TU - V^2) + p_n(T,U,V) - 
\frac{1}{4\pi^3}\sum_{k,l,b\in {\bf Z} \atop
(k,l,b)>0} c_n(4kl-b^2) Li_3({\bf e}[kiT+liU+biV]), 
\nonumber\\
\label{prepstuv}
\eeqa
where 
${\bf e}[x]={\rm exp} 2 \pi i x$.
The first term $ - S(TU-V^2)$ is the tree-level
prepotential of the 
special 
K\"ahler space ${SO(3,2)\over SO(3)\times SO(2)}$;
 $p_n(T,U,V)$ denotes the one-loop cubic polynomial that depends on the
particular instanton embedding $n$

\beqa
p_n(T,U,V)=-\frac{1}{12\pi}U^3+\frac{1}{4\pi}[(\frac{n_1}{2}-6)TV^2+
(\frac{n_1}{2}-5)UV^2-(\frac{4}{3}+n_1-12)V^3] \; .
\eeqa

The condition
$(k,l,b)>0$ means that: either $k > 0, l , b \in {\bf Z}$ or 
$k=0,l>0, b \in {\bf Z}$ 
or $k=l=0, b <0$ (cf. \cite{HM}).
Next, consider truncating an $STUV$ model
to the $STU$ model by setting $V=0$.  Then, the sum over $b$
in (\ref{prepstuv}) yields independently from $n$ the coefficients of the 
three-parameter model,
\beqa
c_{STU}(kl)=\sum_b c_n(4kl-b^2)  \;\;,
\eeqa
as is obvious by the dehatting procedure and
can be checked by explicit comparison.
Therefore the prepotential (\ref{prepstuv}) truncates correctly to the 
prepotential for the $STU$ model.

The (Wilsonian) 
Abelian gauge threshold functions are related (see \cite{dWKLL} for details)
to the second derivatives
of the one-loop prepotential 
$h(T,U,V) = p_n(T,U,V) - \frac{1}{4 \pi^3} \sum_{(k,l,b)>0} c_n(4kl-b^2)
Li_3({\bf e}[kiT+liU+biV])$. 
At the loci of enhanced non-Abelian gauge symmetries, some of the
Abelian
gauge couplings will exhibit logarithmic singularities due
to the additional massless states.
First, consider $\partial_T\partial_U h$.
At the line $T=U$  one $U(1)$ is extended to $SU(2)$ without
additional massless hypermultiplets. 
It can  be easily checked that, as $T \rightarrow U$,
\beqa
\partial_T \partial_U h = -\frac{1}{\pi} \log(T-U) \;\;,
\eeqa
as it should.
The Siegel modular form, which vanishes on the $T=U$ locus and has
modular weight $0$, is given by $\frac{{\cal C}^2_{30}}
{{\cal C}_{12}^5}$.
It can be shown that, as $V \rightarrow 0$,
\beqa
 \frac{{\cal C}^2_{30}}{{\cal C}_{12}^5} \rightarrow (j(T) - j(U))^2\;\;,
\eeqa
up to a normalization constant.
Hence one deduces that
\beqa
\partial_T \partial_U h =
-\frac{1}{2\pi} \log \frac{{\cal C}^2_{30}}{{\cal C}_{12}^5}+ {\rm regular}.
\label{tugaugecoupl}
\eeqa
On the other hand, at the locus $V=0$, a different $U(1)$ gets enhanced to
$SU(2)^{(2)}$, and at the same time $n_H'$ hypermultiplets, 
being doublets of $SU(2)^{(2)}$,
become massless. Using eq. (\ref{nvnh}), $n_V'=2$ and that
$2c_n(-1)=-n_H'\,,\, 2c_n(-4)=n_V'$, it can be checked that, as
 $V\rightarrow 0$,
\beqa
-{1\over 4}\partial_V^2 h  =
\frac{3}{2\pi} (2 + n) \log V=
-{1\over \pi}
\left(1-{1\over 8}n_H'\right)\log V \;\;.
\eeqa
Observe that the factor $(1-{1\over 8}n_H')$ is precisely given
by the $N=2$ $SU(2)$ gauge $\beta$-function coefficient with 
$n_H'/2$ hypermultiplets in the fundamental representation of $SU(2)$.
The Siegel modular form, which vanishes on the $V=0$ locus and has
modular weight $0$, is given by $\frac{{\cal C}_5}
{{\cal C}_{12}^{5/12}}$.
It can be shown that, as $V \rightarrow 0$, 
\beqa
{\cal C}_5 \rightarrow V \left(\Delta(T) \Delta(U) \right)^{\frac{1}{2}}
\;\;,
\eeqa
So we now conclude that
\beqa
-{1\over 4}\partial_V^2 h = \frac{3}{4\pi} (2 + n) \log \left(\frac{{\cal C}_5}
{{\cal C}_{12}^{5/12}}\right)^2+ {\rm regular}.\label{vgaugecoupl}
\eeqa

\newpage

\subsection{Higher gravitational couplings}

Up to now we concentrated on the gauge couplings,
which are lowest order (two derivatives) couplings, 
because of their special analytic properties.
In addition,
there is a class of higher derivative curvature terms in the effective action
\cite{AGNTtop,BCOV,CdWLMR}
whose couplings $g_n$ are also determined
by holomorphic functions $\cF_n(S,M^i)$ of the vector moduli. The
prepotential $\cF$ as well as these higher derivative couplings 
$\cF_n$ arise from chiral integrals (F-terms) in 
$N=2$ superspace. They occur at 1-loop on the heterotic
side; on the type II side they arise at n-loop level, each representing a 
topological partition function of the twisted Calabi-Yau sigma model.
The higher derivative couplings of vector multiplets $X$
to the Weyl multiplet ${\cal W}$ of conformal $N=2$
supergravity can be expressed as a power series \cite{AGNT}
\beqa
F(X,{\cal W}^2)=\sum_{g=0}^\infty F_{g}(X)({\cal W}^2 )^g.\label{expansion}
\eeqa

Specifically for $n=1$ we will consider a  term ($R$ is the Riemann tensor)
\be\label{gndef}
\cL \sim g_1^{-2} R^2 + \ldots\ ,
\ee 
of almost harmonic coupling $g_1$ with
\be\label{Fndef}
g_1^{-2}=\R \cF_1(S,M^i)+\cA_1 .
\ee
At tree level $A_1=0$  
thus $g_1^{-2}$ is a harmonic function. 
However exactly as for the gauge coupling 
the $g_1^{-2}$ cease to be harmonic as soon as
quantum corrections are included, i.e.
a holomorphic anomaly $A_1\neq 0$ is induced \cite{BCOV}.

Thus, one has analogously 
\bea
\cF_1=\cF_1^{(0)}(S,M^i)+\cF_1^{(1)}(M^i)+\ \cF_1^{(np)}(e^{-8 \pi^2 S},M^i)\ .
\eea
 
Furthermore in  a convenient normalization 
\be\label{Ftreehet1}
\cF_1^{(0)}=24\, S .
\ee

At 1-loop one has 

\begin{eqnarray*}
g_1^{-2}=\mbox{Re}F_1+\frac{b_{grav}}{16\pi ^2}(\log \frac{M^2_{Pl}}{p^2}+K) ,
\end{eqnarray*}

which leads to the anomaly 

\begin{eqnarray}
\partial _i\bar{\partial}_{\bar{j}}g_1^{-2}=\frac{b_{grav}}{16\pi ^2}
\partial _i\bar{\partial}_{\bar{j}}K ,
\end{eqnarray}

where\footnote{$\chi$ the Euler characteristic of the dual Calabi-Yau.}
$b_{grav}=2(n_H-(n_V-1)+22)=2((n_H-1)-n_V+24)=48-\chi$ is the 
1-loop coefficient of the 'gravitational' beta-function.

\subsubsection{ST model}

One has 

\beqa
F_1=24S+h_1(T)+\mbox{non-pert.}
\eeqa

and at $T\sim 1$ the singular contribution to $h_1(T)$ coincides with the
one for $\partial ^2_T h$ as no additional gauge singlets become massless

\beqa
h_1=\frac{1}{4\pi ^2}\log (j(iT)-j(i))+\mbox{finite} .
\eeqa

On the other hand the modular transformation properties of $h_1$ follow from
the holomorphic/modular anomaly of this coupling, namely

\beqa
g_1^{-2}=\mbox{Re}F_1+\frac{b_{grav}}{16\pi ^2}(\log \frac{M^2_{Pl}}{p^2}
+K(S,T))
\eeqa

has to be modular invariant. Here one has contrary to the tree level 
Kaehler potential $e^{-K}=(S+\bar{S})(T+\bar{T})^2$ that at one loop

\beqa
e^{-K}=(S+\bar{S})(T+\bar{T})^2-(4(h+\bar{h})-2(T+\bar{T})(\partial _T h+
\partial _{\bar{T}}\bar{h})) ,
\eeqa

i.e.

\beqa
K=-\log(S+\bar{S}-V_{\rm GS})-\log(T+\bar{T})^2
\eeqa

with the Green-Schwarz term $V_{\rm GS}=\frac{4(h+\bar{h})-2(T+\bar{T})
(\partial _T h+\partial _{\bar{T}}\bar{h})}{(T+\bar{T})^2}$.\\
Now with the compensation ansatz

\beqa
F_1=24S^{\rm inv}+\frac{1}{4\pi ^2}\log (j(iT)-j(i))+\frac{x}{4\pi ^2}
\log \frac{1}{\eta^2(iT)}
\eeqa

one gets from the invariance requirement for 
$\mbox{Re}F_1+\frac{b_{grav}}{4\pi ^2}\frac{1}{4}K$ in view of the facts
that $\frac{1}{4}K=\log (2\mbox{Im}(iT))^{-1/2}+stuff$ and 
$\mbox{Im}(iT)$ transforms as absolute value of a modular form of weight
$-2$ that $x=b_{grav}$, i.e.

\beqa
F_1=24S^{\rm inv}+\frac{1}{4\pi ^2}\log (j(iT)-j(i))+
\frac{b_{grav}}{4\pi ^2}\log \frac{1}{\eta^2(iT)} .
\eeqa

\subsubsection{STU model}

At the one-loop level $F_1$ is given by \cite{CLM1,C,KLTh}
\beqa
F_{1} = 24 S^{inv} + 
\frac{1}{4 \pi^2} \log(j(iT)-j(iU))^2 +
\frac{\frac{1}{2}b_{grav}}{4 \pi^2} \log \frac{1}{\eta^2(iT)\eta^2(iU)} .
\label{fgrav}
\eeqa

Note that the square in $(j(iT)-j(iU))^2$ comes from the fact that now
the vanishing of $j(iT)-j(iU)$ in $T-U$ is of first order 
in contrast to the situation in the $ST$ model where $j(iT)-j(i)$ 
vanished in $iT-i$ to the second order. Furthermore the relative $\frac{1}{2}$ 
factor in front of $b_{grav}$ comes from the fact 
that it would on the diagonal exponentiate
the term $\eta ^2 \eta ^2$.\\

For the model we are discussing one has that $b_{grav}=48-\chi=528$.
Later we will see that the one-loop $F_1$ can be written
as a sum over the $N=2$ BPS states.\\

For later use let us also rewrite the result in the variable 
$\bar{y}=e^{-8\pi ^2 S^{\rm inv}}$ 

\beqa
F_1=\frac{6}{4\pi ^2}\log [\bar{y}^{-2}\frac{(j(iT)-j(iU))^{2/6}}
{(\eta ^2(iT)\eta ^2(iU))^{\frac{1}{2}b_{grav}/6}}] \; .
\eeqa

Inserting $S^{inv}$ given in (\ref{sinv}) into
$F_{1}$ yields \cite{CCLMR} 
\beqa
F_{1} &=& 
24 \left( S -\frac{1}{768 \pi^2} \partial_T \partial_U
\left( \tilde{d}^{2,2}_{abc} y^a y^b y^c \right)
- \frac{1}{8 \pi^2} \log (j(T) - j(U)) \right. \nonumber\\
 & &\left. +\frac{1}{8 \pi ^2}
\sum_{k,l \geq 0} klc_1(kl)  Li_1(e^{-2\pi (kT+lU)})  
- \frac{1}{8 \pi^2} Li_1(e^{-2\pi (T-U)})  
\right) \nonumber\\
& & + \frac{1}{4 \pi^2} \log(j(T)-j(U))^2 
+\frac{b_{grav}}{8 \pi^2} \log \frac{1}{\eta^2(T)\eta^2(U)} \nonumber\\
&=& 24  S +{12+6\beta\over\pi}T+{11+6\alpha\over\pi}U
 \nonumber\\
& &-\sum_{k,l}({3\over \pi^2} klc_1(kl) -{5\over2\pi^2} 
c(kl)) \log(1-e^{-2\pi (kT+lU)})    
 \nonumber\\
& &+\frac{1}{4 \pi^2} \log(j(T)-j(U))^2
+\frac{b_{grav}}{8 \pi^2} \log \frac{1}{\eta^2(T)\eta^2(U)} .
\label{fexp}
\eeqa
Thus, $F_1$ is essentially determined by two types of coefficients of
modular functions, namely $c_1(n)$ and $c(n)$.\\

At last let us for convenience of extending the preceeding results to
the $STUV$ model in the next subsection make a little reshuffeling of
the terms.\\
First the perturbative Wilsonian gravitational coupling for the $STU$ model
is given by\footnote{For convenience of 
notation involving the quantity $\tilde{S}$ we will denote here and in
the $STUV$ subsections (3.3.3) and (3.2.3) $\check{S}$ already by $S$, i.e. 
the dilaton is defined to be $S=4 \pi/g^2 - 
i \theta/2 \pi$.} (in the chamber $T > U$)
\beqa
F_1 &=& 24 S^{inv} - \frac{b_{grav}}{\pi} \log \eta(T) \eta(U)
+ \frac{2}{\pi} \log (j(T)-j(U)) \;\;.
\label{f1stu}
\eeqa
Using that \cite{dWKLL}
\beqa
S^{inv} &=& {\tilde S} + \frac{1}{8} L \;\;, \nonumber\\
{\tilde S} &=& S - \frac{1}{2} \partial_T \partial_U h \;\;\; , \;\;\;
L=-\frac{4}{ \pi} \log(j(T)-j(U)) \;\;,
\eeqa
it follows that $F_1$ can be rewritten as 
\beqa
F_1 = 24 \tilde{S} - \frac{1}{\pi} \Big[10 \log (j(T) - j(U))
+ b_{\rm grav} \log \eta(T) \eta(U) \Big] \;\;.
\eeqa
The perturbative 
gravitational coupling is related to the perturbative Wilsonian coupling
by
\beqa
\frac{1}{g^2_{\rm grav}} &=& \mbox{Re} F_1 + \frac{b_{grav}}{4 \pi} K 
= 12(S + {\bar S} + V_{GS}) + \Delta_{\rm grav} \;\;.
\label{fgrav}
\eeqa
This relates the Wilsonian gravitational coupling $F_1$ to the
supersymmetric index, that is to 
$\Delta_{\rm grav}= - \frac{2}{4 \pi} {\tilde I}_{2,2}$ \cite{CCLMR},
where \cite{HM}
(note that $b_{grav}=48-\chi=-2(c_1(0)-24)=-2\tilde{c}_1(0)$)
\beqa
{\tilde I}_{2,2} = \int_{\cal F} \frac{d^2 \tau}{\tau_2}
\Big[ Z_{2,2} \frac{E_4E_6}{\eta^{24}} \left(E_2 - \frac{3}{\pi \tau_2}\right)
- {\tilde c}_1(0) \Big]
\;\;.
\eeqa
It follows from (\ref{fgrav}) that 
\beqa
F_1 &=& 24 S - \frac{2}{\pi} \sum_{r >0} \tilde{c}_1 (-\frac{r^2}{2}) Li_1
\nonumber\\
&=& 24 \tilde{S} - \frac{1}{\pi} 
 \Big[10 \log (j(T) - j(U))
+ b_{\rm grav} \log \eta(T) \eta(U) \Big] \;\;.
\label{stuf1}
\eeqa

Here, we have ignored the issue of ambiguities in (\ref{stuf1})
linear in $T$ and in $U$.

\subsubsection{STUV model}

The classical moduli space of a heterotic $STUV$ model
is locally given by the Siegel upper half plane ${\cal H}_2 = 
\frac{SO(3,2)}{SO(3) \times SO(2)}$. Because of target space 
duality invariance, 
one has to consider modular forms on ${\cal H}_2$, i.e.
Siegel modular forms (cf. appendix A).

The Siegel modular form, which vanishes on the $T=U$ locus and has
modular weight $0$, is given by $\frac{{\cal C}^2_{30}}
{{\cal C}_{12}^5}$.
It can be shown that, as $V \rightarrow 0$,
\beqa
 \frac{{\cal C}^2_{30}}{{\cal C}_{12}^5} \rightarrow (j(T) - j(U))^2\;\;,
\eeqa
up to a normalization constant. On the other hand, 
the Siegel modular form, which vanishes on the $V=0$ locus and has
modular weight $0$, is given by $\frac{{\cal C}_5}
{{\cal C}_{12}^{5/12}}$.
It can be shown that, as $V \rightarrow 0$, 
\beqa
{\cal C}_5 \rightarrow V \left(\Delta(T) \Delta(U) \right)^{\frac{1}{2}}
\;\;,
\eeqa
up to a proportionality constant. Finally, the Siegel form ${\cal C}_{12}$ 
generalizes $\Delta(T) \Delta(U)$, that is
\beqa
{\cal C}_{12} \rightarrow \Delta(T) \Delta(U)
\eeqa 
as $V \rightarrow 0$. \\
Then, in analogy to (\ref{f1stu}), the perturbative Wilsonian gravitational
coupling for an $STUV$ model is now given by (in the chamber $T> U$)
\beqa
F_1 &=& 24 S^{inv} - \frac{b_{grav}}{24 \pi} \log {\cal C}_{12}
+ \frac{1}{\pi} \log \frac{{\cal C}^2_{30}}{{\cal C}_{12}^5} 
- \frac{1}{2 \pi} (n_H'-n_V') \log \left(
\frac{{\cal C}_5}{{\cal C}_{12}^{5/12}}\right)^2 \;\;.
\label{f1stuv}
\eeqa
Here, $n_V'$ and $n_H'$ denote the vector and the hypermultiplets,
which become massless at the $V=0$ locus. Since at $V=0$ there is 
a gauge symmetry restoration $U(1) \rightarrow SU(2)$, we have
$n_V'=2$.\\
The invariant dilaton $S^{inv}$ is given by \cite{dWKLL}
\beqa
S^{inv} &=& {\tilde S} + \frac{1}{10} L \;\;, \nonumber\\
{\tilde S} &=& S - \frac{4}{10}\Big(\partial_T \partial_U - 
\frac{1}{4} \partial_V^2\Big) h \;\;,
\eeqa
where the role of the quantity $L$ is to render $S_{inv}$ free of 
singularities. Using eqs. (\ref{tugaugecoupl}) and (\ref{vgaugecoupl}),
it follows that
\beqa
{\tilde S} = S + \frac{1}{5 \pi} \log \frac{{\cal C}^2_{30}}
{{\cal C}_{12}^5} 
- \frac{3}{10 \pi} (2 + n) \log \left(
\frac{{\cal C}_5}{{\cal C}_{12}^{5/12}}\right)^2
+ {\rm regular}
\eeqa
and, hence,
\beqa
L=-\frac{2}{\pi} \log \frac{{\cal C}^2_{30}}{{\cal C}_{12}^5} 
+ \frac{3}{ \pi} (2 + n) \log \left(\frac{{\cal C}_5}
{{\cal C}_{12}^{5/12}}\right)^2 \;\;.
\eeqa
It follows that the Wilsonian gravitational coupling (\ref{f1stuv})
can be rewritten into
\beqa
F_1 &=& 24 {\tilde S} - \frac{1}{\pi} \Big[
\frac{19}{5} \log \frac{{\cal C}^2_{30}}{{\cal C}_{12}^5} + 
\frac{b_{grav}}{24} \log {\cal C}_{12} \nonumber\\
&+& 
\left( - \frac{72}{10}(2+n) + \frac{1}{2}(n_H'-n_V') \right)
 \log \left(\frac{{\cal C}_5}{{\cal C}_{12}^{5/12}}\right)^2 \Big] \;\;.
\label{f1tildev}
\eeqa
Now recall from (\ref{nvnh}) that $n_H'-n_V'= 12n + 30$.  Inserting 
this into (\ref{f1tildev}) yields 
\beqa
F_1 &=& 24 {\tilde S} - \frac{1}{\pi} \left(
\frac{19}{5} \log {\cal C}^2_{30} +  
  \frac{3}{5}(1-2n) 
 \log {\cal C}_5^2 \right) \;\;.
\label{f1tildestuv}
\eeqa
Note that the $\log {\cal C}_{12}$ terms have completely cancelled out!

Now consider the perturbative 
gravitational coupling, which is again related to 
the perturbative Wilsonian coupling by
\beqa
\frac{1}{g^2_{grav}} &=& \mbox{Re} F_1 + \frac{b_{grav}}{4 \pi} K 
= 12(S + {\bar S} + V_{GS}) + \Delta_{grav} \;\;,
\label{fgravv}
\eeqa
where this time
$\Delta_{grav}= - \frac{2}{4 \pi} {\tilde I}_{3,2}$ with
\beqa
{\tilde I}_{3,2} = \int_{\cal F} \frac{d^2 \tau}{\tau_2}
\Big[ Z_{3,2}\star A_n \Big( E_2 - \frac{3}{\pi \tau_2}\Big)
- d_n(0) \Big]
\;\;.
\label{intws}
\eeqa
Here, we have introduced
\beqa
B_n(\tau) &=& E_2 A_n = \frac{12-n}{24} 
\frac{E_2 E_6 \widehat{ E_{4,1}}}{\Delta} +
\frac{12+n}{24} 
\frac{E_2 E_4 \widehat{ E_{6,1}}}{\Delta} =
\sum_{N \in {\bf Z} \, or \, {\bf Z}+\frac{3}{4}} d_n
(4N) q^{N} \;\;. \nonumber\\
\eeqa
The worldsheet integral (\ref{intws}) can be evaluated 
(cf. \cite{CCL}, App. B)
using the
techniques of \cite{DKL,HM,K1,K2,NEU}. Then we find, 
from (\ref{fgravv}),
that
\beqa
F_1 &=& 24 S - \frac{2}{\pi} \sum_{(k,l,b)>0} 
d_n(4kl-b^2) Li_1 \nonumber\\
&=& 24 {\tilde S} - \frac{1}{\pi} \left(
\frac{19}{5} \log {\cal C}^2_{30}
+ \frac{3}{5} (1 - 2 n)  \log {\cal C}_5^2 \right) \;\;.
\label{cons}
\eeqa

\newpage

\section{The dual Calabi-Yau spaces}

\resetcounter

The massless spectrum of a type II vacuum
compactified on a Calabi--Yau threefold $Y$ 
is characterized by the Hodge numbers 
$h_{1,1}$ and $h_{1,2}$.
For type IIA one finds \cite{CFG,Sei}
$n_V=h_{1,1},\ n_H=h_{2,1}+1$ 
where the extra hypermultiplet counts 
the dilaton multiplet of the universal sector.
For type IIB vacua one has 
$n_V=h_{2,1}$ and  $n_H=h_{1,1}+1$. So the role of $h_{1,1}$ and $h_{1,2}$
is interchanged between type IIA 
and type IIB. Note that type IIA on 
a Calabi--Yau threefold $Y$ is by mirror symmetry equivalent to 
type IIB on the mirror Calabi--Yau $\tilde Y$.

The gauge group is always Abelian and given by $n_V+1$ $U(1)$ factors 
(the extra $U(1)$ is the graviphoton in the universal sector). So

\begin{itemize}
\item
(IIA) $n_V=h^{1,1}$ and $n_H-1=h^{2,1}$ 
\item
(IIB) $n_V=h^{2,1}$ and $n_H-1=h^{1,1}$ 
\end{itemize}

and in both cases one has an additional hypermultiplet corresponding to
the type II dilaton. So for type II vacua ${\cal M}_V$ is dilaton independent
and therefore given by the uncorrected tree level result. 
This tree level prepotential can be computed exactly, i.e. including 
worldsheet instanton corrections, by using mirror symmetry and it shows 
logarithmic singularities (at tree level !) along the conifold locus;
the latter behaviour can be understood \cite{S} as an 1-loop effect
(thanks to the specific string coupling behaviour of RR-fields) due to 
the appearence (at the conifold locus) of massless hypermultiplets 
corresponding to charged black holes in the internal line. \\
More specifically let us consider the IIA string on a Calabi-Yau space X at 
the large radius limit.
 ${\cal M}_V$ is described by
the K\"ahler moduli,i.e. an element in moduli space is given by the 
(complexified) K\"ahler form $B+iJ\in H^2(X,{\bf C})$, which expands in
a basis $(e_{\alpha})$ of integral cohomology $H^2(X,{\bf Z})$ as

\begin{eqnarray*}
B+iJ=\sum_{\alpha}^{h^{1,1}}t_{\alpha}e_{\alpha}  ,
\end{eqnarray*}

where in the complex parameters $t_{\alpha}=B_{\alpha}+iJ_{\alpha}$ (they
are $N=2$ special coordinates) the $B_{\alpha}$ are the moduli corresponding 
to the antisymmetric tensor and $J_{\alpha}$ the moduli which
are associated to the deformation of the metric.
In the large radius limit the moduli $B_{\alpha}$ 
are periodic variables and thus enjoy a discrete 
`PQ-like' symmetry. More precisely, one finds
that the low energy effective theory is invariant under 
$B_{\alpha}\to B_{\alpha}+1$ or equivalently $t_{\alpha}\to t_{\alpha}+1$.

For the holomorphic prepotential (at large radius) one has then

\begin{eqnarray}
{\cal F}=-\frac{i}{6}\sum(D_{\alpha}\cdot D_{\beta}\cdot D_{\gamma})
t_{\alpha}t_{\beta}t_{\gamma}+\frac{1}{(2\pi )^3}\sum _{(d_i)_{i=1,\cdots ,n}}
n_{d_1,\cdots ,d_n}Li_3(\prod _{i=1}^n q_i^{d_i}) \; ,
\end{eqnarray}

with $D_{\alpha}$ the associated divisor to $e_{\alpha}$ and in the 
additionional term involving
the contribution of worldsheet instanton corrections, which vanish
in the limit $t_{\alpha}\rightarrow \infty$,
the $n_{d_1,\cdots ,d_n}$ denote the rational instanton numbers,
which are computed via mirror symmetry ($q_j=e^{2\pi (it_j)})$. 
The coefficients of the cubic part are the 
classical intersection numbers of the $(1,1)$ forms defined by 
$\int_{Y} e_{\alpha}\wedge e_{\beta}\wedge e_{\gamma}$.

The higher derivative couplings $\cF_n$ which 
were defined in eqs.~(\ref{gndef}),(\ref{Fndef})
and which similarly only depend on the vector multiplets,
also obey the type II non-renormalization theorem.
For these couplings one finds that they are proportional to the 
genus $n$ 
topological partition functions of a
twisted Calabi--Yau  $\sigma$--model \cite{BCOV}.
They receive a contribution only at 
fixed string loop order n.
In the large radius limit one has
\beqa\label{FCYhigher}
F_1^{II}=- i\sum_{i=1}^{h}
t_i c_2\cdot  J_i-{1\over\pi}\sum_n\biggl\lbrack
12n_{d_1,\dots ,d_h}^e\log(\tilde\eta(\prod_{i=1}^hq_i^{d_i}))+
n_{d_1,\dots ,d_h}^r\log(1-\prod_{i=1}^hq_i^{d_i})\biggr\rbrack.\label{ftop}
\eeqa
Here $\tilde\eta(q)=\prod_{m=1}^\infty(1-q^m)$, and
the $n_{d_1,\dots ,d_h}^e$
denote the elliptic genus one instanton numbers; 
$c_2$ is the second Chern class of the 
Calabi--Yau manifold.

Now in tests of the proposed string/string duality \cite{KV} one compares
the heterotic ${\cal M}_V$ at weak coupling with the exact ${\cal M}_V$ of the
type II vacuum. In particular one identificaties the heterotic
dilaton with one of the vector moduli, say $t_S=(B+iJ)_S$, of the type IIA
string, namely\footnote{For the normalization compare the shift symmetries
$B_S\rightarrow B_S+1$ and
$\theta \rightarrow \theta +2\pi$} $t_S=4\pi iS$. \\
Now in the considered examples of dual pairs \cite{KV},\cite{KLM} a
certain preference of the dual Calabi-Yau space to have the structure of a
K3 fibration (over $P^1({\bf C})$) was observed - first "experimentally", then
made plausible by noting the property of $D_S$ having zero selfintersection
by \ref{Ftreehet0}, so being fibrelike \cite{KLM}, then
at last stated under precise conditions \cite{AL} using in particular the 
additional property $D_S\cdot c_2(X)=24=\chi_{K3}$ coming from 
equ. (\ref{Ftreehet1}).

By computing the spectrum for suitable choices of the gauge bundles one
gets predictions for the hodge numbers of the dual Calabi-Yau leading to
the proposal of the spaces (the upper postscript indicates the
Euler number, the lower ones the $h^{1,1}$ and $h^{2,1}$) 

\begin{itemize}
\item
$ST$ model:$\;\;\;\;\;\;\; P_{1,1,2,2,6}(12)^{-252}_{2,128}$
\item
$STU$ model:$\;\;\;\; P_{1,1,2,8,12}(24)^{-480}_{3,243}$
\item
$STUV$ model:$\; P_{1,1,2,6,10}(20)^{-372}_{4,190}$
\end{itemize}

\newpage

\subsection{ST model}

The typical defining polynomial

\begin{eqnarray}
p=z_1^{12}+z_2^{12}+z_3^6+z_4^6+z_5^2
\end{eqnarray}

leads to the singular curve C 
(the singularity coming
from the common factor of the weight indices)
of genus 2 (as being isomorphic to
$P_{2,2,6}(12)=P_{1,1,3}(6)$, a 2-fold covering of $P^1=P_{1,1}$
 with 6 branch points) on the Calabi-Yau

\begin{eqnarray}
z_3^6+z_4^6+z_5^2=0\;\;, \;\; z_1=z_2=0
\end{eqnarray}

which is resolved by pointwise insertion of $P^1$'s leading to an 
exceptional divisor E representing a ruled surface over the curve C.
The projective ratio of the first two coordinates gives a base $P^1$
coordinate with K3 fibre (as being a double cover of $P^2$ branched 
along a sextic) $P_{1,2,2,6}(12)=P_{1,1,1,3}(6)$  for our Calabi-Yau space.
Let us for later use introduce the notation L for that fibre class
(the cohomology class of the divisors of the corresponding linear system
$|L|$ defined by the degree 1 polynomials (generated by $z_1$ and $z_2$))
and similarly let us denote the linear system of degree 2 polynomials
by $|H|$. As quadratic polynomials in $z_1$ and $z_2$ are the subsystem
of $|H|$ vanishing on C one has $H=2L+E$. L and H are the relevant
generators of $h^{1,1}$, whose topological intersection numbers will
give later important information concerning the matching of quantities  
in establishing tests of string duality. \\
Note that the mirror, which is \cite{GP} a certain orbifold of the 
generic element in the subclass of spaces

\begin{eqnarray}
p=z_1^{12}+z_2^{12}+z_3^6+z_4^6+z_5^2-12\psi z_1z_2z_3z_4z_5-2\phi z_1^6z_2^6
\end{eqnarray}

(having now only the 2 shown complex deformations but on the other hand
128 independent divisors from resolution of the generated orbifold
singularities, i.e. the mirror exchange of hodge numbers has taken place),
has a conifold singularity along $\Delta _{\mbox{Can}}=(8\psi ^4+\phi)^2-1=0$,
a more complicated singularity along $\phi ^2-1=0$, which we call in view of 
later considerations the strong coupling singularity; let us call (
only slightly abusing the terminology of \cite{CdlOFKM}) these loci
$C_{con}$ and $C_1$ (there are two further discriminant loci \cite{CdlOFKM}, 
of which we want only to mention the locus $C_0$ corresponding to $\psi =0$).\\

The  mirror map for the two complex
structure deformations of the mirror Calabi-Yau coming from

\begin{eqnarray*}
p=z_1^{12}+z_2^{12}+z_3^6+z_4^6+z_5^2+a_0z_1z_2z_3z_4z_5+a_1z_1^6z_2^6
\end{eqnarray*}

is given by ($q_j=e^{2\pi it_j}$)

\begin{eqnarray*}
x=j(q_1)^{-1}+O(q_2), y=g(q_1)q_2+O(q_2^2)
\end{eqnarray*}

(with uniformizing variables at large complex structure
$x=a_1a_0^{-6}$,$y=a_1^{-2}$; also used are the parameters
$\bar{x}=\alpha x,\bar{y}=4y$ with $\alpha =12^3=1728=j(i)$. The parameters of
\cite{KLTh} and \cite{CdlOFKM} are related by $a_0=-12\psi , a_1=-2\phi $).
Note that the discriminant is essentially given by 

\beqa
\Delta=(1-\bar{x})^2-\bar{x}^2\bar{y} .
\eeqa

The investigation of the mirror map proceeds by investigating the 
Picard-Fuchs equations satisfied by the cohomology classes when 
transported along a way in  moduli space surrounding the discriminant
divisors; the corresponding monodromy considerations are relevant when
identifying the large complex structure limit to find the correct 
variables, in which the mirror is to be expressed. According to the 
"regularity theorem" the relevant differential equation describing the 
transport has regular singular points, so that in a suitable basis the 
corresponding matrix has at worst first order poles, whose residue
matrix has rational eigenvalues resp. in fact integer eigenvalues for 
unipotent monodromy. This leads to unique identification of large complex 
structure limit point in (blown up) moduli space \cite{CdlOFKM}. \\
The modular property of the mirror map (at $\bar{y}=0$)
$\bar{x}=\frac{\alpha}{j(q_1)}$
was noted first experimentally \cite{CdlOFKM}, then explained \cite{KLM}
on the basis of the property of the Calabi-Yau being a K3-fibration
together with the known reduction of the mirror map for certain
1-parameter families of K3 to the elliptic case. \\

Then with the expression for the mirror map at weak coupling
($S\approx \infty$, i.e. $y\approx 0$)

\beqa
x=\frac{1}{j(q_1)} \nonumber\\
y=q_2g(q_1)\nonumber
\eeqa

one gets \cite{HKTY} that at weak coupling (so $\Delta=(1-\bar{x})^2$)

\beqa
K_{\bar{x}\bar{x}\bar{x}}=\frac{1}{4\bar{x}^3}\frac{1}{\Delta}=
\frac{j^5}{4\alpha ^3}\frac{1}{(j-\alpha)^2}\nonumber\\
K_{\bar{x}\bar{x}\bar{y}}=\frac{1-\bar{x}}{2\bar{x}^2\bar{y}}\frac{1}{\Delta}=
\frac{j^3}{2\alpha ^2}\frac{1}{j-\alpha}\frac{1}{\bar{y}}\nonumber
\eeqa

respectively with $\partial _{\tau}\bar{x}=-\alpha \frac{j_{\tau}}{j^2}$,
$\partial _{\tau}\bar{y}=\bar{y}\partial _{\tau}\log g$
(listing only the terms contributing in the limit $y\rightarrow 0$ (cf.
\cite{CdlOFKM}))(cf. appendix A1)

\beqa
K_{\tau \tau \tau}
&=&(\frac{d\bar{x}}{d\tau})^3 K_{\bar{x}\bar{x}\bar{x}}+
3K_{\bar{x}\bar{x}\bar{y}}(\frac{d\bar{x}}{d\tau})^2(\frac{d\bar{y}}{d\tau})
\nonumber\\
 &=&-\frac{j_{\tau}^3}{4j(j-\alpha)^2}+
\frac{3}{2}\frac{j_{\tau}^2}{j(j-\alpha)}\partial _{\tau}\log g\nonumber\\
 &\sim &-\frac{j_{\tau}^2}{j(j-\alpha)}[\frac{j_{\tau}}{j-\alpha}-
\frac{3}{2}\partial _{\tau}\log g]\nonumber\\
 &=&4\pi ^2 E_4\partial _{\tau}[\log (j-\alpha)-\frac{3}{2}\log g]\nonumber
\eeqa

or (in $\tau=t_1=iT$)
\beqa
K_{TTT}\sim -4\pi ^2 E_4\partial _T[\log (j-\alpha)-\frac{3}{2}\log g]\nonumber
\eeqa

so that after going to the special gauge where ${\cal F}_0=1$, i.e.
after dividing by $X_0^2$ (which corresponds here to the square of the
fundamental period: $\omega _0^2$, which in turn equals at weak coupling
$E_4(iT)$ (cf. \cite{LY})) and taking a suitable overall normalization

\beqa
\partial ^3_T {\cal F}=\frac{1}{4\pi ^2}\partial _T[\log (j-\alpha)-
\frac{3}{2}\log g]
\eeqa

so
\beqa
\partial ^2_T {\cal F}=S+\frac{1}{4\pi ^2}\log (j-\alpha)-
\frac{3}{2}\frac{1}{4\pi ^2}\log g .
\eeqa

Now concerning the $F_1$ function
by the holomorphic anomaly we will get that (up to an additive constant)
\beqa
F_1^{II}=\log [y^{-\alpha}\frac{(j-\alpha)^{\beta}}{\eta ^2(t_1)^{\gamma}}] .
\eeqa

Namely 
\begin{eqnarray*}
F_1^{II}=\log[(\frac{12\psi}{\omega _0})^{5-\frac{\chi}{12}}
\frac{\partial(\psi ,\chi )}{\partial (t_1 ,t_2)} f ]
\end{eqnarray*}

with $f=\Delta _{\rm {Can}}^a (\phi^2 -1)^b \psi ^c$ ,
$\Delta _{\rm Can}=\frac{\Delta}{\bar{x}^2\bar{y}}
=\frac{(1-\bar{x})^2}{\bar{x}^2\bar{y}}=\bar{y}^{-1}\alpha ^{-1}(j-\alpha)^2$,
$\phi ^2-1=\bar{y}^{-1}-1$,
$\psi \sim y^{-\frac{1}{12}}\frac{E-4^{1/2}}{\eta ^4}$,
$\frac{\partial(\psi ,\phi )}{\partial (t_1 ,t_2)} \sim 
y^{-1}\frac{j_{\tau}}{\psi ^5}$ and $c=1$ and 
$\omega_0$ the fundamental period (note that $\omega _0^2 | _{y=0}=E_4$).\\
One can obtain the values $a=-\frac{1}{6}, b=-\frac{2}{3}$ by
comparing in the large radius limit with the topological intersection
numbers \cite{CdlOFKM}.
This leads in total to the announced expression of $F_1^{II}$ with
\begin{eqnarray*}
\alpha &=& 1+a+b+\frac{1}{12}(c-\frac{\chi}{12})=2\\
\beta  &=& 2a+\frac{1}{2}=\frac{1}{6}\\
\gamma &=& 2(c+3-\frac{\chi}{12})=\frac{1}{6}b_{grav} \; ,
\end{eqnarray*}
i.e.

\beqa
F_1^{II}=
\log [y^{-2}\frac{(j-\alpha)^{\frac{1}{6}}}
{\eta ^2(t_1)^{\frac{1}{6}b_{grav}}}] .
\eeqa

For later use let us come back to the mentioned asymptotic evaluations
\beqa
\alpha=\frac{1}{12} c_2\cdot L=\frac{1}{12} c_2(L)=\frac{1}{12} \chi _{K3}
\eeqa

and\footnote{in going to the second line we used the relation 
$E^3=-8=4\chi_C$ between the singular curve C
of the Calabi-Yau and the ruled surface E of its pointwise resolution
\cite{CdlOFKM} which gives $c_2 \cdot H=\frac{1}{12}[2\chi _{K3}+c_2(E)-E^3]=
\frac{1}{12}[2\chi _{K3}+2\chi _C-4\chi _C]$ and in going to the third line we
used that because of the $K3$-fibration
$\chi =(\chi _{P1}-12)\chi _{K3}+12(\chi _C +1)$ one has $-\frac{\chi}{12}=
-4+\chi _{K_3}-\chi _C -1$ ( the $+1$ comes from the possibility of having
$z_3=z_4=z_5=0$)}
\begin{eqnarray*}
\beta +\frac{\gamma}{12}&=&\frac{1}{12}c_2 \cdot H\\
                        &=&\frac{1}{12}2(\chi _{K3}-\chi _C) \\ 
                        &=&\frac{1}{12}2(4-\frac{\chi}{12}+1) \; ,
\end{eqnarray*}

so (as $2(4-\frac{\chi}{12})(12\frac{\beta}{\gamma}+1)=
\gamma (12\frac{\beta}{\gamma}+1)=2(4-\frac{\chi}{12})+2$)
\beqa
\frac{\gamma}{\beta}=48-\chi=b_{grav}
\eeqa

showing also $\frac{\alpha}{\beta}=\frac{\chi _ {K3}}{2}$.\\

Furthermore the consideration of the monodromy around the conifold\footnote{
Actually one considers a slightly modified quantity; namely one
has still to correct by the (simple structured) monodromy around $C_0$,
which just corresponds to the operation $(\psi ,\phi)\rightarrow (\mu \psi,
-\phi)$.} in its operation on the periods leads via mirror map to
its operation on the coordinates $t_1,t_2$ in $H^2=H^{1,1}$ (remember
$B+iJ=t_1H+t_2L$). This in turn leads then again to an corresponding 
operation on $H_2$, i.e. on classes of the embedded instanton
worldsheets. This leads to the symmetry $n_{jk}=n_{j,j-k}$ in the 
instanton expansions around the large complex structure limit point 
of the Yukawa couplings (which of course in the end are to be
identified (after a certain 'gauging' by the fundamental period) with
the corresponding third derivatives of the prepotential on the heterotic
side \cite{KV}), for example in \cite{CdlOFKM}
($q_r=e^{2\pi it_r}$,$F_{abc}^0$ the topological intersection numbers) 
\begin{eqnarray}
F_{111}=F_{111}^{(0)}+\sum _{j,k} j^3n_{jk}\frac{q_1^jq_2^k}{1-q_1^jq_2^k}
\end{eqnarray}

These coefficients are similarly relevant in the later considered 
expression (with the modified Dedekind function $\eta (q)=\prod
_{n=1}^{\infty}(1-q^n)$ and the genus one instanton contributions $d_{jk}$
besides the usual genus zero contributions)

\begin{eqnarray}
F_1^{top}=-\frac{2\pi i}{12}c_2\cdot (B+iJ)-\sum_{j,k}[2d_{jk}\log \eta 
(q_1^jq_2^k)+\frac{n_{jk}}{6}\log (1-q_1^jq_2^k)]
\end{eqnarray}

This transformation - in the end caused by the conifold monodromy ! - has
led to the proposal \cite{KLM} that - as the symmetry in the instanton
coefficients translates to the symmetry $q_1\rightarrow q_1q_2,
q_2\rightarrow 1/q_2$ in the expansion variables - one has actually
to identify $q_2=q_S/q_T$ 
leading to the nonperturbative exchange symmetry S-T ! \\
As for later use in tests of S-T exchange symmetry let us state a further 
property of the $n_{jk}$ and at the same time let us look at the problem 
from a more explicit point of view, which makes the instanton properties 
- as they are the cornerstone of the nonperturbative exchange symmetry - 
more directly visible (cf. \cite{CdlOFKM}). \\
To understand $n_{jk}=n_{j,j-k}$ let $f:X\rightarrow X^{\prime}$ the 
resolution of the Calabi-Yau with $f:E\rightarrow C$ the contained resolution 
of the singular curve by the surface E as $P^1$ bundle over C, so the fibers
of E over C are smooth rational curves $l$ with $l\cdot E=l\cdot (H-2L)=-2$ 
(as $l\cdot H=0,l\cdot L=1$) and let
$\Gamma \subset X$ an irredicuble rational 
embedded instanton curve. If
$\Gamma \not\subset E$ (otherwise $\Gamma$ is one of the
fibers $l$), then $\Gamma$ intersects E in $n_{\Gamma}=
\Gamma \cdot E\ge 0$ points $p_i$ and taking the union of $\Gamma$ with the 
fibers $l_i$ through these points one gets a connected curve
$\Gamma ^{\prime}$ homologous to $\Gamma +nl$,a (contributing) degenerate
instanton. Now if you already started with a connected degenerated instanton
$\Gamma$, where the possibility to contain (at an intersection point with E)
the corresponding fibre is realized for a subset of the intersection 
points ($\Gamma$ containing, say, m fibers as components, so 
$\tilde{\Gamma}=\Gamma-\sum_{i=1}^ml_i$ is still connected, $m\le n=
n_{\tilde{\Gamma}}\ge 0$), throwing away these fibers and plugging in the 
fibers at the other intersection points will give you an involution 
$\Gamma \rightarrow \Gamma ^{\prime}$ on the
set of degenerate instantons (with the exception of the fibers)
, where $\Gamma ^{\prime}$, which is homologous to $\Gamma +(\Gamma \cdot E)l=
\Gamma +((\tilde{\Gamma}+ml)\cdot E)l=\Gamma +(n-2m)l=\Gamma -ml+(n-m)l$ 
(showing here something like the mirror reflected Picard-Lefshetz monodromy),
contributes now to $n_{j,j-k}$ when $\Gamma$ contributed to $n_{jk}$:

\begin{eqnarray}
\Gamma ^{\prime} \cdot H &=& \Gamma \cdot H+0=j \\
\Gamma ^{\prime} \cdot L &=& \Gamma \cdot L+(\Gamma \cdot E)l\cdot L=
k+\Gamma \cdot E\nonumber\\
 &=& k+\Gamma \cdot (H-2L)=k+j-2k=j-k
\end{eqnarray}

Similarly, to understand the later also used property  $n_{jk}=2\delta_{j0}
\delta_{k1}$ for $j>k$, note that $n_{01}=2=-\chi _C$ and that 
(let $r:=\tilde{\Gamma} \cdot L$) for $j>0$ one has $k=\Gamma \cdot L=
\tilde{\Gamma} \cdot L+ml\cdot L=r+m\le j-r\le j$ as
$0\le m\le \tilde{\Gamma} \cdot E=\tilde{\Gamma} \cdot (H-2L)=j-2r$. \\
  
\newpage

\subsection{STU model}

The defining polynomial of $P^4_{1,1,2,8,12}(24)^{-480}_{3,243}$ is

\begin{eqnarray*}   
p=z_1^{24}+z_2^{24}+z_3^{12}+z_4^3+z_5^2+a_0z_1z_2z_3z_4z_5+a_1(z_1z_2z_3)^4
+a_2(z_1z_2)^{12} \; ,
\end{eqnarray*}

which shows already the deformations of the mirror Calabi-Yau 
with uniformizing variables at large complex 
structure (also used are the parameters 
$a_0=-12\psi_0,\, a_1=-2\psi_1,\, a_2=-\psi_2$ 
and $\bar{x} =\frac{\alpha}{4}x, \; \bar{y} =4y, \;
 \bar{z} =4z $, where $\alpha = j(i) = 1728$)

\begin{eqnarray*}
x=-\frac{2}{\alpha^2}\psi_0^{-6}\psi_1, \;
y=\psi_2^{-2}, \;
z=-\frac{1}{4}\psi_1^{-2}\psi_2 \; .
\end{eqnarray*}

For later use let us also record the inverse relations
\begin{eqnarray*}
\psi_0 &\sim& \bar{y}^{-\frac{1}{24}}(\bar{x}^2\bar{z})^{-\frac{1}{12}}  \\
\psi_1 &\sim& (\bar{y}^{-\frac{1}{2}}\bar{z}^{-1})^{\frac{1}{2}}  \\
\psi_2 &\sim& \bar{y}^{-\frac{1}{2}}.
\end{eqnarray*}

Like \cite{CdlOFKM} I will use discriminant factors shifted
compared to \cite{HKTY,KLM} (we are  in the limit $\bar{y}\rightarrow 0$)

\begin{eqnarray*}
\tilde{\Delta}_1 &=& \bar{y}^{-1}(\frac{1-\bar{z}}{\bar{z}})^2  \\
\tilde{\Delta}_2 &=& \bar{y}^{-1}(\bar{x}^2\bar{z})^
{-2}((1-\bar{x})^2-\bar{x}^2\bar{z})^2  \\
\tilde{\Delta}_3 &=& \frac{1-\bar{y}}{\bar{y}}\approx\bar{y}^{-1} \; .
\end{eqnarray*}

In the  limit $\bar{y}\rightarrow 0$ the mirror map is given by 
($q_1:=q_T, \; q_3:=q_U/q_T; \; j:=j(iT), \; k:=j(iU)$) \cite{KLM,LY}

\begin{eqnarray*}
\bar{x} &=& \frac{\alpha}{2}\frac{j+k-\alpha}{jk+\sqrt{j(j-\alpha)}
                     \sqrt{k(k-\alpha)}}
                =q_1+\sum_{m+n>1}a_{mn}q_1^mq_3^n  \\
\bar{y} &=& q_sf_y(q_1,q_3)+O(q_2^2)\qquad\qquad
                =q_2\sum_{m+n\ge 1}c_{mn}q_1^mq_3^n+O(q_2^2) \\
\bar{z} &=& (\frac{\alpha}{2})^2\frac{1}{jk\bar{x}^2}\qquad\qquad\qquad\qquad
               =q_3+\sum_{m+n>1}b_{mn}q_1^mq_3^n \; .
\end{eqnarray*}

One has $t_1=t_T=iT, \; t_2=4\pi iS, \; t_3=t_U-t_T,\;t_U=iU$;
so $q_s=e^{-8\pi^2S}$, where S is the tree-level dilaton of the
heterotic string, and analogously to the $ST$ case again
$\bar{y}=e^{-8\pi^2S^{inv}}$ with the modular
invariant dilaton \cite{dWKLL,KLTh}. \\
Note that in the divisor picture one has here besides the hyperplane section
$J$ first again a ruled surface $E$ corresponding to the necessary resolution
caused by the common factor $2$ of the last three weights, and then still
a further divisor $F$ corresponding to the higher common divisibility of the 
last two weights by $4$ introducing the Hirzebruch surface $F_2$ over which
the resolved $CY$ is eliptically fibered (the trace of which projection map
in the space $P_{1,1,2,8,12}(24)$ is seen by projection to the first three
coordinates). One has then here the relation $J=4L+2E+F$ with the $K3$ fibre
class $L$. \\

The Yukawa couplings are given by
\cite{HKTY}
\beqa
F^{II}_{klm} = {\cal F}^0_{klm} + \sum_{d_1,...,d_h} 
\frac{n^r_{d_1,...,d_h}d_k d_l d_m}{1-\prod_i^h q_i^{d_i}} 
\prod_{i=1}^h q_i^{d_i} ,
\eeqa
where $q_i = e^{- 2 \pi t_i}$.  The ${\cal F}^0_{klm}$ denote the
intersection numbers, whereas the $n^r_{d_1,...,d_h}$ denote the rational 
instanton numbers of genus zero.  These instanton numbers are expected to be 
integer numbers.  We will, in the following, work inside the K\"ahler cone 
$\sigma(K)=\{\sum_i t_i J_i | t_i > 0\}$. For points inside the K\"{a}hler 
cone $\sigma(K)$, one has for the degrees $d_i$ that $d_i \geq 0$.
Integrating back yields that
\beqa
{F}_0^{II} = {\cal F}^0 - \frac{1}{(2 \pi)^3} \sum_{d_1,...,d_h}
n^r_{d_1,...,d_h} Li_3(\prod_{i=1}^hq_i^{d_i})
\label{ftype2}
\eeqa
up to a quadratic polynomial in the $t_i$. ${\cal F}^0$ is cubic in the $t_i$.
Thus we have three K\"ahler moduli $t_1$, $t_2$, $t_3$ and 
instanton numbers $n^r_{d_1,d_2,d_3}$. The classical Yukawa couplings 
${\cal F}^0_{klm}$ on the type II side are given by \cite{HKTY}
\beqa
{\cal F}^0_{t_1t_1t_1}&=& 8 \;\;,\;\; 
{\cal F}^0_{t_1t_1t_2}= 2 \;\;,\;\;
{\cal F}^0_{t_1t_1t_3}= 4 \;\;,\;\; \nonumber\\
{\cal F}^0_{t_1t_2t_3}&=& 1 \;\;,\;\;
{\cal F}^0_{t_1t_3t_3}= 2 .
\eeqa
It follows that 
\beqa
{\cal F}^0 = \frac{4}{3} t_1^3 + t_1^2 t_2 + 2 t_1^2 t_3 + t_1 t_2 t_3 +
t_1 t_3^2 .
\label{fzero}
\eeqa
Some of the instanton numbers $n^r_{d_1,d_2,d_3}$ can be found in 
\cite{HKTY}. When investigating the prepotential $F_0^{II}$ \cite{KLM},
two symmetries  become manifest, namely
\beqa 
t_1\rightarrow t_1+t_3,\; t_3\rightarrow -t_3
 \qquad {\rm for}\quad t_2=\infty ,\label{tu}
\eeqa
and 
\beqa
t_2\rightarrow -t_2,\; t_3\rightarrow t_2+t_3.\label{st}
\eeqa
These symmetries are true symmetries of $F_0^{II}$, since the world-sheet 
instanton numbers $n^r$ enjoy the remarkable properties \cite{KLM}
\beqa
n^r_{d_1,0,d_3}=n^r_{d_1,0,d_1-d_3} \qquad {\rm and}\quad 
n^r_{d_1,d_2,d_3}=n^r_{d_1,d_3-d_2,d_3}.\label{inssym}
\eeqa
Observe that ${\cal F}^0$ is completely invariant under the
symmetry (\ref{st}).\\

Concerning $F_1$
one gets from the holomorphic anomaly of $F_1$
\cite{BCOV} that up to an additive constant (cf. \cite{CdlOFKM,C,KLTh})

\begin{eqnarray*}
F_1^{II} = \log[(\frac{24\psi_0}{\omega_0})^{3+3-\frac{\chi}{12}}
\frac{\partial(\psi_0,\psi_1,\psi_2)}{\partial(T,S,U)}f]
\end{eqnarray*}

with a holomorphic function 
$f=\tilde{\Delta}_1^a \tilde{\Delta}_2^b \tilde{\Delta}_3^d \psi_0^c $
with $c=3$
and the fundamental period at weak coupling (as in all what follows)
$\omega_0^2=(E_4(T)E_4(U))^{\frac{1}{2}}$.\\
 
One again gets the values $a=b=-\frac{1}{6},d=-\frac{1}{2}$ by
comparing in the large radius limit with the topological intersection numbers.
Furthermore one can eventually
find the crucial relation (cf. \cite{C}) 
$\tilde{\Delta}_1\tilde{\Delta}_2 \sim \bar{y}^{-2}(j-k)^4$,
which shows that generically there are no more complicated enhancement
loci than those corresponding to $T=U$. Then one gets with
$f=\psi _0^3 (\tilde{\Delta}_3)^{-1/2}(\tilde{\Delta}_1\tilde{\Delta}_2)^{-1/6}
\sim \bar{y}^{\frac{17}{24}}(jk)^{\frac{1}{4}}(j-k)^{-\frac{2}{3}}$,
$\psi _0 \sim \bar{y}^{-\frac{1}{24}}(jk)^{\frac{1}{12}}$,
$\frac{\psi _0}{\omega_0}\sim \bar{y}^{-\frac{1}{24}}
\frac{1}{\eta ^2(t_T)\eta ^2(t_U)}$,
$\frac{\partial(\psi_0,\psi_1,\psi_2)}{\partial(T,S,U)} \sim
\bar{y}^{-\frac{3}{4}}\frac{j-k}{\sqrt{jk}}\omega_0^2\psi _0$ 
that the factors besides $\bar{y}$ powers and $(j-k)$ powers 
collect (with $\tilde{\psi _0} :\sim (jk)^{\frac{1}{12}}$) as follows:
$(\frac{\tilde{\psi _0}}{\omega_0})^{6-\frac{\chi}{12}}
\frac{\omega_0^2 \tilde{\psi _0}^4}{\sqrt{jk}}=
(\frac{\tilde{\psi _0}}{\omega_0})^{4-\frac{\chi}{12}}=
(\frac{1}{\eta ^2(t_T)\eta ^2(t_U)})^{\frac{1}{12}b_{grav}}$,
so that one has finally

\beqa
F_1^{II}=\log [\bar{y}^{-2}\frac{(j-k)^{1/3}}
{(\eta ^2(t_T)\eta ^2(t_U))^{b_{grav}/12}}] \; .
\eeqa

\subsection{STUV model}

In the dual type IIA description, based
on compactifications on four-parameter Calabi--Yau threefolds $X_n$,
the Euler numbers are $\chi(X_n) = 2(h_{1,1} - h_{2,1}) = 24 n -420$
and the Hodge numbers are given by $h_{1,1} = n_V = 4$, $h_{2,1}= n_H-1=
214 -12 n$. 
For the Hodge numbers compare  the second column
of table A.1 in \cite{CF}.
The $n=2$ Calabi--Yau threefold $X_2$, for instance, is given by
the space $P_{1,1,2,6,10}(20)$ of \cite{BKKM}.
The Calabi--Yau spaces 
$X_0$ and $X_1$ 
are given in \cite{CF,LSTY}.

Furthermore
\footnote{Note that, since the heterotic perturbative gauge group is 
reflected, on the dual
type IIA side, in the (monodromy invariant part of the) Picard group of 
the generic $K3$ fibre of the Calabi--Yau \cite{A,AL}, 
the discussion presented here agrees
precisely with the one of \cite{GN1}
concerning the zero divisor of the period map for the (mirror of the) $K3$.
${\cal D}^+$ can be matched with the domain of the period map $\Phi(z)$.},
for the $K3$-fibre $P_{1,1,3,5}(10)$ of $X_n$,
one finds that (cf. \cite{BKKM} for $n=2$) in the basis $j_1,j_3,j_4$
(where we denote the intersections of the CY divisors with the 
$K3$ ($J_2$) by
small letters) the intersection form is given by 
${\tiny \left(\begin{array}{ccc}2&1&4\\1&0&2\\4&2&6\end{array}\right)}$, 
which is equivalent (over {\bf Z}) to $-\Lambda$ under the base change 
$f_2=j_1-j_3$, $f_{-2}=j_3$ and $f_3=2j_1-j_4$. 
The enhancement loci become the 
conditions $t_3=0$ and $t_4=0$ for the K\"ahler moduli on the type II 
side .

The cubic parts of
the type II prepotentials of $X_0$, $X_1$ and $X_2$
are given in \cite{BKKM,LSTY} and can be written in a universal,
$n$-dependent function as follows:
\beqa
F^{II}_{\rm cubic}&=&t_2(t_1^2+t_1t_3+4t_1t_4+2t_3t_4+3t_4^2)\nonumber \\
&+&{4\over 3}t_1^3+8t_1^2t_4+{n\over 2}t_1t_3^2
+\left(1+{n\over 2}\right)t_1^2t_3 +2(n+2)t_1t_3t_4\nonumber \\
&+&nt_3^2t_4
+(14-n)t_1t_4^2 
+ (4+n)t_3t_4^2+(8-n)t_4^3 . \label{cubicf}
\eeqa

Note that for $t_4=0$, $F^{II}_{\rm cubic}$ 
precisely reduces to the cubic prepotential
of the three parameter models \cite{HKTY,LSTY}. 

Next, let us consider the contributions of the worldsheet 
instantons to the type II prepotential of a four-parameter model.
Generically, they are given by 
\beqa
F^{II}_{\rm inst} =-{1\over ( 2\pi)^3}\sum_{d_1,\dots , d_4}n^r_{d_1,\dots ,
d_4}Li_3\left(\prod_{i=1}^4q^{d_i}\right) \;\;.
\label{instanton}
\eeqa
The $n^r_{d_1,d_2,d_3,d_4}$ denote the rational instanton numbers.
The heterotic weak coupling limit $S \rightarrow \infty$ corresponds
to the large K\"ahler class limit $t_2 \rightarrow \infty$.  In this
limit, only the instanton numbers with $d_2=0$ contribute
in the above sum.  Using the identification 
$kT+lU+bV=d_1t_1+d_3t_3+d_4t_4$, it follows that (independently of $n$)
\begin{eqnarray}
k&=&d_3 \;\;, \nonumber\\
l&=&d_1-d_3   \;\;,\nonumber\\
b&=&d_4-2d_1  \;\;.
\end{eqnarray}
Then, (\ref{instanton}) turns into
\beqa
F^{II}_{\rm inst}=-{1\over (2\pi)^3}\sum_{k,l,b}n^r_{k,l,b}
Li_3(e^{-2\pi(kT+lU+bV)}) \;\;.
\label{largespr}
\eeqa
Later we will see by comparison with (\ref{prepstuv}) that the 
rational instanton
numbers have to satisfy the nontrivial constraint
\beqa
n^r_{k,l,b}=n^r(4kl-b^2) \; .
\label{constraint}
\eeqa

\newpage

\section{BPS Spectral Sums}

\resetcounter

\subsection{BPS States}

BPS states - the states saturating the Bogomolny bound $m\ge |Z|$ 
between mass and charge -
play an important role in many difficult dynamical questions as they 
provide in some respects a rigid skeleton of a theory. This can be related
to the issue of quantum uncorrectedness of their masses in certain
situations or to their
property of describing (via a certain index) something like a topological 
subsector of a theory. \\
For example they are relevant for
questions such as strong/weak coupling duality - say in tests of the 
nonperturbative $SL(2,{\bf Z})_S$ duality of the $N=4$ theory in 4D given by
the heterotic string on $T^6$ \cite{Sen}, where one can make checks of 
nonperturbative predictions as one expects the masses of these states to be
uncorrected in the quantum theory. Another important framework is the
conjectured  4D string/string/string triality \cite{DLR}: namely in the 
intermediate step in six dimensions the heterotic string on $T^4$ is dual
to a type IIA string \cite{W}; in completing the compactification process
in going to 4D on
a further $T^2$ the heterotic string aquires beside the aforementioned 
$SL(2,{\bf Z})_S$ strong/weak coupling duality an duality relating to the 
moduli of $T^2$, namely the $SO(2,2,{\bf Z})$ target space duality
consisting of $SL(2,{\bf Z})_T\times SL(2,{\bf Z})_U$ and the exchange 
symmetry T-U. As the 6D duality leads to an exchange of S and T in 4D one
gets the IIA string with fields T,S,U (similar considerations lead to a
4D type IIB string with fields U,T,S). Under this S-T exchange the BPS
spectra of the
heterotic and the type IIA string get mapped into each other, whereas the 
BPS spectra of the individual strings don't have this symmetry. The latter
fact is due to the property of the BPS masses in the (4D) $N=4$ situation as
being given by the maximum of the 2 central charges $|Z_1|$ and$|Z_2|$ of the
$N=4$ algebra. So there exist two kinds of massive BPS multiplets called 
short and intermediate (see below).
Now the point is that the states , which from the $N=4$ point of view are 
intermediate, are short from the $N=2$ point of view \cite{CCLMR}. So one can 
expect an S-T symmetry of the BPS spectrum of certain $N=2$ heterotic string
compactifications. \\
A further area, in which BPS states play a central role, is the study of
1-loop threshold corrections to gauge and gravitational couplings in  $N=2$
heterotic string compactifications \cite{HM}. In cases, where the 
contributions are due to BPS states only, one can then also expect a S-T
exchange symmetry of the concerned couplings. \\
Lastly a related theme is the resolution of the conifold singularity 
\cite{S}, in whose description the BPS states also play an essential role,
as we will see. \\

\subsubsection{The $N=4$ situation}
Before we come to the main  issue of this report concerning $N=2$ string 
compactifications let us point to some features of the $N=4$ situation to 
put some questions into a broader perspective. \\
Consider the $N=4$ heterotic string got by compactification on $T^6$. The 
two central charges of the $N=4$ algebra are given by the 6-dimensinal 
electric/magnetic charge vectors $\vec{Q},\vec{P}$ as
\begin{eqnarray*}
|Z_{1,2}|^2=\vec{Q}^2+\vec{P}^2\pm 2\sqrt{\vec{Q}^2\vec{P}^2-(\vec{Q}\cdot
 \vec{P})^2}
\end{eqnarray*}
Now let us focus on the discussion of the moduli dependence of a $T^2$ 
subsector. One gets 
\begin{eqnarray*}
|Z_{1,2}|^2=\frac{1}{4}\frac{1}{(S+\bar{S})(T+\bar{T})(U+\bar{U})}
|{\cal M}_{1,2}|^2
\end{eqnarray*}
with\footnote{$\check{P}^0=T+U,\check{P}^1=i(1+TU),\check{P}^2=T-U,
\check{P}^3=-i(1-TU)$}
${\cal M}_1=(\check{M}_I+iS\check{N}_I)\check{P}^I$ and 
${\cal M}_2=(\check{M}_I-i\bar{S}\check{N}_I)\check{P}^I$
the integers $\check{M}_I$ ($\check{N}_I$) being the electric (magnetic) 
charge quantum numbers ($I=0,...3$) of the gauge group $U(1)^4$.
We will work in a rotated basis with $P=(1,-TU,iT,iU)^T$ and corresponding 
quantum numbers $M_I,N_I$. \\
Now as the $BPS$ masses are 
being given by the maximum of the 2 central charges $|Z_1|$ and$|Z_2|$ of the
$N=4$ algebra there exist two kinds of massive BPS multiplets called 
\begin{itemize}
\item
short for 
\begin{eqnarray*}   
m_S^2=|Z_1|^2=|Z_2|^2
\end{eqnarray*}
with the associated soliton background solution preserving 1/2 of the
supersymmetries of $N=4$ , the multiplet containing maximal spin one
\item
intermediate for 
\begin{eqnarray*}   
m_I^2=\max(|Z_1|^2,|Z_2|^2)
\end{eqnarray*}
with the associated soliton background solution preserving 1/4 of the
supersymmetries of $N=4$ , the multiplet containing maximal spin 3/2
\end{itemize}
The mass formula is invariant under the perturbative T-duality group 
$SL(2,{\bf Z})_T\times SL(2,{\bf Z})_U\times {\bf Z}_2^{T\leftrightarrow U}$ 
as well as under the nonperturbative S-duality group $SL(2,{\bf Z})_S$. \\
Now the fundamental distinction for a $N=4$ BPS state is whether it is short
or intermediate, i.e. whether the (S-independent) expression $\Delta Z^2=
|Z_1|^2-|Z_2|^2$ is generically zero or not. So the shortness condition 
means the parallelness $\vec{Q}\| \vec{P}$, i.e. $sM_I=pN_I$ with $s,p\epsilon
{\bf Z}$. This means for the holomorphic mass of a short multiplet that it 
factorizes\footnote{via the identification $M=s{\cal V},N=p{\cal V}$
(${\cal V}$ the vector ${\cal V}=(m_2,n_2,n_1,-m_1)$ of the $T^2$ lattice)} 
into a S-dependent and a T,U-dependent term
\begin{eqnarray*}
{\cal M}={\cal M}_S{\cal M}_{T,U}
\end{eqnarray*}
with
\begin{eqnarray*}
{\cal M}_S &=& s+ipS \\
{\cal M}_{T,U} &=& m_2-im_1U+in_1T-n_2TU
\end{eqnarray*}
So here the quantum numbers $m_i \,(n_i)$ are the momentum (winding) numbers 
associated to the $T^2$, s (p) is the electric (magnetic) quantum number
associated with the $U(1)_S$. Examples are the elementary electric heterotic
string states with magnetic charge $p=0$, where $M^2\sim p_R^2$ with the
right-moving $T^2$ lattice momentum $p_R$ and level matching $\frac{p_L^2}{2}-
\frac{p_R^2}{2}=m_1n_1+m_2n_2=N_R+h_R-N_L+1/2=1-N_L$. For $S\rightarrow \infty$
an infinite number of these states becomes massless. Similarly for 
$S\rightarrow 0$ an infinite tower of magnetic monopoles with $s=0$
(p unrestricted) becomes arbitrary light. \\
One has two subclasses (again respected by the duality group)

\begin{itemize}
\item 
$m_1n_1+m_2n_2=0$: KK excitations of elementary states and KK monopoles 
(no massless states for finite values of T and U) 
\item
$m_1n_1+m_2n_2=1$: containing elementary states becoming massless along
critical lines/points within T,U moduli space (the cases $T=U,T=U=1,T=U=\rho$ 
of classically enhanced gauge symmetries $SU(2),SU(2)^2,SU(3)$); also 
containig H-monopoles
\end{itemize}

\subsubsection{The $N=2$ situation}
The BPS masses are given by $m_{BPS}^2=|Z|^2$ with $Z$ the (complex) central 
charge of the
supersymmetry algebra. As a function of the $n_V+1$ massless abelian vector
multiplets $X^I$ of the abelian gauge group $U(1)^{n_V+1}$ 
(the graviphoton
$X^0$ will be set to one)
one has with the holomorphic
prepotential ${\cal F}(X^I)$ and ${\cal M}=M_IX^I+iN^I{\cal F}_I$ 
(with the electric resp. magnetic quantum numbers as coefficients) 
the mass formula
\begin{eqnarray*}
m_{BPS}^2=e^K|{\cal M}|^2=\frac{|M_IX^I+iN^I{\cal F}_I|^2}
{X^I\bar{{\cal F}}_I+\bar{X}^I{\cal F}_I}
\end{eqnarray*}
with the expression $e^{-K}=X^I\bar{{\cal F}}_I+\bar{X}^I{\cal F}_I$ 
involving the
Kaehler potential. The relevant sector for us will then be given by
\footnote{In the standard heterotic normalization $S=\frac{1}{g^2}-i
\frac{\theta}{8\pi ^2}$; the Peccei-Quinn symmetry associated with the axion
is given by the shift $\theta \rightarrow \theta +2\pi$,$S\rightarrow S-
\frac{i}{4\pi}$.} 
$S=i\frac{X^1}{X^0},T=-i\frac{X^2}{X^0},U=-i\frac{X^3}{X^0}$.
With the tree level prepotential\footnote{But note that the period vector 
$(X^I,i{\cal F}_I)$ following from ${\cal F}^{(0)}$ doesn't lead to the 
behaviour that all classical gauge couplings
become small as $S\rightarrow \infty$ 
\cite{AFGNT,CdAFvP,dWKLL}: whereas the couplings for 
$U(1)_T\times U(1)_U$ are proportional to ReS, the couplings involving $U(1)_S$
are constant or even growing; to get a period vector with all gauge couplings
proportional to ReS one has to make the symplectic transformation 
$(X^I,i{\cal F}_I)\rightarrow (P^I,iQ_I)$ with $P^1=i{\cal F}_1,Q_1=iX^1$ 
(the rest unchanged).}
${\cal F}^{(0)}=i\frac{X^1X^2X^3}{X^0}=-STU$
one gets for the classical period vector
\begin{eqnarray}
\Omega =(1,TU,iT,iU|iSTU,iS,-SU,-ST)^T
\end{eqnarray}
The electric period fields $P^I$ are only T,U-dependent, whereas the magnetic
period fields $Q_I$ qre proportional to S. \\
In this basis the holomorphic BPS mass becomes
\begin{eqnarray}
{\cal M}=M_0+M_1TU+iM_2T+iM_3U+iS(N_0TU+N_1+iN_2U+iN_3T)
\end{eqnarray}

\newpage

\subsection{BPS spectral sums}

In this and the next subsection 
we review some contributions made over the years to the study and
interpretation of quantities involving sums over BPS states \cite{FKLZ,HM,V}.

\subsubsection{$F_1$ function and the conifold}                          
                          
Let us start from the 
phenomenon associated with the so called conifold point in the Calabi-Yau
moduli space. As 
one approaches such a point p a 3-cycle $C_p$ together with its period
$\int _{C_p} \Omega$ (which represents a coordinate value $z(p)=0$ of the
complex structure deformations) and thus together with the corresponding mass
will vanish ($m^2=|z|^2$ after convenient normalization of $\Omega$).
Here the associated classical state was at first constructed by Strominger
\cite{S} by wrapping a 3-brane around the cycle, leading to a 4D extreme
black hole  carrying the relevant Ramond-Ramond U(1) gauge field (whereas
 the
fundamental string excitations are neutral under this U(1))
saturating the Bogomolny bound and representing thus a stable BPS state
(an N=2 hypermultiplet).  \\
The assumption in \cite{S} (necessary for  resolution of the conifold
singularity)
that there are no stable black holes corresponding to integer multiples of the 
vanishing cycle was established by Bershadsky, Sadov and Vafa \cite{BSV2}
using new insights \cite{BSV1},\cite{OV} gained in the rapid development
taking place immediately after the beginning \cite{P},\cite{W2} of the
D-brane revolution. \\
It is now of interest to  develop further the idea of resolution of the
conifold singularity by the existence of the (becoming) massless 
black hole state in the full
theory in that it gives an alternative explanation of divergent string
amplitudes.
To go ahead it is of course usefull to consider a (pertubatively and
nonpertubatively)
uncorrected quantity. Moreover this will in the end bring us back to the
free energy (represented as a BPS sum). \\
Concretely we will study a term for the type II string on the 
Calabi-Yau, which in turn is related to a topological amplitude. 
Namely let us 
consider again the moduli dependent topological 1-loop amplitude $g_1^{-2}$, 
which gives terms for the vector
multiplet related to gravitational couplings (R is the Riemann tensor)

\begin{eqnarray*}
\int g_1^{-2} R^2 .
\end{eqnarray*}

So our aim is now to understand the singular structure of $g_1^{-2}$ 
near a conifold point in terms of the
full theory including the nearly massless black hole.  \\
In the studied examples the singularity was found to be

\begin{eqnarray*}
g_1^{-2}=-\frac{1}{12}\log |z|^2+\cdots
\end{eqnarray*}

(the precise coefficient fixed by considering other limits), which gives
\footnote{After considerations of normalization \cite{V} using
$\frac{1/2}{128\pi ^2}\int  \epsilon ^{abij}\epsilon ^{cdkl}R_{abcd}R_{ijkl}=
\frac{\chi}{2}$} for the correction of the effective action near the
conifold point

\begin{eqnarray*}
\delta S=-\frac{\chi}{24}\log m^2
\end{eqnarray*}

But exactly such a term would come from a 1-loop computation \cite{V}
in the theory with a light black hole of mass m as the contribution of exactly 
$one$ hypermultiplet, thus (again) confirming the consistency of Stromingers 
proposal. \\
Now in the effective theory near the conifold point a 1-loop term (proportional
to $\chi$) is generated not only by the light hypermultiplet, but by all of
them.
So $g_1^{-2}$ conjecturally \cite{V} contains (there could be further
$\chi$-proportional terms in the effective theory near the conifold point)
a term
like (suitable regularization understood; also there is the jumping
phenomenon which can destroy the (for $F_1$ needed) modular invariance of the
BPS spectrum)

\begin{eqnarray*}
-\frac{1}{12}\sum_{BPS}\log m^2
\end{eqnarray*}

(may be one has to restrict to a subspace of vanishing cycles 
and to discuss the jumping phenomenon).
Note that in this setup the $BPS$ mass formula $m_{BPS}^2=e^K|{\cal M}|^2$
becomes 
\beqa
m^2=\frac{\int _C \Omega \cdot \int _C \bar{\Omega}}
{\int \Omega \wedge \bar{\Omega}} \; .
\eeqa

In the (complex) one dimensional situation 
(with the true winding states on $T^2$) one finds (cf. the next subsection; 
appendix \cite{FKLZ}) such a relation

\begin{eqnarray*}
F=\log Z=\sum_{n,m}\log\frac{|n+\tau m|^2}{\tau _2}=-\log (\det D^2)=g_1^{-2}
\end{eqnarray*}

giving $g_1^{-2}$ as a partition function for BPS states
(D being the laplacian acting on the 
fields in the multiplet).\\

A more precise indication of what to expect comes from an anomaly
comparison: the proposed BPS sum as an expression for $g_1^{-2}$ 
would give for 
$\delta \bar{\delta}g_1^{-2}$ something proportional to $\delta \bar{\delta}K$ 
(as we had $e^{-K}=\int \Omega \wedge \bar{\Omega }$), but not the full
anomaly as one has \cite{BCOV,CdWLMR}

\begin{eqnarray*}
\partial _i\bar{\partial}_{\bar{j}}g_1^{-2}=
\frac{1}{2}(3+h^{1,1}-\frac{\chi}{12})G_{i\bar{j}}-\frac{1}{2}R_{i\bar{j}}
\end{eqnarray*}

leading to (say in the STU-model) the dependence in T,U
(using the anomaly factor 
\cite{BCOV} in the first contribution to this sum and the moduli metric 
$\partial \bar{\partial}K$ in the second; 
$e^{-\hat{K}}=(T+\bar{T})(U+\bar{U})$)

\begin{eqnarray*}
g_1^{-2}&=&
\log |F_{hol}|^2+(3+h^{1,1}-\frac{\chi}{12})\hat{K}-
(h^{1,1}-1)\log \frac{1}{(T+\bar{T})(U+\bar{U})}\nonumber\\
&=&2\mbox{Re}\log F_{hol}+\frac{b_{grav}}{12}\hat{K}
\end{eqnarray*}
       
\subsubsection{The topological free energy}

This proposal was similar in spirit to the one made 
in \cite{FKLZ}, where a moduli dependent quantity 
(there called "topological free
energy") $F=\log Z$ was defined as a certain nonholomorphic partition function
for supersymmetric string compactifications. It
corresponds to the generating functional of the effective action after
integrating out the massive modes coming from the compactification
(so in the toroidal case the Kaluza-Klein states and the winding
states are concerned, but not the massive oscillator states, which are 
also present in the uncompactified theory; this is meant by the
qualification as "topological"). 
It is given as (the log of) the determinant of the chiral
mass matrix for the massive compactification modes and gives the effective
description for amplitudes with external light states and possible massive
(in the qualified sense) states in the loops. Its moduli dependence leads to an
expression of F in terms of automorphic functions (or forms) of an duality
group: may it be the T-duality group for toroidal and/or orbifold 
compactifications or the symplectic duality group coming from the special
Kaehler geometry of an $N=2$ situation.\\
More concretely the process of integrating out the massive modes leads to
\begin{eqnarray*}   
e^{F_{Bos}}=\int{\cal D}\phi \; e^{\phi M^2_{\phi}\phi^{\dagger}}+\cdots
\end{eqnarray*}
where we have neglected higher orders (and derivative terms) in $\phi$
around the $\phi=0$ vacuum.
Because of the space time supersymmetry one has then
\begin{eqnarray*}
e^F=\det M^{\dagger}M
\end{eqnarray*}
with the fermionic mass matrix, an expression, which can further be 
evaluated in the form
$(\mbox{nonholomorphic part})\times |\mbox{holomorphic piece}|^2$,
namely (the nonholomorphic piece comes from an $e^K$ rescaling
in the supergravity expression for $\det M^{\dagger}M$; $W$ is the 
superpotential, $K$ the Kaehler potential)
\begin{eqnarray*}
e^K|\det W_{ij}|^2
\end{eqnarray*}
For example in the familiar case of compactification on (3 copies of)
$T^2$, where one has
$e^K=\frac{1}{S+\bar{S}}\frac{1}{(T+\bar{T})(U+\bar{U})}=\frac{1}{S+\bar{S}}
e^{\hat{K}}$,
one has explicitely (the prime on the 
sum means $(m_1,n_1,m_2,n_2)\neq (0,0,0,0)$ and the constraint comes
from $\vec{p}_L^2-\vec{p}_R^2=m_1n_1+m_2n_2$ and the exclusion of 
oscillator states)
\begin{eqnarray*}
F=\sum^{\prime}_{m_1n_1+m_2n_2=0}\log e^{\hat{K}}|{\cal M}_{T,U}|^2
\end{eqnarray*}
(with a regularization of the infinite sum understood, which respects the
T duality group $SL(2,{\bf Z})_T \times SL(2,{\bf Z})_U \times {\bf Z}_2$). 
Note that the constraint decomposes into two orbits: $n_2=m_1=0$ and 
$n_2=n_1=0$. So
\begin{eqnarray*}
F=\sum_{m_2,n_1}\log \frac{|m_2+in_1T|^2}{T+\bar{T}}+
  \sum_{m_2,m_1}\log \frac{|m_2+im_1U|^2}{U+\bar{U}}
\end{eqnarray*}
or after the $\zeta$-function regularization
\begin{eqnarray}
F=\log(T+\bar{T})|\eta (T)|^4 + \log (U+\bar{U})|\eta (U)|^4
\end{eqnarray}
Note that exactly at the boundary points of moduli space, $T,U=0,\infty$, 
where additional KK- resp. winding-modes become massless (so that one
expects an infinite contribution to the free energy), the expression
diverges.\\
Or to argue in an other way (and writing in terms of the bosonic quantities
now): the interpretation of the result
\begin{eqnarray*}
Z=e^K|e^{\cal F}|^2=\frac{|e^{\cal F}|^2}{(S+\bar{S})(T+\bar{T})(U+\bar{U})}
\end{eqnarray*}
coming from the "holomorphic free energy" 
\beqa
{\cal F}=\sum_{BPS}\log {\cal M}
\eeqa
and $m^2_{BPS}=e^K |{\cal M}|^2$ in the relation
\begin{eqnarray*}
\log Z=\sum_{BPS}\log m^2_{BPS}
\end{eqnarray*}
shows that the appearence of the explicit expression above was to be 
expected because for $Z$ to be invariant $e^{\cal F}$ has to be a modular
form of weight -1 with respect to the $SL(2,{\bf Z})$ groups in S,T and U;
so one gets up to an invariant function that
$e^{\cal F}=\frac{1}{\eta ^2(S)\eta ^2(T)\eta ^2(U)}$.\\
This can be applied to the "toroidal sector" of the heterotic
string on $K3\times T^2$.
In the setup of $N=2$ supersymmetry the relevant expressions 
for the BPS masses and associated quantities 
in terms of the
prepotential (not to be confused
with the holomorphic free energy)  ${\cal F}$, and the special coordinates
$X_I$ are $e^{-K}=X^I\bar{{\cal F}}_I+\bar{X}^I{\cal F}_I$ and
\begin{eqnarray*}
\log Z=\sum_{M_I,N^I}\log \frac{|M_IX^I+iN^I{\cal F}_I|^2}{X^I\bar{{\cal F}}_I+\bar{X}^I{\cal F}_I}
\end{eqnarray*}

Note that the sum goes over the restricted subset of integers satisfying
the topological level matching condition $\vec{p}_L^2-\vec{p}^2_R=0$; and 
as the duality group $\Gamma \subset Sp(2n+2)$ leaves K invariant only up to a
Kaehler transformation so then $\log Z$ is invariant only if also a symplectic
transformation on the integers $M_I$ and $N^I$ (accompanying the one on the
$X^I$ and $i{\cal F}_I$) is made , i.e. the sum runs over symplectic orbits. \\To see the full analogy to the toroidal situation let us consider the
Calabi-Yau example in greater detail. For definitiveness let us restrict
ourselves first to the thpe IIA case. The chiral masses we want to describe
come from the couplings $X^IHH^{\prime}$ between one N=2 vector multiplet and
two N=2 hypermultiplets , i.e. the relevant massive string states are 
hypermultiplets with masses given by vev's of the moduli scalars in the (1,1)
vector multiplets ; so the chiral N=2 masses are just given by the holomorphic
functions $X^I$. Obviously by considering these field theoretical masses
$X^I$ alone the spectrum can't be invariant as the $X^I$ are mixed under the
symplectic duality transformations with the periods $i{\cal F}_I$ . So the 
relevant ansatz for the hypermultiplet masses is $M=M_IX^I+iN^I{\cal F}_I$ 
with the integers satisfying symplectic constraints. Also the interpretation
becomes analogous by considering first the field theory, i.e. large radius,
limit . Let us describe the situation in the special gauge (special coordinate
system) with $T^i=-iX^i/X^0$ and $f(T)=(X^0)^{-2}{\cal F}(X)$ , where
$K=-\log [\sum_i(T^i+\bar{T}^i)(f_i+\bar{f}_i)-2(f+\bar{f})]$ and
\begin{eqnarray*}
\log Z=\sum_{M_I,N^I}
\log \frac{|M_0+M_iT^i+iN^0f+iN^if_i|^2}
{(T^i+\bar{T}^i)(f_i+\bar{f}_i)-2(f+\bar{f})}
\end{eqnarray*}
With the harmonic (1,1) forms $V_i$ and $J=\sum_i(T^i+\bar{T}^i)V_i$ one has
$e^K=
\int J\wedge J\wedge J=d_{ijk}(T^i+\bar{T}^i)(T^j+\bar{T}^j)(T^k+\bar{T}^k)$,
where $f(T^i)=d_{ijk}T^iT^jT^k$ and the $d_{ijk}$ coincide in the field theory 
limit with the topological intersection numbers:
$d_{ijk}=\int V_i\wedge V_j\wedge V_k$ .
With the expression
\beqa
m^2_{M_0}=\frac{M_0^2}{\int J\wedge J\wedge J}
\eeqa
for those masses having only nonvanishing $M_0$
one recognizes (as now these masses are - in this field theory limit - 
inversely proportional to the volume of the Calabi-Yau space) these masses as
the ordinary field theoretical Kaluza-Klein masses. So the states with only
nonvanishing $M_0$ give here the momentum spectrum. \\
On the other hand for small radii nonperturbative instanton configurations
giving nontrivial maps of the world-sheet into the Calabi- Yau space give 
quantum
modifications to the aforementioned results. Also now the generalized winding
states
become light. As the duality invariance forces one to include states with
nonvanishing $M_i$ and $N^I$ , the masses given by $M_iT^i$ resp.
$N^I{\cal F}_I$;
i.e. we have now also identified the spectrum of generalized winding states.

This 
makes the toroidal analogy complete.  \\
To include now also the type IIB string in the discussion note first that
there the
moduli in the vector multiplets are given by (2,1) moduli representing the
complex
structure deformations of the underlying Calabi-Yau space. These are described
now
by the (true) periods (, which are furthermore now quantum uncorrected by
world-sheet instantons)
\beqa
X^I=\int_{A^I}\Omega \;\;, \;\; i{\cal F}_J=\int_{B_J}\Omega
\eeqa
of the holomorphic 3-form $\Omega$ integrated over a canonical symplectic
homology basis $(A^I,B_I)$. So now these periods are the chiral masses
of the massive hypermultiplets 
(here $e^{-K}=i\int \Omega \wedge \bar{\Omega}$).

\newpage

\subsection{threshold corrections and couplings}

The occurence of a sum over the BPS spectrum here is intimately related
- as we will see - to the appearence of a similar expression in the 
study of threshold corrections \cite{HM}. Note first that the product
atructure $K3\times T^2$ of the internal space of the heterotic string is 
reflected on the CFT level by the decomposition of the right-moving internal
$c=9$ superconformal algebra

\begin{eqnarray}
A_{int}=A^{N=2}_{c=3}\oplus A^{N=4}_{c=6}
\end{eqnarray}

Let us denote the $U(1)$ current of the $c=3$ theory by $J^{(1)}$; the
representations of the $c=6$ theory are labeled by the conformal weight
and the representation of the (level one) $SU(2)$ Kac-Moody algebra as
$(h,I)$. The total $U(1)$ current of the $c=9$ theory is then
$J=J^{(1)}+J^{(2)}$, where $J^{(2)}=2J^3$ with the $SU(2)$ Cartan current.
Furthermore the (local) factorization of the moduli space into vector multiplet
and hypermultiplet contributions is in the $c=6$ algebra reflected by the 
massless NS representations with $(h=0,I=0)$ and $(h=1/2,I=1/2)$ respectively
(these are connected by the spectral flow of the $N=4$ theory to massless 
Ramond representations). Namely as the (complex) central charge is determined
by the right-moving momenta $p_R$
carried by the two free dimension 1/2 superfields of the $c=3$ theory and as
the mass of a NS state is ($N_R$ the right-moving oscillator number
coming from (spacetime-sector)$\oplus ((c=3)-\mbox{sector})$)

\begin{eqnarray*}
M^2=(N_R-1/2)+\frac{1}{2}p_R^2+h
\end{eqnarray*}

one has (corresponding to the massless NS representations of $A^{N=4}_{c=6}$) 
the description of the (bosonic) BPS states (with the slight abuse
of language as in \cite{HM})

\begin{eqnarray*}
M^2=\frac{1}{2}p_R^2\Rightarrow \left\{ \begin{array}{r@{\quad,\quad}l}
 \mbox{vector}:\;N_R=1/2 & (h=0,I=0) \\ \mbox{hyper}:\;N_R=\;\;\;0 & (h=1/2,I=1/2) \end {array} \right.
\end{eqnarray*}

Let us insert a remark here on 
the jumping phenomenon known in $N=2$ theories
\cite{CFIV,CV,SW} with their $complex$ central charge, where
in moving in moduli space along a path joining two given points one can't
avoid passing through configurations with collinear Bogomolny charges
leading to soliton number jumping (BPS states suddenly (dis)appear under
infinitesimal moduli perturbation) related \cite{V} to the intersection
numbers of the vanishing cycles, resp. (in view of the Picard-Lefshetz 
formula \cite{L}) to the corresponding monodromy.
Now with regard to this phenomenon, which seems at first
to contradict the philosophy that the BPS spectrum gives some 'rigid 
topological skeleton' of a theory note \cite{HM} that as usual in such 
'topological' connections one has to consider an index-like difference
to get a deformation invariant quantity: namely these 'chaotic' BPS states
always appear in vectormultiplet, hypermultiplet $pairs$; as one moves away
from the special points they pair into long representations being no longer 
saturated. \\
This brings us to the connection with the threshold corrections. 
Note that such an index-like difference structure of BPS sums is  
reasonable to expect there: namely firstly there are no corrections in $N=4$
and indeed $N=4$ BPS states split up into a $N=2$ hypermultiplet and a $N=2$
vectormultiplet; and secondly one should expect at all only BPS states,
i.e. short $N=2$ representations as massive long $N=2$ representations
(of the extended supertranslation algebra) are short as $N=4$ representations,
so noncontributing. \\
This is reflected in the threshold corrections to the gauge
coupling
given by the 1-loop renormalization 
(for gauge group G,$b_G$ the 1-loop beta function
coefficient, Q a generator of G) \cite{KL}

\begin{eqnarray*}
g^{-2}=\mbox{Re}S+\frac{b_G}{16\pi ^2}\log
 \frac{M^2_{string}}{p^2}+\frac{1}{16\pi ^2}\Delta^{univ}+
\frac{1}{16\pi ^2}\Delta^{index}
\end{eqnarray*}

where we have seperated the universal part of the threshold correction, which
is related to the Green-Schwarz term and so can be expressed in terms of the 
1-loop correction $h^1$ of the prepotential

\begin{eqnarray*}
\frac{1}{16\pi ^2}\Delta^{univ}=\frac{1}{2}V_{GS}=\frac{1}{(T+\bar{T})
(U+\bar{U})}\mbox{Re}(2h^1-(T+\bar{T})\partial _Th^1-(U+\bar{U})\partial _Uh^1)
\end{eqnarray*}

and the further piece ($J_0$ the mentioned total $U(1)$ charge)

\begin{eqnarray*}
\Delta^{index}=\int_{{\cal F}}\frac{d^2\tau}{\tau _2}[\frac{-i}{\eta ^2}
Tr_{{\cal H}^{int}_R}\{J_0(-1)^{J_0}q^{L_0-\frac{22}{24}}\bar{q}^{\bar{L}_0-
\frac{9}{24}}[Q^2-\frac{1}{8\pi \tau _2}]\}-b_G]
\end{eqnarray*}

involving the supersymmetric index \cite{CFIV,CV} described by 
the trace over the Ramond sector of the internal $(c,\bar{c})=(22,9)$ CFT
\begin{eqnarray*}
Tr_{\rm Ramond}J(-1)^Jq^{L_0-\frac{c}{24}}\bar{q}^{\bar{L}_0-\frac{\bar{c}}
{24}}
\end{eqnarray*}

To put this definition into perspective 
recall that defines the usual partition
function as
$Trq^{L_0-\frac{c}{24}}\bar{q}^{\bar{L}_0-\frac{\bar{c}}{24}}$
and the 'elliptic genus' \cite{Le,SchW,Wi}
as 
$Tr_{\rm Ramond}(-1)^Jq^{L_0-\frac{c}{24}}\bar{q}^{\bar{L}_0-\frac{\bar{c}}
{24}}$.

Now according to the decomposition (0.15) and as $J=J^{(1)}+J^{(2)}$ one is 
left with the index in the (left)$\times (c=3)$ part times the elliptic genus
in the $N=4$ part as the index for the $N=4$ part vanishes since the 
eigenvalues of $J^3$ come in opposite pairs and $J^{(2)}=2J^3$.
Furthermore in the 
$(c=3)$-part the BPS vectors come (in the Ramond sector) with U(1)-charge
$\pm \frac{1}{2}$ leading to a 
$\frac{1}{2}e^{i\pi /2}-\frac{1}{2}e^{-i\pi /2}=i$ contribution to
$Je^{i\pi J}$ while one has twofold this content for a BPS hypermultiplet
 leading
to a 2i contribution; this is, including the elliptic genus factor from the 
$N=4$ part, corrected to -2i resp. 2i showing

\begin{eqnarray*}
\frac{1}{\eta ^2}
Tr_{\rm Ramond}J(-1)^Jq^{L_0-\frac{c}{24}}\bar{q}^{\bar{L}_0-\frac{\bar{c}}
{24}}=
-2i\Big( \sum_{\mbox{vector}}^{BPS}q^{\Delta}\bar{q}^{\bar{\Delta}}
- \sum_{\mbox{hyper}}^{BPS}q^{\Delta}\bar{q}^{\bar{\Delta}}\Big)
\end{eqnarray*}

So one gets essentially
\begin{eqnarray*}
g^{-2}\sim \frac{2}{16\pi ^2}
\Big(\sum_{\mbox{vector}}^{BPS}Q^2\log m^2-
\sum_{\mbox{hyper}}^{BPS}Q^2\log m^2\Big)
\end{eqnarray*}

Now as we described the pay-off for the prepotential already in section 3
let us discuss here the connection with the gravitational $F_1$ coupling.
The (nonperturbative) holomorphic free energy ${\cal F}$, defined 
as a BPS sum, will diverge at those loci in the (nonperturbative) moduli
space where additional BPS states become massless, i.e. along the loci
of massless magnetic monopoles and dyons and along other singular
strong coupling loci. By string/string duality this nonperturbative
holomorphic free energy is given by the corresponding classical quantity
of type II string involving a sum over classical BPS states, which is 
singular precisely along the discriminant locus $\Delta$ of the (mirror)
Calabi-Yau. \\
Now the holomorphic free energy
${\cal F}$ will be identified with ${\cal F}^{grav}$ (our ${\cal F}_1$), 
which is at tree level

\begin{eqnarray*}
{\cal F}_{grav}=24S\;\;,\;\;g^{-2}_{grav}=\mbox{Re}{\cal F}_{grav}=24\mbox{Re}S
\end{eqnarray*}

and at 1-loop (the last term reflects the
classical symmetry enhancement)

\begin{eqnarray*}
{\cal F}_{grav}=24S^{inv}+
\frac{1}{4\pi ^2}\log (j(T)-j(U))^2+
\frac{\frac{1}{2}b_{grav}}{4\pi ^2}\log \frac{1}{\eta ^2(T)\eta ^2(U)}
\end{eqnarray*}

which has the correct modular weight to render\footnote{an expression, which 
has in the general case to be supplemented by the quantity
$12\frac{3-n_v}{16\pi ^2}\log (S+\bar{S})$, which vanishes for the
3 parameter model; $K$ is again the tree level Kaehler potential with 
$e^{-K}=(S+\bar{S})(T+\bar{T})(U+\bar{U})$}

\begin{eqnarray}
g^{-2}_{grav}=\mbox{Re}{\cal F}_{grav}+\frac{b_{grav}}{16\pi ^2}(\log
\frac{M^2_{Pl}}{p^2}+K)
\end{eqnarray}

invariant under T-duality. \\
As we saw (by explicit evaluation in the case of the Dedekind-function
resp. the symmetry enhancement argument in case of the j-function) 
the term involving the Dedekind- resp. j-functions arises
from the orbit $m_1n_1+m_2n_2=0$ resp. 1. So it is natural to identify
(${\cal M}=M_IP^I+iN^IQ_I$)

\begin{eqnarray}
{\cal F}_{grav}\sim {\cal F}=\sum^{BPS}_{\mbox{vector}}\log {\cal M}-
\sum^{BPS}_{\mbox{hyper}}\log {\cal M}
\end{eqnarray}

where here in the sum over $M_I$ and $N^I$ the classical period vector
$\Omega=(P^I|iQ_I)^T=(1,TU,iT,iU|iSTU,iS,-SU,-ST)^T$ is concerned.  \\
So even the tree level part ${\cal F}_{grav}=24S$ should arise from BPS
states in (6.2) at $S\rightarrow \infty$, say from the term ($\check{S}=
4\pi S$)

\begin{eqnarray}
\log \frac{1}{\eta ^2(\check{S})}=\sum_{(s,p)\neq (0,0)}\log (s+ip\check{S})
\end{eqnarray}

whose occurence could be made plausible from the expression
${\cal M}=(s+ipS){\cal M}_{T,U}$ for the holomorphic
mass of the short $N=4$ multiplet. \\
Now let us be more concrete about the 1-loop corrected gravitational 
coupling , which we rewrite (with the true heterotic 
loop counting parameter $2(\mbox{Re}S+\frac{1}{2}V_{GS})$) as

\begin{eqnarray}
g^{-2}_{grav}=24(\mbox{Re}S+\frac{1}{2}V_{GS})+\frac{b_{grav}}{16\pi ^2}
\log \frac{M^2_{string}}{p^2}+\Delta _{grav}
\end{eqnarray}

with 
($e^{-\hat{K}}=(T+\bar{T})(U+\bar{U})$,$M^2_{Pl}\sim \frac{1}{g^2}
M^2_{string}$)

\begin{eqnarray}
\Delta _{grav} &=& 24(\mbox{Re}\sigma -\frac{1}{2}V_{GS})+
\frac{b_{grav}}{16\pi ^2}\hat{K}
+\mbox{Re}[\frac{\frac{1}{2}b_{grav}}{4\pi ^2}
\log \frac{1}{\eta ^2(T)\eta ^2(U)}\nonumber\\
 & & +\frac{1}{4\pi ^2}\log (j(T)-j(U))^2]
\end{eqnarray}

In the $s=0$ model of \cite{HM} with generic gauge group $E_8\times E_7\times
U(1)^4$ one can explicitely compute \cite{CCLMR} the 1-loop corrected 
gravitational coupling ${\cal F}_{grav}$ obtaining in the limit $T\rightarrow 
\infty$ (U finite) as result the tree level holomorphic gravitational 
coupling: ${\cal F}_{grav}\rightarrow 24S$, showing a $N=4$ like behaviour
of this model in the decompactification limit $T\rightarrow \infty$ at weak
coupling: $\mbox{Re}S>\mbox{Re}T\rightarrow \infty$. \\
Now as $N=4$ compactifications of the heterotic and type II strings are
related through exchange symmetries $S\leftrightarrow T,U$ by string/
string/string triality \cite{DLR} there could be such a nonperturbative
exchange symmetry in the $s=0$ model by the possible (in view of
$\mbox{intermed}_{N=4}\rightarrow \mbox{short}_{N=2}$) S-T symmetry of the 
BPS spectrum of certain $N=2$ heterotic compactifications. Since 
contributions to the holomorphic gravitational coupling ${\cal F}_{grav}$ 
arise from BPS states only (as shown at 1-loop \cite{HM}) let us 
discuss that symmetry for the object ${\cal F}_{grav}$ ! \\

\subsubsection{ST model}

First, in the 2 parameter model (cf. also \cite{AFT}), where we already 
observed the exchange symmetry S-T, the holomorphic gravitational coupling
${\cal F}^{II}_1=\frac{2\pi ^2}{3}{\cal F}_{grav}$ is (with $t_1=-iT,
t_2=-(i\check{S}-iT);c_2\cdot (B+iJ)=24t_2+52t_1=-i(24\check{S}+28T)$)

\begin{eqnarray}
F_1^{II}=-\frac{2\pi i}{12}c_2\cdot (B+iJ)-\sum_{j,k}[2d_{jk}\log \eta 
(q_1^jq_2^k)+\frac{n_{jk}}{6}\log (1-q_1^jq_2^k)]
\end{eqnarray}

So in the weak coupling limit $\mbox{Re}T<\mbox{Re}S\rightarrow \infty$
(so $q_1^jq_2^k\rightarrow 0$)

\begin{eqnarray}
F_1^{II}\rightarrow -\frac{2\pi i}{12}c_2\cdot (B+iJ)=
\frac{2\pi }{12}(24\check{S}+28T)
\end{eqnarray}

whereas in the strong coupling limit $\mbox{Re}S<\mbox{Re}T\rightarrow \infty$
(so $q_1^jq_2^k\neq 0 \Rightarrow k>j$) where then
$n_{jk}=2\delta_{j0}\delta _{k1}$ and $d_{jk}=0$, so
$\sum_{j,k\ge 0}\frac{n_{jk}}{6}\log (1-q_1^jq_2^k)=
\frac{2}{6}\log (1-q_2)=\frac{2}{6}\log (q_2)=\frac{4\pi i}{6}\log t_2$

\begin{eqnarray}
F_1^{II}\rightarrow -\frac{2\pi i}{12}c_2\cdot (B+iJ)+\frac{4\pi}{6}(\check{S}
-T)=\frac{2\pi i}{12}(28\check{S}+24T)
\end{eqnarray}

so that for $S,T\rightarrow \infty$

\begin{eqnarray}
F_1^{II}\rightarrow \frac{2\pi i}{12}[(24\check{S}+28T)\theta(\check{S}-T)+
(28\check{S}+24T)\theta(T-\check{S})]
\end{eqnarray}

showing $\check{S}\leftrightarrow T$ exchange symmetry and the corresponding
chamber dependence (cf. \cite{HM}). \\

\subsubsection{STU model}

Second, in the $s=0$ model with corresponding exchange symmetry assumed, one
gets with the aid of the explicit expressions that analogously \cite{CCLMR}

\begin{eqnarray}
F_{grav}\rightarrow \frac{6}{\pi}[T\theta(T-\check{S})+
\check{S}\theta(\check{S}-T)]
\end{eqnarray}

\subsubsection{STUV model}

Lastly note again in the $STUV$ model the 
important role in the computation of the Wilsonian
gravitational coupling $F_1$ is played by BPS states 
\cite{CCLMR,FKLZ,HM,V},
\beqa
F_1 \propto \log {\cal M}  \;\;,
\eeqa
where ${\cal M}$ denotes the moduli-dependent holomorphic mass
of an $N=2$ BPS state.  For the $STUV$ models
under consideration, the tree-level mass ${\cal M}$ is given by
\cite{CLM2,K1,NEU}
\beqa
{\cal M} = m_2 - i m_1 U + in_1 T + n_2 (-UT + V^2) + i b V  \;\;.
\eeqa 
Here, $l=(n_1,m_1,n_2,m_2,b)$ denotes the set of integral quantum numbers 
carried by the BPS state. 
The level matching condition for 
a BPS state reads 
\beqa
2(p^2_L - p^2_R) = 4 n^Tm + b^2  \;\;.
\eeqa
Of special relevance to the computation of perturbative
corrections to $F_1$ are those BPS states, whose tree-level mass vanishes
at certain surfaces/lines in the perturbative
moduli space ${\cal H}_2=\frac{SO(3,2)}{SO(3)\times SO(2)}$.
Note that ${\cal M}=0$ is the
condition (see appendix \ref{hump}) for a rational quadratic 
divisor
\beqa
{\sc H}_l=\{\left( \begin{array}{cc}i T &i V \\i V &i U \end{array} \right)
\in {\cal H}_2| m_2 - i m_1 U + in_1 T + n_2 (-UT + V^2) + i b V =0 \}
\eeqa 
of discriminant
\beqa
{\rm D}(l)= 2(p^2_L - p^2_R) = 
4m_1n_1+4n_2m_2 + b^2 \;\;.
\eeqa
Consider, for instance, BPS states becoming massless at the surface $V=0$,
the so-called Humbert surface $H_{\rm 1}$ (cf. appendix \ref{hump}).
They lay on the orbit ${\rm D}(l)=1$, i.e. on the orbit $n^Tm=0, 
b^2=1$.  On the other hand, BPS states becoming massless
at $T=U$, the Humbert surface $H_{\rm 4}$, lay on the orbit ${\rm D}(l)=4$,
which means that  
they carry quantum numbers satisfying $n^T m=1, b^2=0$ \cite{CCL}.

\newpage

\section{The duality}

\resetcounter

The assertion of duality states 
that $N=2$ heterotic vacua are  quantum 
equivalent to $N=2$ type IIA vacua \cite{KV}. 
Of course this can be combined - as is done in the actual procedure
of establishing the duality checks - with the perturbative 
equivalence of IIA with IIB on the mirror $CY$.\\
Note first of all, for heterotic vacua the rank
of the gauge group is bounded by the central charge
to be less than 24 (eq.~(\ref{rankhet}))
while in type II vacua the rank can certainly be
much larger since both Hodge numbers easily exceed
22.
Furthermore, we saw that the heterotic vacua have large 
non-Abelian gauge groups at special points in their
moduli space while type II A vacua only have an
Abelian gauge group.
However, the analysis of Seiberg and Witten \cite{SW}
taught us that asymptotically free
non-Abelian gauge groups
generically do not survive non-perturbatively but instead 
are broken to their Abelian 
subgroups.
Conversely one has to show that in a particular limit
corresponding to the heterotic weak coupling limit
a type II vacuum can have a non-Abelian enhancement
of its gauge group \cite{A,Wi}.

Furthermore one has to keep in mind that
the prepotential on the heterotic side 
$\cF_{het}$ is only known perturbatively,
that is in a weak coupling expansion. 
For a heterotic vacuum weak coupling
corresponds to large $S$ and hence there 
has to be a type II modulus in a
vector multiplet (so this cannot be the type II dilaton which 
sits in a hypermultiplet)
which is identified
with the heterotic dilaton; i.e. it has to be one of the $h_{1,1}$
K\"ahler deformations of the Calabi--Yau threefold.
Thus one is interested in identifying 
this dual type II partner $t^s$
of the heterotic dilaton $S$.
From the discrete PQ-symmetries 
one immediately infers
that the relation must be 
\be
t^s\equiv 4 \pi i S\ .
\ee

Once $t^s$ has been identified one can expand
the type IIA prepotential $\cF_{IIA}$ in a 
$t^s$ perturbation expansion around large $t^s$,
i.e.
\be
\cF_{II}=\cF_{II}(t^s,t^i)+\cF_{II}(t^i)
+\cF_{II}(e^{-2\pi i t^s},t^i) \; .
\ee

This expansion can be compared to the 
perturbative expansion of the heterotic prepotential.
In particular one has to find\footnote{There can still be 
non-perturbative inequivalencies
\cite{AG,BKKM,LSTY,MV}.}
\be\label{dualcheck}
\cF_{II}(t^s,t^i)+\cF_{II}(t^i)
= \cF_{\rm het}^{(0)}(S,\phi^I)
 +\cF_{\rm het}^{(1)}(\phi^I)
\ee
(up to an appropriate overall normalization).
Note that the left hand side is determined at tree level 
whereas the right hand side sums 
perturbative contributions at tree level and at one-loop.

Now generic properties should  involve the
heterotic dilaton since it couples
universally for all heterotic vacua.
Indeed, from 
eqs.~(\ref{Ftreehet0}),(\ref{dualcheck}) 
one infers that the 
Calabi--Yau intersection numbers of a dual
type IIA vacuum have to respect that $D_s\cdot D_s=0$ and that
${\rm sign}(d_{sij})=(+,-,\ldots,-)={\rm sign}(\eta_{ij})$.
In addition, eqs.~(\ref{Ftreehet1}), (\ref{FCYhigher}) imply
$\int e_S\wedge c_2(Y)=24$.
These conditions imply that the Calabi--Yau manifold
has to be a $K3$-fibration \cite{AL,KLM}, i.e.
the Calabi--Yau is fibred
over a $P^1$ base with $K3$ fibers. 
The size of the $P^1$ is parameterized
by the modulus $t^s$ which is the type II dual of the
heterotic dilaton.
Over a finite number of points on the base,
the $K3$ fibre can degenerate to a singular $K3$.
The other K\"ahler moduli $t^i$ are either moduli 
of the $K3$ fibre or of the singular fibers.
In general one finds
${\rm sign}(d_{sij})=(+,-,\ldots,-,0,\ldots,0)$,
where the non-vanishing entries correspond to moduli
from generic $K3$ fibers while
the zeros arise from singular fibers.
Since a $K3$ has at most 20 moduli
the non-vanishing entries have to be less than
$20$.\footnote{There is a possible subtlety
here since this counts only geometrical $K3$ moduli.
However, it is conceivable that quantum effects
raise this number up to 22 \cite{AM}.}\\
Note that the $K3$ fibration is intuitively understood to arise by the argument
of adiabatic extension of the $6D$ duality between the heterotic string on 
$T^4$ and type IIA on $K3$ \cite{VW}. Namely the heterotic compactification
space $K3 \times T^2$ can be seen as fibration of $T^2_{K3}\times T^2$ over
$P^1$ corresponding then to the type IIA picture.\\
Now the puzzle concerning non-abelian gauge symmetry enhancement 
in type IIA compactifications is easily understood. The 
non-abelian gauge symmetry enhancement arises for the heterotic string
at weak coupling by the Frenkel-Kac-Segal mechanism. As the heterotic
dilaton parameter corresponds to the size of the base $P^1$ of the Calabi-Yau
on the type IIA side the weak coupling regime of the heterotic string
is mapped to the 6D decompactification limit, i.e. effectively to type IIA
on $K3$. There the enhanced gauge symmetry can be understood as arising
from the $BPS$ soliton states given membranes wrapped around 2-cycles of
$K3$. Namely the mass of theses $BPS$ states is proportional to the area 
of the 2-cycle, so as the 2-cycle shrinks to zero size the new state appears
in the massless spectrum. Furthermore the local description of these
degenerate $K3$ surfaces follows the pattern of $ADE$ singularities, 
so that the sought for heterotic gauge symmetry structure can arise from the 
intersection lattice $E_8 \oplus E_8 \oplus H \oplus H \oplus H$ of $K3$
(with the negative definite $E_8$ lattice and 
$H={\tiny \left(\begin{array}{cc}  0&1\\1&0\end{array}\right)}$
the hyperbolic plane).\\
So the following picture about the matching of the vector moduli arises.
$S_{het}$ corresponds to the cohomology class of the base $P^1$ of the 
$K3$ fibration of the Calabi-Yau; eqivalently it corresponds to the 
4-cycle represented by the generic $K3$ fibre. The perturbative
heterotic gauge fields correspond to the 4-cycels which arise from 
2-cycels of the generic $K3$ which are invariant under the monodromy group
$\Gamma$
of the base $P^1$ (the fundamental group of the base $P^1$ with the finitely
many points over which the $K3$ degenerates deleted) and so can be extended to
4-cycels of the Calabi-Yau. Lastly the nonperturbative heterotic gauge fields
correspond to classes coming from the degenerate fibers.\\
Let us follow the connection of the classical heterotic gauge group and 
the classes in the Picard lattice
${\rm Pic}:={\rm Pic}(K3)^{\Gamma}$ more closely 
(${\rm Pic}(K3)=H^2(K3,{\bf Z})\cap
H^{1,1}(K3)$); we will denote the rank of ${\rm Pic}$ by $\rho$, the socalled
Picard number. The moduli space of K\"ahler forms is then \cite{A} given by
the Grassmanian of space-like two-planes in $\Lambda \otimes_{\bf Z} {\bf R}$
where the lattice $\Lambda$ is defined by 
$\Lambda:={\rm Pic} \oplus H \in H^*(K3,{\bf Z})$ 
with the $H$-part corresponding to $H^0 \oplus H^4$.
Now as the signature of ${\rm Pic}$ is $(1,\rho -1)$ one has
(globally, i.e. up to division by a discrete group)
\beqa
{\cal M}_{\mbox{K\"ahl}}(K3)=\frac{O(2,\rho)}{O(2)\times O(\rho)}
\eeqa
and the roots of the gauge group are giben by
$\{ \alpha \in \Lambda |\alpha ^2=-2 \; \mbox{and}\; \alpha \in {\cal V} \}$, 
where 
${\cal V}$ is the space-like two-plane in $\Lambda \otimes_{\bf Z} {\bf R}$.\\
For example in our three-parameter model $P_{1,1,2,8,12}(24)$ with
$K3$ given by $P_{1,1,4,6}(12)$ one has $\rho =2$ coming from the 
generic hyperplane section and the blow-up curve $C$ of the ${\bf Z}_2$
singularity due to the common factor in the last two weights (correspondingly
one has for the $K3$'s of the two- and four-parameter model that $\rho =1$
and $\rho =3$). One finds $\Lambda =H \oplus H$ and 
${\cal M}=\frac{O(2,2)}{O(2)\times O(2)}$ (up to $O(2,2;{\bf Z})$ modding)
and as the size parameter of $C$ goes to zero a $SU(2)$ arises which
corresponds to the heterotic $SU(2)$ arising for $T-U\rightarrow 0$.

\newpage

\subsection{ST model}

Now the vector multiplet 
couplings are known for the type II vacuum \cite{CdlOFKM,HKTY},
whereas for the heterotic vacuum they have been studied pertubatively only
\cite{AFGNT,dWKLL}. Recall that in the weak coupling expansion

\begin{eqnarray}
{\cal F}^{het}=\frac{1}{2}ST^2+h(T)+h^{np}(e^{-8\pi ^2S},T)
\end{eqnarray}

the dilaton independent 1-loop correction was found to transform (up to
a quartic polynomial) under the $SL(2,{\bf Z})_T$ as a modular form
of weight -4. The enhancement at $T=1$ of the gauge group $U(1)^3$ 
(including the graviphoton) to $SU(2)\times U(1)^2$ is reflected in 
the singularity of h at $T=1$ (besides the one at $T=\infty$); because
of the modular property one gets then from the logarithmic singularity
$\partial ^2_Th\sim \frac{1}{2\pi ^2}\log (T-1)$ of $\partial ^2_Th$
that

\begin{eqnarray}
\partial ^2_Th=\frac{1}{4\pi ^2}\log (j(iT)-j(i))+\mbox{finite}
\end{eqnarray}

As the dilaton is in contrast to the tree level situation no longer
invariant under T-duality one defines a corrected modular invariant
coordinate with the correction term $\sigma =S^{inv}-S$ given by (up
to a constant)
      
\begin{eqnarray}
\sigma =\frac{1}{3}(\partial ^2_Th-\frac{1}{4\pi ^2}\log (j(iT)-j(i)))
\end{eqnarray}

Now the fact that the discriminant factor (conifold part)
\begin{eqnarray*}
(1-\bar{x})^2-\bar{x}^2\bar{y}
\end{eqnarray*}
 
shows a complete square in $\bar{x}$ splitting as one moves $\bar{y}$
away from 0 was one of the identification properties used to match the 
variables in the string dual pair \cite{KV} as this is precisely the 
stringy realization of the expected Seiberg-Witten behaviour for
a $SU(2)$ $N=2$ gauge theory, where the isolated singular point of
classical $SU(2)$ restauration is split into two singular points
of massless monopoles in the full quantum theory. 
So one was led at first \cite{KV} to
identify $\bar{y}=e^{-S}$ as $\bar{y}=0$ corresponds to large S and
(at least there) the classical enhancement point $T=1$ with $\bar{x}=1$. 

Now the 1-loop corrections of the prepotential can be matched 
\cite{KV} on the basis of the common occurence of the 
j-function on both sides of
the proposed dual pair: coming on the heterotic side from the behaviour
under T-duality, and on the Calabi-Yau side from the mentioned $K3$ fibre
structure.
We get the identification of the modular invariant
dilaton (cf. also \cite{dWKLL}) by matching the two prepotentials
at weak coupling

\beqa
\partial ^2_T {\cal F}^{het}=S+\partial ^2_T h \;\;\;\;\;\;\;\;\;\;\;\nonumber
\eeqa
and
\beqa
\partial ^2_T {\cal F}^{II}=S+\frac{1}{4\pi ^2}\log (j-\alpha)-
\frac{3}{2}\frac{1}{4\pi ^2}\log g \nonumber
\eeqa

so that we have in view of

\beqa
S^{\rm inv}&=&
S+\frac{1}{3}[\partial ^2_T h-\frac{1}{4\pi ^2}\log (j-\alpha)]\nonumber\\
 &=&S-\frac{1}{8\pi ^2}\log g
\eeqa

or with $\check{S}^{\rm inv}=4\pi S^{\rm inv}$

\begin{eqnarray*}
t &=& t_1=iT\\
y &=& q_2g(q_1)=e^{-8\pi^2S}g=e^{-8\pi^2S^{\rm inv}}=q(i\check{S}^{\rm inv}) .
\end{eqnarray*}

with $t_1=iT$ and $t_2=i\check{S}^{inv}=4\pi iS^{inv}$.\\

Now after matching (at weak coupling) the prepotentials describing gauge
couplings one can match
the $F_1$ functions describing gravitational couplings.
One gets with
$b_{grav}=48-\chi=300$

\begin{eqnarray*}
F_1&=&24S^{inv}+\frac{1}{4\pi ^2}\log (j(iT)-j(i))
-\frac{300}{4\pi ^2}\log \eta ^2(iT)\nonumber\\
 &=&\frac{6}{4\pi ^2}\log 
[y^{-\alpha}\frac{(j-\alpha)^{\beta}}{\eta ^2(t_1)^{\gamma}}]
\end{eqnarray*}

with $\alpha =2,\beta =\frac{1}{6},\gamma =\frac{1}{6}b_{grav}$, i.e.
$\alpha /\beta =12, \gamma /\beta =b_{grav}$ showing 
(after a suitable overall normalization)
the coincidence
with the type II side (the normalization constant $24$ of $S^{inv}$ 
corresponds to the Euler number of $K3$, the modularity compensation
constant $b_{grav}$ is matched).

\newpage

\subsection{STU model}

Let us now compare the heterotic and the type II prepotentials and identify
the $t_i$ ($i=1,2,3$) with $S$, $T$ and $U$.
In the following,  we will actually match $ - 4 \pi { F}_0^{het}$
with ${F}_0^{II}$.

First compare the cubic terms in (\ref{ftype2}) and (\ref{fhet}).
By assuming that the $t_i$ and $S$, $T$ and $U$ are linearly related,
the following identification 
between the K\"ahler class moduli and the heterotic moduli is enforced
by the cubic terms
\beqa
t_1 &=& U  \nonumber\\
t_3 &= &T-U\nonumber\\
t_2 &=& 4\pi S^{inv}_\infty=\check{S} +\beta T + \alpha U 
\label{t2}
\eeqa
where $\check{S}=4 \pi S$. 
Recall that we are working inside the K\"ahler cone
$\sigma(K)=\{\sum_i t_i J_i | t_i > 0\}$.  
Now, in the heterotic
weak coupling limit one has that indeed $t_2 >0$. Demanding
$t_3 >0$ implies that one is choosing the chamber $T>U$ on the heterotic side.
The identification of $t_1$ and $t_3$ agrees, of course, with the one of 
\cite{KLM}. The identification of $4\pi S^{inv}_\infty$ with the K\"ahler
variable $t_2$ becomes very natural when performing the map to
the mirror Calabi-Yau compactification with complex structure
coordinates $x,y,z$. Here, since $y$ is invariant under
the CY monodromy group, $y$ should be identified \cite{KLM}
with $e^{- 8 \pi^2 S^{inv}}$.
Thus, equation (\ref{sinv}) provides the explicit mirror map;
for large $T,U$ the K\"ahler variable $q_2 = e^{- 2 \pi t_2}$ and the 
complex structure field $y$ completely agree.

Next, consider the exponential terms in the prepotential $F_0$.
In the heterotic weak coupling limit $S \rightarrow \infty$, one has that
$t_2 \rightarrow \infty$ and, hence, $q_2 = e^{-2 \pi t_2} \rightarrow 0$.
Then, (\ref{ftype2}) becomes 
\beqa
{F}_0^{II} = {\cal F}^0 - \frac{1}{(2 \pi)^3} \sum_{d_1,d_3}
n^r_{d_1,0,d_3} Li_3(q_1^{d_1}q_3^{d_3}) .
\label{ftype2weak}
\eeqa
Some of the instanton coefficients contained in (\ref{ftype2weak}) are
as follows \cite{HKTY}
\beqa
n^r_{d_1,0,0} &=&n^r_{d_1,0,d_1}= 480 = -2 (-240) \;\;,\;\; \nonumber\\
n^r_{0,0,1} &=& -2 \;\;\;,\;\;
n^r_{0,0,d_3} = 0 \;\;\;,\;\;\; d_3 = 2, \dots , 10 \;\;;\;\; \nonumber\\
n^r_{2,0,1} &=& 282888 = -2 (-141444) \;\;,\;\; \nonumber\\
n^r_{3,0,1} &=& n^r_{3,0,2} = 17058560 = -2 (-8529280) \;\;,\;\; \nonumber\\
n^r_{4,0,1} &=& 477516780 = -2 ( - 238758390) \;\;.\;\;
\eeqa
Note that the fact that $n^r_{d_1,0,0} = n^r_{d_1,0,d_1}$ is a reflection of 
the $T \leftrightarrow U$ exchange symmetry.
Now rewriting $kT+lU = (l+k)U + k (T-U)= (l+k) t_1 + k t_3$ and matching 
${ F}_0^{II} = - 4 \pi  {F}_0^{het}$ yields the following identifications
\beqa
d_1 &=& k + l  \;\;,\;\;  d_3 = k 
\eeqa
and so 
\beqa
n^r_{d_1,0,d_3} &=& n^r_{k+l,0,k} = -2 c_1(kl) .
\label{prepins}
\eeqa
Note that $d_3 = k \ge 0$ for points inside the K\"ahler cone.  Also, if 
$d_3=k=0$,
then $d_1 = l >0$.  On the other hand, if $d_3=k >0$, then 
$d_1 \ge 0$, that is $l \ge -k$.

Comparison of the instanton coefficients listed above
with the $c_1$-coefficients
ocurring in the $q$-expansion of $F(q)=\frac{E_4 E_6}{\eta^{24}}$ in 
equation (\ref{fq}),
shows that the relation (\ref{prepins}) is indeed satisfied.

Let us now determine the action of the symmetries (\ref{tu})
and (\ref{st}) on the heterotic variables.
Clearly the perturbative symmetry (\ref{tu}) corresponds
to the exchange $T\leftrightarrow U$ for $S\rightarrow\infty$.
The non-perturbative symmetry (\ref{st})
corresponds to 
\beqa 
S&\rightarrow & -(1+\beta )S-{\alpha (2+\beta)\over 4\pi}
U-{\beta(2+\beta)\over 4\pi}T\nonumber\\
T&\rightarrow &4\pi S+(1+\beta )T+\alpha U\nonumber\\
U&\rightarrow &U .
\label{hetst}
\eeqa
There is one very convenient choice for the parameters $\alpha$ and
$\beta$, in which the non-perturbative symmetry (\ref{hetst})
takes a very simple suggestive form. Namely, for $\alpha=0$ and 
$\beta=-1$, this transformation becomes
\beqa
4\pi S\leftrightarrow T,\label{stsim}
\eeqa
that is, it just describes the exchange of the heterotic dilaton $S$ with
the K\"ahler modulus $T$ of the two-dimensional torus.
This choice for $\alpha$ and $\beta$ is very reasonable,
since it is only in this case
that the real 
parts of $S$ and $T$ remain positive after the exchange 
(\ref{hetst}).
At the end of this paper, by considering \cite{AlFoIbaQue2}
some  six-dimensional one-loop gauge couplings, we will give some further 
arguments indicating that the choice
$\beta=-1$ is the physically correct one.
So, for the time being, we will set $\alpha=0$ and $\beta=-1$ 
and discuss a few issues related to  the exchange symmetry $4\pi S
\leftrightarrow T$.

The non-perturbative
exchange symmetry $4\pi S\leftrightarrow T$ 
is true for arbitrary $U$ in the chamber $S,T >U$.
As already discussed in detail in \cite{KLM}, 
at the fixed point $t_2=S^{inv}_\infty=0$ of 
this transformation, one has that $S=T>U$, the complex structure
field $y$ takes the value $y=1$, and the discriminant of the 
Calabi-Yau model vanishes. The locus
$S=T>U$ corresponds to a strong coupling
singularity with additional massless
states. In the model based on the Calabi-Yau space $P_{1,1,2,8,12}(24)$,
a non-Abelian gauge symmetry enhancement with an equal number of
massless vector and hypermultiplets takes place at $S=T>U$,
such that the non-Abelian $\beta$-function vanishes \cite{KaMoPle,KM}.

On the other hand, the non 
perturbative exchange symmetry $4\pi S\leftrightarrow
T$ implies that for  $T\rightarrow\infty$ there is a `perturbative' 
$4\pi S\leftrightarrow U$ exchange symmetry. This symmetry is nothing
but the $T-S$ transformed perturbative symmetry (\ref{tu}).
Furthermore, for $T\rightarrow\infty$, there is a modular symmetry
$SL(2,{\bf Z})_S\times SL(2,{\bf Z})_U$ and the corresponding 
`perturbative' monodromy
matrices of the prepotential can be computed in a straightforward way. 
Hence, for $T\rightarrow\infty$, there is a `perturbative'
gauge symmetry enhancement of either $U(1)^2$ to $SU(2)\times U(1)$ or
to $SU(2)^2$ or to $SU(3)$ at the points $S=U$, $S=U=1$ or $S=U=e^{i\pi /6}$,
respectively, with no additional massless hyper multiplets \cite{CCLMR}.\\

Let us now investigate the gravitational coupling $F_1$. 
The non-exponential piece, which dominates for large $t_i$,
reads \cite{KM}
\beqa
- i \sum_{i=1}^3t_i c_2\cdot J_i=92 t_1+24 t_2+48 t_3.\label{largef}
\eeqa
This expression is explicitly invariant under the non-perturbative
symmetry (\ref{st}).
Furthermore, by also explicitly
 checking some of the elliptic instanton numbers $n_{d_1,d_2,d_3}^e$,
one discovers that, just like in the case of $n^r$,  
\beqa
n^e_{d_1,0,d_3}=n^e_{d_1,0,d_1-d_3} \qquad {\rm and}\quad 
n^e_{d_1,d_2,d_3}=n^e_{d_1,d_3-d_2,d_3}.\label{inssyme}
\eeqa
It follows that $F_1^{II}$ is symmetric under the two exchange
symmetries (\ref{tu}) and (\ref{st}).\\

Let us now compare the heterotic and the type II gravitational
couplings \cite{CCLM}.
We will  match $4\pi F_{1}^{het}$ with $F_1^{II}$.
First take the decompactification limit to $D=5$,
i.e. the limit $T,U\rightarrow\infty$ ($T>U$). This eliminates all
instanton contributions, i.e. all exponential terms.
In the heterotic case we get in this limit
\beqa
F_{1}^{het}\rightarrow F_{1}^\infty=
24S_{inv}^\infty+{12\over \pi}T+{11\over\pi}U=
24S+{12+6\beta\over \pi}T+{11+6\alpha\over 4\pi}U.
\label{fgravinf}
\eeqa
By comparing this expression with the type II large $t_i$ 
limit given in (\ref{largef}), one finds that (\ref{largef}) and
(\ref{fgravinf}) match up precisely for 
the identification given in (\ref{t2}) between heterotic and type II
moduli.  When
choosing $\alpha=0$ and $\beta=-1$, it follows that
$F_{1}^\infty$ is symmetric under
the exchange $4\pi S\leftrightarrow T$.\footnote{In \cite{CCLMR}
a different choice was made for these two parameters, namely 
$\alpha=-11/6$ and $\beta=-2$. Hence it follows 
that $F_{1}^\infty=24 S$.}
This symmetry implies that in the limit $T\rightarrow\infty$, $F_{1}^{het}$
can be written  \cite{CCLMR}
in terms of $SL(2,{\bf Z})_S$ 
modular functions $j(4\pi S)$ and $\eta (4\pi S)$ 
by just replacing $4\pi S$ with $T$ in equation (\ref{fexp}).

Next, let us compare the exponential terms in $F_1^{II}$ and $F_{1}^{het}$.
In the type II case we 
have to consider the weak coupling limit $q_2\rightarrow 0$;
hence only the terms with the instanton numbers $n_{d_1,0,d_3}^{r,e}$
contribute to the sum. We will see that, when comparing with the heterotic
expression, one gets a very interesting relation between the
rational and elliptic instanton numbers for $d_2=0$.
In order to do 
this comparison, we have to recall that $Li_1(e^{-2\pi(kT+lU)})=
-\log(1-e^{-2\pi(kT+lU)})$.  The difference $j(T) -j(U)$ can
be written in the following useful form (in the chamber $T>U$) 
\cite{BORCH,HM}
\beqa
\log(j(T)-j(U))=2\pi T+\sum_{k,l}c(kl)\log(1-e^{-2\pi(kT+lU)}),
\label{diffj}
\eeqa
where the integers $k$ and $l$ can take the following values \cite{HM}:
either $k=1, l=-1$ or $k>0, l=0$ or $k=0, l>0$ or $k>0,l>0$.
First consider the terms with $k=1,l=-1$ on the heterotic side.  
Matching the term $\log(1-e^{-2\pi{(T-U)}})$ contained in $4 \pi 
F^{het}_{1}$ with $F_1^{II}$ requires that
\beqa
10 c(-1) - 12 c_1 (-1) = 12 n^e_{0,0,1} + n^r_{0,0,1} .
\eeqa
This is indeed satisfied, since $c(-1) = c_1(-1) = 1$ and
$n^e_{0,0,1}=0, n^r_{0,0,1}=-2$.

Next, consider the terms in the sum with $k >0,
l=0$ (and analogously $k=0, l >0$). Since $c(0)=0$, only the term
${b_{grav}\over 8\pi^2}\log\eta^{-2}(T)$ contributes on the heterotic
side ($b_{grav}=528$). Matching $4\pi F_{1}^{het}$ with $F_1^{II}$
yields the following relation among the instanton numbers
($d_1=d_3=k$):
\beqa
12\sum_{i=1}^sn_{k_i,0,k_i}^e+n_{k,0,k}^r= b_{grav}=528.
\label{relf}
\eeqa
The $k_i$ ($i=1,\dots , s$) are the divisors of $k$ ($k_1=k$, $k_s=1$).
Using Klemm's list of 
explicit instanton numbers one checks that this relation 
is indeed true ($n_{1,0,1}^e=4$, $n_{k,0,k}^e=0$ for $k>1$,
$n_{k,0,k}^r=-\chi=480$).

Finally, consider the case where $k>0,l>0$. By 
comparing the heterotic and type
II expressions we derive the following interesting
relation ($d_1=k+l$, $d_3=k$):
\beqa
12\sum_{i=1}^sn_{d_1^i,0,d_3^i}^e
&=&-n_{k+l,0,k}^r+10c(kl)+12klc_1(kl)=\nonumber\\ 
&=&10c(kl)+(12kl+2)c_1(kl).
\label{relt}
\eeqa
Here $s$ is the number of common divisors $m_i$
($i=1,\dots ,s$) of $d_1=k+l$ and $d_3=k$
with $d_1^i=d_1/m_i$ and $d_3^i=d_3/m_i$ (where $m_1=1$).
Again we can explicitly check the non-trivial relation (\ref{relt})
for the first few terms. For example, for $k=l=1$
one has
\beqa
n_{2,0,1}^e=-948 \qquad{\rm and}\qquad n_{2,0,1}^r=282888,
\eeqa
which, together with equations (\ref{fq}) and (\ref{cthree}), confirms the
above relation.
For $k=2$ and $l=1$ one finds that 
$12 n_{3,0,2}^e  +  n_{3,0,2}^r = 10 c(2) + 24 c_1(2)$
is indeed satisfied, since
\beqa
n_{3,0,2}^e=-568640 \qquad {\rm and}\qquad n_{3,0,2}^r=17058560.
\eeqa
And finally, for $k=l=2$ for instance, one finds that the relation
$12 \left(n_{4,0,2}^e  +  n_{2,0,1}^e\right) + 
n_{4,0,2}^r = 10 c(4) + 48 c_1(4)$ indeed holds due to
\beqa
n_{4,0,2}^e=-1059653772 \;\;\;,\;\;\;
n_{2,0,1}^e = -948
\qquad {\rm and}\qquad n_{4,0,2}^r=8606976768.
\eeqa
One can (cf. appendix A1)
rewrite  equation (\ref{relt}) as
\beqa
12\sum_{i=1}^sn_{d_1^i,0,d_3^i}^e+n_{d_1,0,d_3}^r
=10c(kl)+12klc_1(kl)=-2\tilde{c}_1(kl)\;,\; k>0, l>0.
\eeqa

At the end let us 
briefly comment on the relation
to the heterotic/heterotic duality in six dimensions 
\cite{AlFoIbaQue2,DuMinWit}
with the 6-dimensional heterotic string compactified on $K_3$.
(The decompactification limit from $D=4$ to $D=6$ is obtained 
by sending $T\rightarrow \infty$ with $U$ finite;
as discussed in \cite{Duff,DuMinWit}, the 
$D=6$ heterotic/heterotic duality becomes an
exchange symmetry of $S$ with $T$ in $D=4$.)
We will concentrate on CY's 
with Hodge numbers (3,243), which are elliptic fibrations
over ${\bf F}_n$ with $n=0,2$ \cite{MV,SeiWit}. 
Being elliptic fibrations, they can be used to
compactify $F$-theory to six dimensions.
In  the $D=6$ heterotic string, 
the integer $n$ is related to the number $s$ of $SU(2)$ instantons
in one of the two $E_8$'s by $n=s-12$ \cite{MV,SeiWit}.
First consider the case of an elliptic fibration
over ${\bf F}_0$, corresponding to  the symmetric embedding of
the $SU(2)$ bundles with equal instanton numbers $s=s'=12$ into $E_8\times
E_8'$. This leads to a $D=6$ heterotic model with
gauge group $E_7\times E_7'$ with $\tilde v_\alpha
=\tilde v_\alpha '=0$ \cite{DuMinWit}. 
There are 510 hypermultiplets transforming as $4(56,1)+4(1,56)+62(1,1)$.
The heterotic/heterotic duality originates
from the existence of small instanton configurations \cite{Wit2}. 
The model is, however, not self-dual, since the non-perturbative gauge groups
appear in different points of the hyper multiplet moduli space than the
original gauge groups \cite{DuMinWit}. 
For generic vev's of of the hyper multiplets the gauge group is
completely broken and one is left with 244 hyper multiplets and no vector
multiplets.
Upon compactification to $D=4$ on $T_2$ one arrives
at the heterotic string with gauge group $U(1)^4$, 
which is the dual to the considered 
type II string on the CY $P_{1,1,2,8,12}(24)$.
Semiclassically, at special points in
the hypermultiplet moduli space, this gauge group can be enhanced to
a non-Abelian gauge group, inherited from the $E_7\times E_7'$
with $N=2$ $\beta$-function coefficient 
$b_{\alpha}=12(1+{\tilde v_{\alpha}\over v_{\alpha}})=12$ \cite{AlFoIbaQue2}.

Next, consider embedding the $SU(2)$ bundles in an asymmetric way into the
two $E_8$'s \cite{AlFoIbaQue1,AlFoIbaQue2}: $s=14$, $s'=10$. 
This corresponds to the elliptic
fibration over ${\bf F}_2$ \cite{MV,SeiWit}. 
Note that ${\bf F}_0$ and ${\bf F}_2$
are connected by deformation
\cite{MV}, as
we will discuss in the following.
Then, in this case, one has \cite{AlFoIbaQue2} a gauge group
$E_7\times E_7'$ with $\tilde v_\alpha=1/6$ and $\tilde v_\alpha '=-1/6$
and hyper multiplets transforming as
$5(56,1)+3(1,56)+62(1,1)$.
The second $E_7$ can be completely Higgsed away, leading to a 
$D=6$ 
heterotic model
with gauge group $E_7$ and hypermultiplets $5(56)+97(1)$.
As explained in \cite{AlFoIbaQue2},
this model also possesses a heterotic/heterotic
duality, however without involving non-perturbative small instanton
configurations. Hence in this sense, this model is really self-dual.
Just like in the case of the symmetric embedding, the gauge group $E_7$
is spontaneously broken for arbitrary vev's of the gauge non-singlet
hyper multiplets and one is again left with 244 hyper multiplets and
no vector multiplet. When compactifying on $T_2$ to $D=4$, one obtains
the same
heterotic string model with $U(1)^4$ gauge group as before.
For special values of the hyper multiplets a non-Abelian gauge group
is obtained, now however with 
$\beta$-function coefficient 
$b_\alpha=12(1+{\tilde v_\alpha\over v_\alpha})=24$ \cite{AlFoIbaQue2}.

In summary, the symmetric (12,12) model and the asymmetric (14,10) model
should be considered as being the same \cite{AlFoIbaQue2,MV},  since both 
are related by
the Higgsing and both lead
to the same heterotic string
in $D=4$. 

As already mentioned, we  would like to provide a six-dimensional
argument for why  $\beta=-1$ is the physically correct choice
for one of the cubic parameters.
We will 
directly follow the discussion given in \cite{AlFoIbaQue2} 
and consider the one-loop gauge
coupling for the enhanced non-Abelian gauge groups that are inherited
from the six-dimensional gauge symmetries.
Specifically, the gauge kinetic function
is of the form \cite{AlFoIbaQue2,dWKLL}
\beqa
f_\alpha=S^{inv}-{b_\alpha\over 8\pi^2}\log(\eta(T)\eta(U))^2.
\label{gaugec}
\eeqa
Using equation (\ref{largetsinv}) this then
in the decompactification limit
$T\rightarrow\infty$ to $D=6$
becomes
$f_\alpha\rightarrow S+{1+\beta+{\tilde v_\alpha\over v_\alpha}\over 4\pi}T$.
By comparing this expression 
with the six-dimensional gauge coupling \cite{Sagn}, it then follows that
$\beta = -1$.

\newpage

\subsection{STUV model}

In order to match (\ref{cubicf})
with the cubic part of the heterotic prepotential given in 
(\ref{prepstuv}), we will perform the following identification of
type II and heterotic moduli (which differs from the one 
given in \cite{BKKM}):
\beqa
t_1&=&U-2V,\qquad t_2= S-{n\over 2}T-\left(1-{n\over 2}\right)U,\nonumber\\
t_3&=&T-U,\qquad t_4=V  \;\;,
\label{coordin}
\eeqa
which is valid in the chamber $T>U>2V$.  
Then, (\ref{cubicf}) turns into
\beqa
F^{II}_{\rm cubic}= - F^{\rm het}_{\rm cubic} = S(TU-V^2)
+{1\over 3}U^3+\left({4\over 3}+n\right)V^3-\left(1+{n\over 2}\right)UV^2
-{n\over 2}
TV^2 \;.\label{cubicfa}
\eeqa
Note that, using the heterotic moduli, the prepotential is independent of $n$
in the limit $V=0$.

The heterotic weak coupling limit $S \rightarrow \infty$ corresponds
to the large K\"ahler class limit $t_2 \rightarrow \infty$.  In this
limit, only the instanton numbers with $d_2=0$ contribute
in the above sum.  Using the identification 
$kT+lU+bV=d_1t_1+d_3t_3+d_4t_4$, it follows that (independently of $n$)
\begin{eqnarray}
k&=&d_3 \;\;, \nonumber\\
l&=&d_1-d_3   \;\;,\nonumber\\
b&=&d_4-2d_1  \;\;.
\end{eqnarray}
Then, (\ref{instanton}) turns into
\beqa
F^{II}_{\rm inst}=-{1\over (2\pi)^3}\sum_{k,l,b}n^r_{k,l,b}
Li_3(e^{-2\pi(kT+lU+bV)}) \;\;.
\label{largespr}
\eeqa

The rational instanton numbers have now to satisfy
\beqa
n^r_{k,l,b} = - 2 c_n(4kl-b^2) \;\;.
\label{insmod}
\eeqa
Note that 
the constraint (\ref{constraint}) is non-trivial.
Also note that $2c_n(0)=\chi(X_n)$ and 
recall that $2c_n(-1)=-n_H' \,,\, 2c_n(-4)=n_V'$.\\

For concreteness, let us now check the above relations 
for the four-parameter model of
\cite{BKKM}, which has a dual type II description based on the
Calabi--Yau space $X_2=P_{1,1,2,6,10}(20)$. 
Using the instanton numbers given in \cite{BKKM}, it can be
checked that both (\ref{constraint}) and (\ref{insmod}) for $c_2$
indeed hold, as can be seen from the second 
table in appendix A.6 and the table given below.\\

\hspace{0.5cm}

\begin{center}
\begin{tabular}{|ccc|ccc||c|c|} \hline
$d_1$ & $d_3$ & $d_4$ & $k$ & $l$ & $b$ & $N=4kl-b^2$ &$n_{d_1,0,d_3,d_4}$\\ 
\hline\hline
  0   &   0   &   3   & 0 & 0 & 3 &      --9     &          0        \\ \hline
  0   &   1   &   0   & 1 & --1 & 0 &      --4     &         --2        \\
  0   &   0   &   2   & 0 & 0 & 2 &      --4     &         --2        \\
  1   &   0   &   0   & 0 & 1 &--2 &      --4     &         --2        \\
  1   &   0   &   4   & 0 & 1 &2 &      --4     &         --2       \\  \hline 
  0   &   0   &   1   & 0 & 0 & 1 &      --1     &         56        \\
  1   &   0   &   3   & 0 & 1 & 1 &      --1     &         56        \\
  1   &   0   &   1   & 0 & 1 &--1 &      --1   &         56        \\  \hline
  1   &   0   &   2   & 0 & 1 & 0 &       0     &        372        \\  \hline
  2   &   1   &   3   & 1 & 1 &--1 &       3     &      53952        \\  \hline
  2   &   1   &   4   & 1 & 1 & 0 &       4     &     174240        \\  \hline
\end{tabular}
\end{center}

\hspace{0.5cm}

The truncation to the three-parameter Calabi--Yau model is made by setting 
$V=0$. The instanton numbers $n^{r}_{k,l}$ of the $S$-$T$-$U$ model are then 
given by \cite{BKKM}
\beqa
n^{r}_{k,l} = 
\sum_{b} n^r(4kl-b^2) \;\;,\label{truncsum}
\eeqa
where the summation range over $b$ is finite.  
For example, $n^r_{0,1}= -2 + 56 + 372 +56 -2 = 480$ \cite{BKKM}.\\

\newpage

\section{Summary and Outlook}

We have studied the duality between two different string theories,
the heterotic string on $K3 \times T^2$ and type IIA string on $CY$,
leading to $N=2$ SUSY in $D=4$. For this we picked specific $CY$'s
to match the spectra of the heterotic theories. Actually one can
extend such a study of different and related (by Higgs transitions)
models on each side to the matching of whole webs \cite{CF} 
of compactifications for both string theories.
Then we have presented quite explicit evidence that the proposed dual models
are indeed dual even on the level of the couplings $F_0$ and $F_1$.
This was possible by the use of the special functions occuring on the 
heterotic side, which are specific modular forms thanks to the operation
of the T-duality group, and the instanton expansions on the type II side.\\
Actually one can also make a closer connection to the corresponding field
theory results of Seiberg and Witten by taking the appropriate
field theory limit \cite{KKLMV,KLMVW}. Also in string theory similar 
comparisons are possible between the type I string and the $SO(32)$ 
heterotic string \cite{ABFPT}.\\
Especially fruitfull is furthermore the reformulation in $D=6$ of the 
duality considered here in $D=4$, which is possible thanks to
$F$-theory \cite{V-F,MV}. Namely, as type IIA on a $CY$ can be understood
as $F$-theory on $T^2 \times CY$, one can cancel the $T^2$ part on both
sides (if one has not used it in a special way like in the $ST$ model).
This leads to the duality between the heterotic string on $K3$ and 
$F$-theory on (an elliptically fibered) $CY$. \\
This makes it possible
to directly 'connect' these $N=2$ studies to the more complicated models  
of $N=1$ in $D=4$, which is of course 
the next step, especially in view of coming closer to
phenomenological relevance.  
There again one can study dual models: now heterotic on $CY$ and
on the type II side one has to use F-theory to make sense out of 
a non-$CY$ base (a $CY$ base would already give $N=2$). One possibility
\cite{CL}
is to orbifold known dual $N=2$ pairs by ${\bf Z}_2$. In this way one can
even again match the relevant holomorphic coupling, now given by the
superpotential, which is in the considered case again modular \cite{DGW}.\\
Another path one can follow already in the $N=2$ setup
is a closer study of the higher gravitational 
couplings and the geometrical information about numbers of higher genus curves
on the $CY$ contained in them \cite{AGNTtop,AGNTtop1,BCOV,CdWLMR}.

\newpage

\appendix

\section{Modular forms \label{modforms}}
\setcounter{equation}{0}

\subsection{Ordinary modular forms}

A modular form $F_r(T)$ of weight $r$ obeys the transformation law
\be\label{modform}
F_r(T) \to (icT+d)^r F_r(T) \ .
\label{mformdef}
\ee
One can show that there are no modular forms of weight 0 and 2 
while at weight 4 and 6 one has the Eisenstein functions
\bea
E_4 (q)&=&  1 + 240 \sum_{n=1}^\infty \frac{n^3 q^n}{1-q^n} \nonumber\\
        &=& 1 + 240 q + 2160 q^2 \ldots\ , \\
E_6 (q) &=&  1 - 504 \sum_{n=1}^\infty \frac{n^5 q^n}{1-q^n} \nonumber\\
        &=& 1 - 504 q -16632 q^2 \ldots\ , \nonumber
\label{Eisenstein}
\eea
where $q=e^{2\pi i(iT)}$.
Both function have no pole (including  $T=\infty$)
on the entire fundamental domain;
$E_4$ has exactly one simple zero at $T=\rho$ while 
$E_6$ has one simple zero at $T=1$. 
One can construct modular forms of arbitrary even weight from
products of these two Eisenstein functions.

The unique cusp form of weight $12$ is $\eta^{24}$, where
\be
\eta(q)=q^{1\over 24} \prod_{n=1}^\infty (1-q^n)\; ,
\label{etadef}
\ee
the Dedekind $\eta$-function
($\eta$ does not vanish at $\rho$ or 1).
The modular invariant 
$j$ function defined by
\bea
j(q)&=&  {E_4^3\over \eta^{24}} \nonumber\\
    &=& q^{-1} + 744 + 196884 q + \ldots
\label{jdef}
\eea
has a simple pole at $T=\infty$ and a triple zero at $T=\rho$.

One also defines a function
\beqa
E_2=1-24\sum_{n=1}^{\infty}\frac{nq^n}{1-q^n} 
\eeqa
which is not quite automorphic.\\
Likewise the derivative of a modular form does again not quite transform 
like a modular form, instead a corresponding correction term arises.\\
This leads to the relations 
$24\frac{1}{2\pi i}\partial _{\tau}\log \eta=E_2$,
$E_6/E_4=E_2-\frac{3}{2\pi i}\partial _{\tau}\log E_4$ 
from which it follows that $\frac{1}{2\pi i}\partial _{\tau}\log j=-E_6/E_4$
so that with $E_4^3-E_6^2=\alpha \eta^{24}\Rightarrow j-\alpha=
\frac{E_6^2}{\eta ^{24}}$ one has 
$\frac{1}{2\pi i}\partial _{\tau}\log j=-\frac{E_6E_4^2}{\eta ^{24}}
\Rightarrow (\frac{1}{2\pi i})^2 j_{\tau}^2=E_4j(j-\alpha)$, i.e.

\beqa
\frac{j_{\tau}^2}{j(j-\alpha)}=-4\pi ^2E_4 \; .
\eeqa

Note that one has despite the general difficulties with derivatives
nevertheless the following: 
$\partial ^{k+1} f_{-k}$ is a modular form of weight $k+2$
for $f_{-k}$ a modular form of weight $-k$.

The coefficients $c_1(n)$ are defined by
\beqa
{E_4E_6\over\eta^{24}} = 
\sum_{n \ge -1} c_1(n) q^n &=& \frac{1}{q} - 240 - 141444 q -
8529280 q^2 - 238758390 q^3 \nonumber\\
& &-4303488384q^4 + \dots\label{fq}
\eeqa

The coefficients $\tilde c_1(n)$ are defined
as follows \cite{HM}
\beqa 
E_2\frac{E_6E_4}{\eta^{24}}(q) 
=  \sum_{n=-1}^{\infty}\tilde c_1(n)q^n=\frac{1}{q}-264 -135756 q -
5117440 q^2 + 
\dots \label{tilc}
\eeqa

We will also make use of the remarkable Borcherds identity \cite{HM}
\beqa
\log(j(T)-j(U))=2\pi T+\sum_{k,l}c(kl)\log(1-e^{-2\pi(kT+lU)}),
\label{diffj}
\eeqa
where the universal constants $c(n)$ are defined as follows:
\beqa
j(q)-744 =
\sum_{n=-1}^\infty c(n)q^n  &=&{1\over q}+196884q+21493760q^2+
864299970q^3 \nonumber\\
& &+ 20245856256q^4 +\dots \label{cthree}
\eeqa

Now consider the relation $\biggl( \frac{E_6E_4}{\eta^{24}}
\biggr) '=-\frac{2\pi}{6}(\frac{E_2E_4E_6}{\eta^{24}}+2\frac{E_6^2}
{\eta^{24}}+3\frac{E_4^3}{\eta^{24}})$. 
From this we can, for $n >0$,
derive the useful equation 
$12nc_1(n)+10c(n)=-2\tilde{c}_1(n)$.

Furthermore we will use the polylogarithmic functions

\beqa
Li_1(x)&=&\sum_{j=1}^{\infty}\frac{x^j}{j}=-\log (1-x) \nonumber\\
Li_2(x)&=&\sum_{j=1}^{\infty}\frac{x^j}{j^2}=\int _0^1\frac{dt}{t}
\frac{1}{1-xt} \nonumber\\
Li_3(x)&=&\sum_{j=1}^{\infty}\frac{x^j}{j^3}=-\int _0^1\frac{dt}{t}
\int _0^1\frac{ds}{s}\log (1-xts) \nonumber \; .
\eeqa

Note that $\partial_{t_k} \partial_{t_l} Li_3=
(-2\pi)^2 d_k d_l Li_1$, where $Li_1(x) = - \log(1-x)$.

\subsection{Siegel modular forms \label{hump}}

The classical moduli space of a heterotic $STUV$ model
is locally given by the Siegel upper half-plane
${\cal H}_2=\frac{SO(3,2)}{SO(3)\times SO(2)}$ 
(note the exceptional isomorphism
$SO(5)=B_2=C_2=Sp(4)$, here in a non compact formulation).  The standard
action of $Sp(4,Z)$ on an element $\tau$ of the Siegel upper 
half-plane ${\cal H}_2$ is given by 
\beqa
M \rightarrow M\cdot \tau =(a\tau +b)(c\tau +d)^{-1} \;\;,
\eeqa
where $\tau ={\tiny \left( \begin{array}{cc}
\tau_1 & \tau_3 \\ \tau_3 & \tau_2
\end{array} \right)}
={\tiny \left( \begin{array}{cc}
iT & iV \\ iV & iU
\end{array} \right)},
M={\tiny \left( \begin{array}{cc} a & b \\ c & d \end{array} \right)} \;
\in \;G=Sp(4,Z)$
and where $\mbox{det}\;\mbox{Im}\tau=\mbox{Re}T\mbox{Re}U-(\mbox{Re}V)^2>0$.
Note that $a,b,c$ and $d$ denote $2 \times 2$ matrices.
A Siegel modular form $F$ of even weight $k$ transforms as
\begin{eqnarray}
F(M\cdot \tau )=\det(c\tau +d)^k  F(\tau)
\end{eqnarray}
for every $M \in  \;G=Sp(4,Z)$, whereas a modular form of odd weight $k$
transforms as 
$F(M\cdot \tau )=\varepsilon (M)\det(c\tau +d)^k F(\tau)$.
Here $\varepsilon :G\rightarrow G/G(2)=S_6\rightarrow \{\pm 1\}$ is the sign
of the permutation in $S_6$; $G(2)$ denotes the principal
congruence subgroup of level 2.

The Eisenstein series are given by
\begin{eqnarray}
{\cal E}_k=\sum \det (c\tau +d)^{-k} \;\; .
\end{eqnarray}

Now, recall that the usual 
modular forms of $Sl(2,{\bf Z})$ are generated by the 
(normalized) Eisenstein series $E_4$ and $E_6$.  These are 
related to the two
modular forms $E_{12}$ and $\Delta$
of weight $12$ by
\begin{eqnarray}   
aE_4^3+bE_6^2 &=& (a+b)E_{12}  \;\;,  \nonumber\\
E_4^3-E_6^2 &=& \alpha \Delta \;\; ,
\end{eqnarray}
where $\Delta=\eta^{24}$ is the cusp form, and where 
$a=(3\cdot 7)^2, b=2\cdot 5^3, c=a+b=691,
\alpha =2^6\cdot 3^3=1728$.

Similarly,
the ring of Siegel modular forms is generated by the (algebraic independent) 
Eisenstein series
${\cal E}_4, {\cal E}_6, {\cal E}_{10}, {\cal E}_{12}$ and by 
one further
cusp form of odd weight ${\cal C}_{35}$, whose square can again be expressed
in terms of the even generators.
Alternatively, instead of using
${\cal E}_{10}$ and ${\cal E}_{12}$, one can also use  
the cusp forms ${\cal C}_{10}$ and ${\cal C}_{12}$.

A Siegel cusp form is defined as follows.
Since a modular
form $f$ 
is invariant under the translation group ${\tiny U=\{ \left( \begin{array}{cc}
 1 & b \\ 0 & 1 \end{array} \right) \in G\}}$, where the integer-valued
$2\times 2$- matrix $b$ is symmetric, it has a Fourier expansion 
$F=\sum_M a(M)e^{2\pi i trM\tau}$.  Here, the summation extends over all 
symmetric half-integral $2\times 2$ matrices (that is
over symmetric matrices that have 
integer-valued diagonal entries and half-integer-valued off-diagonal
entries). The
Fourier coefficient $a(M)$ depends only on the class of $M$ under
conjugation by $Sl(2,{\bf Z})$, 
and it is zero unless $M$ is positive semidefinite. 

Now, consider the Siegel operator $\Phi$, which associates, to every Siegel 
modular
form $F$ 
with Fourier coefficients $a(M)$, the ordinary $SL(2,{\bf Z})$ 
modular form $\Phi F$ with Fourier
coefficients 
$a(n)=a({\tiny \left( \begin{array}{cc} n & 0 \\ 0 & 0 \end{array} \right)})$.
This yields a surjective homomorphism of graded rings of 
modular forms. The forms 
in the kernel are the cusp forms. Thus,
identities between ordinary modular forms
lead to Siegel cusp forms, as follows:
\begin{eqnarray}
E_4E_6=E_{10} &\rightarrow& {\cal E}_4{\cal E}_6-{\cal E}_{10}=:
p \, {\cal C}_{10}     \;\;,   \nonumber\\
aE_4^3+bE_6^2
=cE_{12}&\rightarrow& a{\cal E}_4^3+b{\cal E}_6^2-c{\cal E}_{12}=:
\alpha ^2\frac{ab}{c}{\cal C}_{12}  \;\;, 
\end{eqnarray}
where $p$ denotes a normalization constant given by 
$p=\frac{2^{10}\cdot 3^5
\cdot 5^2\cdot 7\cdot 53}{43867}$.  We will
drop this normalization constant in the following, for notational
simplicity.

Next, consider restricting the Siegel modular forms to the 
diagonal $D=\{ {\tiny \left( \begin{array}{cc} \tau_1 & 0 
\\ 0 & \tau_2 \end{array} \right) \}}$ (corresponding to the embedding
$\frac{SO(2,2)}{SO(2)\times SO(2)} 
\rightarrow \frac{SO(3,2)}{SO(3)\times SO(2)}$). Then, interestingly,
\begin{eqnarray}
{\cal E}_k \Big(
{\tiny \left( \begin{array}{cc} 
\tau_1 & 0 \\ 0 & \tau_2 \end{array} \right)} \Big)
=E_k(\tau_1)E_k(\tau_2) \;\;.
\end{eqnarray}
Specifically
\begin{eqnarray}
{\cal E}_4 &\rightarrow& E_4(\tau_1)E_4(\tau_2) \;\;, \nonumber\\
{\cal E}_6 &\rightarrow& E_6(\tau_1)E_6(\tau_2) \;\;,  \nonumber\\
{\cal C}_{10} &\rightarrow& 0 \;\;, \nonumber\\
{\cal C}_{12} &\rightarrow& \Delta(\tau_1)\Delta(\tau_2) \;\;.
\end{eqnarray}
More precisely, one finds that, up to a normalization
constant, ${\cal C}_{10} \rightarrow \tau_3^2 \Delta(\tau_1)\Delta(\tau_2)$
as $\tau_3 \rightarrow 0$.

Now,  consider the 
behaviour on $D$ of the 
odd generator ${\cal C}_{35}$.  Since ${\cal C}_{35}$ 
is a more complicated object,  one first reexpresses 
its square in terms of 
the other, even generators.  Namely, by using the results in \cite{I},
one finds that
 on the diagonal $D$, ${\cal C}_{35} =0$ as well as 
\begin{eqnarray}
\alpha ^2\frac{{\cal C}_{35}^2}{\C _{10}}=
{\cal C}_{12}^5 (j(\tau_1) - j(\tau_2))^2 
=(\eta ^2(\tau _1)\eta ^2(\tau _2))^{60}(j(\tau_1)-j(\tau_2))^2 \;\;,
\label{c35c10}
\end{eqnarray}
where
$j(\tau)=E^3_4/\Delta$.  Then, using ${\cal C}_5$ and ${\cal C}_{30}$,
which are related to the forms already defined
by ${\cal C}_{10} = {\cal C}_5^2$ and ${\cal C}_{35} = 
{\cal C}_{30}{\cal C}_{5}$, respectively, it follows that
\beqa
\alpha^2 {\cal C}_{30}^2 \rightarrow \Delta^5(\tau_1) 
\Delta^5(\tau_2) (j(\tau_1)-j(\tau_2))^2  
\label{c30t}
\eeqa
on the diagonal $D$.

A rational quadratic divisor of ${\cal H}_2$ is, by definition \cite{GN2},
the set 
\beqa
{\sc H}_l=\{{\tiny \left( \begin{array}{cc}  iT & i V \\ i V & i U 
\end{array} \right)}
\in {\cal H}_2|i n_1T+ i m_1U+ i bV+n_2(- TU + V^2)+m_2=0\} \;\;,
\eeqa
where $l=(n_1,m_1,b,n_2,m_2) \in {\bf Z}^5$ is a primitive (i.e.
with the greatest commom divisor equals $1$) integral vector.
The number 
${\rm D}(l)=b^2-4m_1n_1+4n_2m_2$ is called the discriminant of ${\sc H}_l$.
This divisor determines the Humbert surface $H_{\rm D}$ 
in the Siegel threefold
$Sp_4({\bf Z})\setminus {\cal H}_2$.
The Humbert surface $H_{\rm D}$ 
is (the image in $Sp_4({\bf Z})\setminus {\cal H}_2$
of) the union of all ${\sc H}_l$ of
discriminant ${\rm D}(l)$. 
Each Humbert surface ${\sc H}_{\rm D}$ can be
represented by a linear relation in $T$, $U$ and $V$. 
For instance, the divisor of ${\cal C}_{5}$ is the diagonal 
${\sc H}_{\rm 1}=\{Z={\tiny \left( \begin{array}{cc} 
iT & 0 \\ 0 &i U \end{array} \right)}
\in Sp_4({\bf Z})\setminus {\cal H}_2\}$. 
Similarly, the divisor of the Siegel
modular form ${\cal C}_{30}$ is the surface
${\sc H}_{\rm 4}=\{Z={\tiny \left( 
\begin{array}{cc}i T &i V \\i V &i U \end{array} \right)}
\in Sp_4({\bf Z})\setminus {\cal H}_2|T=U\}$. \
The divisor of the Siegel modular form ${\cal C}_{35}$, on the other hand,
is the sum (with multiplicity $1$) of the surfaces 
${\sc H}_{\rm 1}$ and ${\sc H}_{\rm 4}$.

\subsection{Jacobi forms}

A Siegel modular form $F(T,U,V)$ of weight $k$
has a Fourier expansion with respect to its variable $iU$
\begin{eqnarray}
F(T,U,V)=\sum_{m=0}^{\infty}\phi_{k,m}(T,V)s^m  \;\;,
\end{eqnarray}
where $s={\bf e}[iU], {\bf e}[x] = {\rm exp}2 \pi i x$.
Each of the $\phi_{k,m}(T,V)$ is a Jacobi form
of weight $k$ and index $m$ \cite{EZ}.  That is, for each 
${\tiny \left( \begin{array}{cc} a & b \\ c & d \end{array} \right)}\in
Sl(2,{\bf Z})$ and $\lambda,\mu \in {\bf Z}$
\begin{eqnarray}
\phi_{k,m}\Big(\frac{aT-ib}{icT+d},\frac{V}{icT+d}\Big) &=&
(icT+d)^k e^{2\pi im\frac{c(iV)^2}{icT+d}}\phi(T,V) \;\;,\nonumber\\
\phi_{k,m}(T,V+\lambda T+\mu) &=& e^{-2\pi im(\lambda^2 iT+2\lambda iV)}
\phi_{k,m}(T,V)\;\;.
\end{eqnarray}
A Jacobi form $\phi_{k,m}(T,V)$ of index $m$ has 
in turn 
an
expansion
\begin{eqnarray}
\phi(T,V)=\sum_{n\ge 0}\sum_{l\epsilon {\bf Z}}c(n,l)q^nr^l \;\;,
\end{eqnarray}
where $q={\bf e}[iT], r = {\bf e}[iV]$.
Of special relevance are the Jacobi forms $\phi_{k,1}$ of index $1$.
The summation in $l$ extends
in the usual
case, and for the generators introduced above, over $4n-l^2\ge 0$;
for the forms
divided by $\Delta$, $4n-l^2 \geq  -1 \, {\rm or}\,  -4$, depending on whether
the form is a cusp form or not.
Furthermore
\begin{eqnarray}
c(n,l)=c(4n-l^2) \;\;.
\end{eqnarray}

Consider, for instance, the Eisenstein series, which have 
the expansion
\begin{eqnarray}
{\cal E}_k(T,U,V) = E_k(T) - \frac{2k}{B_k}\, E_{k,1}(T,V)\, s + {\cal O}(s^2)
\;\;.
\end{eqnarray}
Here, the $B_k$ denote the Bernoulli numbers.
Thus, for instance, 
\begin{eqnarray}
{\cal E}_4&=&E_4+240E_{4,1}s+\cdots \;\;, \nonumber\\
{\cal E}_6&=&E_6-504E_{6,1}s+\cdots \;\;.
\end{eqnarray}
The Jacobi forms $E_{4,1}(T,V)$ and $E_{6,1}(T,V)$ 
of index $1$ have the expansion (the expansion coefficients are listed
in the first table of appendix A.6)
\begin{eqnarray}
E_{4,1}&=&1+(r^2+56r+126+56r{-1}+r^{-2})q \nonumber\\
&+&(126r^2+576r+756+576r^{-1}+126r^{-2})q^2+\cdots \;\;, \nonumber\\
E_{6,1}&=&1+(r^2-88r-330-88r^{-1}+r^{-2})q \nonumber\\
&+&(-330r^2-4224r-7524-4224r^{-1}-330r^{-2})q^2+\cdots \;\;.
\end{eqnarray}
Note that 
$E_{k,1}\rightarrow E_k$ as $V \rightarrow 0$.

Similarly, the cusp forms ${\cal C}_{10}(T,U,V)$ and ${\cal C}_{12}(T,U,V)$
have the expansion
\beqa
{\cal C}_{10}(T,U,V) &=& \phi_{10,1}(T,V) s + {\cal O}(s^2) \;\;,\nonumber\\
{\cal C}_{12}(T,U,V) &=& 
\Delta(T) + \frac{1}{12} \phi_{12,1} (T,V) s +  {\cal O}(s^2) \;\;,
\label{c10c12}
\eeqa
where
\begin{eqnarray}
\phi_{10,1}&=&\frac{1}{144}(E_6E_{4,1}-E_4E_{6,1})
\rightarrow \;\;\;\;0 \;\;, \nonumber\\
\phi_{12,1}&=&\frac{1}{144}(E_4^2 E_{4,1}-E_6E_{6,1})\rightarrow 12\Delta
\;\;.
\end{eqnarray}
Here, we have indicated the behaviour under the truncation $V \rightarrow 0$.
The Jacobi forms $\phi_{10,1}$ and $\phi_{12,1}$ of index $1$ have the 
following expansion (the expansion coefficients are listed in the
first table in appendix A.6):
\begin{eqnarray}
\phi_{10,1}&=&(r-2+r^{-1})q+(-2r^2-16r+36-16r^{-1}-2r^{-2})q^2+\cdots \;\;,
\nonumber\\
\phi_{12,1}&=&(r+10+r^{-1})q+(10r^2-88r-132-88r^{-1}+10r^{-2})q^2+\cdots
\;\;.
\end{eqnarray}

\subsection{Product expansions \label{prex}}

The Siegel modular forms ${\cal C}_5$ and 
${\cal C}_{30}={\cal C}_{35}/{\cal C}_{5}$ have the following
product expansion \cite{GN2}
\begin{eqnarray}
{\cal C}_{5}=(qrs)^{1/2}\prod_{n,m,l\in {\bf Z} \atop (n,m,l)>0}
(1-q^nr^ls^m)^{f(4nm-l^2)} \;\;, \nonumber\\
{\cal C}_{30}=(q^3rs^3)^{1/2}(q-s)\prod_{n,m,l\in {\bf Z} \atop (n,m,l)>0}
(1-q^nr^ls^m)^{f'_2(4nm-l^2)} \;\;,
\label{prodexp}
\end{eqnarray}
where the condition $(n,m,l)>0$ means that $n\geq 0,m\geq 0$
and either $l \in {\bf  Z}$
if $n+m>0$, or $l<0$ if $n=m=0$.
The coefficients $f(4nm-l^2)$ and $f'_2(4nm-l^2)$, which are
listed in the first table in appendix A.6,
are defined as follows \cite{GN2}.  Consider the
expansion of
\beqa
\phi_{0,1}:=\frac{\phi_{12,1}}{\Delta (T)}
=\sum_{n\ge 0}\sum_{l\epsilon {\bf Z}}
f(n,l)q^nr^l  \;\;,
\eeqa
where the sum over $l$ is restricted to $4n-l^2\geq -1$.  Then,
$f(N)=f(n,l)$ if $N=4n-l^2\geq-1$, and $f(N)=0$ otherwise.
The coefficients $f_2'(N)$ are then given by
$f'_2(N)=8f(4N)+(2\left (\frac{-N}{2}\right )-3)f(N)+f(\frac{N}{4})$.
Here, $(\frac{D}{2})=1,-1,0$ depending on whether $D\equiv 1 \,
{\rm mod} \, 8, 5\, {\rm mod}\, 8,0\,{\rm mod}\, 2$.

Using the product expansions (\ref{prodexp}), we can perform a check
on the expansion (\ref{c10c12}) of ${\cal C}_{10}=
qrs\prod(1-q^nr^ls^m)^{2f}$.
Namely, consider the term in ${\cal C}_{10}$ with $n=m=0,l=-1$.
It gives rise to 
$qs r(1-r^{-1})^2=qs(r-2+r^{-1})$, which indeed
matches the $q$-term of $\phi_{10,1}$.

Similarly, we can perform a check on (\ref{c30t}).  Setting
$r=1$ in (\ref{prodexp}), we see that the terms with $m=0$ have
$f'_2(0)=60$; they thus match $\Delta ^{5/2}(T) = \eta^{60}$
occurring in 
${\cal C}_{30} \propto \Delta ^{5/2}(T)\Delta ^{5/2}(U)(j(T)-j(U))$.
The sum over $l$ for the terms with $m=n=1$, on the other hand,
yields
$f'_2(4)+2 (f'_2(3)+f'_2(0))=196884$, which matches the $q$-term
in the expansion of $j-744 = q^{-1}+196884q+\cdots$.

\subsection{Theta functions and Jacobi forms \label{thefjaf}}

The 
standard Jacobi theta functions are defined as follows ($z=iV$)
\begin{eqnarray}
\th_1(\tau,z)&=&i\sum_{n\in{\bf Z}}(-1)^nq^{\frac{1}{2}
(n-\frac{1}{2})^2}r^{n-\frac{1}{2}} \;\;,\nonumber\\
\th_2(\tau,z)&=&\sum_{n\in {\bf Z}} 
q^{\frac{1}{2}(n-\frac{1}{2})^2}r^{n-\frac{1}{2}} \;\;,
\nonumber\\
\th_3(\tau,z)&=&\sum_{n\in {\bf Z}} q^{\frac{1}{2}n^2}r^n \;\;,\nonumber\\
\th_4(\tau,z)&=&\sum_{n\in {\bf Z}}(-1)^nq^{\frac{1}{2}n^2}r^n \;\;.
\end{eqnarray}
It is useful to introduce
\begin{eqnarray}
\th_{0,1}(\tau,z)&=&\th_3(2\tau,z)=\sum_{n \in {\bf Z}} q^{n^2}r^n \;\;,
\nonumber\\
\th_{1,1}(\tau,z)&=&\th_2(2\tau,z)=\sum_{n \in {\bf Z}} 
q^{(n-\frac{1}{2})^2}r^{n-\frac{1}{2}}
\end{eqnarray}
as well as
\begin{eqnarray}
\th_{ev}(\tau,z)&=&\th_{0,1}(\tau,2z)=\sum_{n\equiv 0(2)} q^{n^2/4}r^n \;\;,
\nonumber\\
\th_{odd}(\tau,z)&=&\th_{1,1}(\tau,2z)
=\sum_{n\equiv 1(2)} q^{(n-\frac{1}{2})^2}r^{n-\frac{1}{2}} \;\;.
\end{eqnarray}
Next, consider setting $z=0$.  The $\theta_i (\tau,0)$ will
be simply denoted by $\theta_i$, whereas the 
 $\theta_i (2\tau,0)$ will be denoted by $\theta_i(2 \cdot)$
($i=1,\dots ,4$).
It is well known that
$\theta_1 = 0$ and that 
$\th_3^4=\th_2^4+\th_4^4$ as well as $\th_2\th_3\th_4=2\eta^3$.
Also
\beqa
E_4&=&\frac{1}{2} \left( \th_2^8 + \th_3^8 + \th_4^8\right) \;\;,\nonumber\\
E_6&=& \frac{1}{2}\left(
\th_2^4+\th_3^4)(\th_3^4+\th_4^4)(\th_4^4-\th_2^4)\right)\nonumber\\
    &=&\frac{1}{2}\left(
-\th_2^6(\th_3^4+\th_4^4)\th_2^2+\th_3^6(\th_4^4-\th_2^4)\th_3^2+
\th_4^6(\th_2^4+\th_3^4)\th_4^2 \right) \;\;.
\eeqa
Additional useful identities are given by
\begin{eqnarray}
2\th_2(2\cdot)\th_3(2\cdot)&=&\th_2^2 \;\;,\nonumber\\
\th_2^2(2\cdot)+\th_3^2(2\cdot)&=&\th_3^2 \;\;,\nonumber\\
\th_3^2(2\cdot)-\th_2^2(2\cdot)&=&\th_4^2 \;\;,\nonumber\\
2\th_2^2(2\cdot)&=&\th_3^2-\th_4^2 \;\;, \nonumber\\
2\th_3^2(2\cdot)&=&\th_3^2+\th_4^2 \;\;,\nonumber\\
\th_4^2(2\cdot)&=&\th_3\th_4 \;\;.
\end{eqnarray}
Now consider 
Jacobi forms 
$f(\tau,z)=\sum_{n\ge 0 \atop l\in {\bf Z}}c(4n-l^2)q^nr^l$
of weight $k$ and index $1$.  The following examples provide 
useful identities 
between Jacobi forms of
index $1$ and Jacobi theta functions
\begin{eqnarray}
\phi_{10,1}&=&\;\;\;\;\;\;\;\;\;\;\;\;\;\;\;\;\;\;\;\;
\;\;\;\;\;\;\;\;\;\;\;\;\;\;\;\;\;\;\;-\eta^{18}\th_1^2(\tau,z) \;\;,
\nonumber\\
\phi_{12,1}&=&12\eta^{24}\frac{\th_3^2(\tau,z)}{\th_3^2}+(\th_4^4-\th_2^4)
[-\eta^{18}\th_1^2(\tau,z)] 
\end{eqnarray}
as well as
\begin{eqnarray}
E_{4,1}&=& \frac{1}{2} \left(
\th_2^6\th_2^2(\tau,z)+\th_3^6\th_3^2(\tau,z)+\th_4^6\th_4^2(\tau,z) 
\right) \;\;, \\
E_{6,1}&=& \frac{1}{2} \left(
-\th_2^6(\th_3^4+\th_4^4)\,\th_2^2(\tau,z)+
\th_3^6(\th_4^4-\th_2^4)\,\th_3^2(\tau,z)+
\th_4^6(\th_2^4+\th_3^4)\,\th_4^2(\tau,z)\right) .  \nonumber
\end{eqnarray}
A Jacobi form of index $1$ 
has the following decomposition \cite{EZ,K1,K2}
\begin{eqnarray}
f(\tau,z)=f_{ev}(\tau)\th_{ev}(\tau,z)+f_{odd}(\tau)\th_{odd}(\tau,z)  \;\;,
\end{eqnarray}
where 
\begin{eqnarray}
f_{ev}&=&\sum_{N\equiv 0(4)}c(N)q^{N/4} \;\;, \nonumber\\
f_{odd}&=&\sum_{N\equiv -1(4)}c(N)q^{N/4} \;\;.
\end{eqnarray}
Consider, for instance, $E_{4,1}$.  It has the decomposition \cite{K2}
\beqa
E_{4,1\,ev }&=& \th_3^7(2\cdot) + 7 \th_3^3(2\cdot)\th_2^4(2\cdot) \;\;,
\nonumber\\  
E_{4,1\,odd }&=& \th_2^7(2\cdot) + 7 \th_2^3(2\cdot)\th_3^4(2\cdot) \;\;.
\eeqa
Furthermore one has (with
 $\theta_{ev} \equiv \theta_{ev}(\tau,z)$
and $\theta_{odd} \equiv \theta_{odd}(\tau,z)$)

\beqa
\th_1^2(\tau,z)&=&\th_2(2\cdot)\th_{ev}-\th_3(2\cdot)\th_{odd}\;\;,\nonumber\\
\th_2^2(\tau,z)&=&\th_2(2\cdot)\th_{ev}+\th_3(2\cdot)\th_{odd}\;\;,\nonumber\\
\th_3^2(\tau,z)&=&\th_3(2\cdot)\th_{ev}+\th_2(2\cdot)\th_{odd}\;\;,\nonumber\\
\th_4^2(\tau,z)&=&\th_3(2\cdot)\th_{ev}-\th_2(2\cdot)\th_{odd} \;\;.
\eeqa
Next, consider the elliptic genus $Z(\tau,z)$ of $K3$,
which is a Jacobi form of weight $0$ and index $1$, given by \cite{KYY}
\beqa
Z(\tau,z)= 2 \frac{\phi_{12,1}}{\Delta}
= 24\frac{\th_3^2(\tau,z)}{\th_3^2}-2\frac{\th_4^4-\th_2^4}{\eta^4}
\frac{\th_1^2(\tau,z)}{\eta^2} \;\;.
\eeqa
It has the decomposition 
\begin{eqnarray}
Z_{ev}&=&24\frac{\th_3(2\cdot)}{\th_3^2}-2\frac{\th_4^4-\th_2^4}{\eta^4}
\frac{\th_2(2\cdot)}{\eta^2} = 20 + 216 q + 1616q^2 + \cdots
\;\;,\nonumber\\
Z_{odd}&=&24\frac{\th_2(2\cdot)}{\th_3^2}+2\frac{\th_4^4-\th_2^4}{\eta^4}
\frac{\th_3(2\cdot)}{\eta^2} = 2 q^{-\frac{1}{4}} - 128 q^{\frac{3}{4}}
- 1026q^{\frac{7}{4}} + \cdots
\;\;.
\end{eqnarray}
Now we introduce the hatted modular function $\widehat{f(\tau,z)}$ as
\beqa
\widehat{f(\tau,z)}=f_{ev}(\tau)+f_{odd}(\tau)\;\;.
\label{fefo}
\eeqa
Hence the hatted modular function corresponds in a one-to-one way
to the index 1 Jacobi form. In particular, the Jacobi form $f(\tau,z)$ and
its hatted relative $\widehat{f(\tau,z)}$ possess identical power
series
expansion coefficients $c(N)$:
\beqa
f(\tau,z)=\sum_{n,l}c(4n-l^2)q^nr^l, \qquad\widehat{f(\tau,z)}=
\sum_{N\in 4{\bf Z} \, or \, 
4{\bf Z} + 3} c(N)q^{N/4} \;\;. \label{hatjacobif}
\eeqa 
Note that an ordinary modular form (that is a form not having any 
$z$-dependence), if 
occurring as a multiplicative
factor in front of a proper Jacobi form, is left
untouched by the hatting procedure (\ref{fefo}).
Thus, for instance,
\beqa
\widehat{E_{4,1}}&=&
\frac{1}{2} \left(\th_2^6\,
\widehat{\th_2^2(\tau,z)}+\th_3^6\,\widehat{\th_3^2(\tau,z)}+
\th_4^6\,\widehat{\th_4^2(\tau,z)}\right) \\
&=&
\frac{1}{2}\left(
\th_2^6[\th_2(2\cdot)+\th_3(2\cdot)]+
\th_3^6[\th_2(2\cdot)+\th_3(2\cdot)]+
\th_4^6[\th_3(2\cdot)-\th_2(2\cdot)]\right) \;\;,\nonumber\\
\widehat{E_{6,1}}&=& \frac{1}{2} \left(
-\th_2^6(\th_3^4+\th_4^4)\,\widehat{\th_2^2(\tau,z)}+
\th_3^6(\th_4^4-\th_2^4)\,\widehat{\th_3^2(\tau,z)}+
\th_4^6(\th_2^4+\th_3^4)\,\widehat{\th_4^2(\tau,z)}\right) \;\;, \nonumber
\eeqa
and similarly
\beqa
{\hat Z} = Z_{ev} + Z_{odd}
=24\frac{\th_2(2\cdot) + \th_3(2\cdot)}{\th_3^2}
-2\frac{(\th_4^4-\th_2^4)}{\eta^4}
\frac{(\th_2(2\cdot)-\th_3(2\cdot))}{\eta^2} 
\;\;.
\eeqa

Furthermore, consider introducing
\beqa
\tilde{f}=\hat{f}(4\cdot)=\sum_{N\in 4{\bf Z} \, or \, 
4{\bf Z} + 3} c(N)q^N \;\;.
\eeqa
Note that $\tilde{f}$ is the $\Gamma_0(4)$
modular form of half-integral weight $k-1/2$ 
associated to a Jacobi form of weight $k$ and index $1$ \cite{EZ}.

\subsection{Lie algebra lattices and Jacobi forms}

The relation between Lie algebra lattice sums 
and Jacobi forms will
be established in three steps. We start by reviewing the well-known
relationship between the Lie algebra lattice $E_8$ and the Eisenstein
series $E_4$. Then we go on to show the relation between the 
Lie algebra lattice $E_7$ and the Jacobi Eisenstein series $E_{4,1}$.
Finally, we will relate the processes of splitting off an $A_1$ and the
hatting procedure. This will explain the relation between turning on a
Wilson line and the hatting procedure.

First the relation between the Eisenstein series $E_4$ and the 
partition function of the 
$E_8$ lattice $\Lambda=\{x\in{\bf Z}^8\cup\pi+{\bf Z}^8|(x,\pi)\in{\bf Z}\}$
is well known ($\pi=(1/2,\cdots,1/2)\in{\bf Z}^8$) and reads
\begin{eqnarray}
E_4=\sum_{x\in\Lambda}q^{\frac{1}{2}x^2}=\frac{1}{2}(\th_2^8+\th_3^8+\th_4^8)
\;\;.
\end{eqnarray}
Because of the lattice relation 
$\Lambda_{E_8}=\Lambda_{D_8^{(0)}}+\Lambda_{D_8^{(S)}}$, this also 
shows that
the fermionically computed partition function  
$P_{D_8^{(0)}}+P_{D_8^{(S)}}$ of $E_8$ is identical to 
the bosonically computed one, if one recalls the relation
between the bosonic conjugacy class picture and the 
fermionic boundary condition picture:
\begin{eqnarray}
P_{D_n^{(0)}}&=&\frac{\th_3^n+\th_4^n}{2}=\frac{NS^+ +NS^-}{2} \;\;,\nonumber\\
P_{D_n^{(V)}}&=&\frac{\th_3^n-\th_4^n}{2}=\frac{NS^+ -NS^-}{2} \;\;,\nonumber\\
P_{D_n^{(S/C)}}&=&\frac{\th_2^n}{2}=\frac{R^+}{2} \;\;.
\end{eqnarray}
Now consider the Jacobi form $E_{4,1}(\tau,z)=\sum c(4n-l^2)q^nr^l$.
Since the expression $\sum_{x\in\Lambda}q^{\frac{1}{2}x^2}r^{(x,\pi)}$
has the correct weights (and truncation), and since the space in question is 
one-dimensional, this 
represents $E_{4,1}$. If one considers the $l=0$ resp. $l=1$ sector, one
finds $\sum_{(x,\pi)=0}q^{\frac{1}{2}(x,x)}=\sum_n c(4n)q^n=
\sum_{N\equiv 0(4)}c(N)q^{N/4}$ resp. $\sum_{(x,\pi)=1}q^{\frac{1}{2}(x,x)}=
\sum_n c(4n-1)q^n=\sum_{N\equiv -1(4)}c(N)q^{\frac{N+1}{4}}$, i.e.
$E_{4,1ev}=\sum_{(x,\pi)=0}q^{\frac{1}{2}x^2}$ and
$E_{4,1odd}=q^{-1/4}\sum_{(x,\pi)=1}q^{\frac{1}{2}x^2}=
\sum_{x\in -\frac{\pi}{2}+\Lambda \atop (x,\pi)=0}q^{\frac{1}{2}x^2}$.
Thus, 
\begin{eqnarray}
E_{4,1 \,ev}=\sum_{(x,\pi)=0}^{E_8}q^{\frac{1}{2}x^2}&=&
\sum_{x\in (0)}^{E_7}q^{\frac{1}{2}x^2}=P_{E_7^{(0)}} \;\;, \nonumber\\
E_{4,1 \,odd}=\sum_{x\in -\frac{\pi}{2}+\Lambda \atop (x,\pi)=0}^{E_8}
q^{\frac{1}{2}x^2}&=&\sum_{x\in (1)}^{E_7}q^{\frac{1}{2}x^2}=P_{E_7^{(1)}}\;\;,
\end{eqnarray}
where the lattice sums 
$P_{E_7^{(i)}}=\sum_{x\in (i)}q^{\frac{1}{2}x^2}$ 
run over vectors within the conjugacy class (i).\\
Besides this lattice theoretic argument, this can also be checked explicitly
\begin{eqnarray}
E_{4,1 \,ev}&=&\th_3^3(2\cdot)
(\th_3^4(2\cdot)+7\th_2^4(2\cdot))=\th _3(2\cdot)\th _3^2(2\cdot)
(\th _4^4(2\cdot)+8\th _2^4(2\cdot))\nonumber\\
&=&\th _3(2\cdot)\frac{\th_3^2+\th_4^2}{2}[\th_3^2\th_4^2+2(\th_3^2-\th_4^2)^2]
=\th _3(2\cdot)[\th_3^6+\th_4^6-\frac{\th_3^2\th_4^2}{2}
(\th_3^2+\th_4^2)]\nonumber\\
&=&\th _3(2\cdot)\frac{1}{2}[\th_3^6+\th_4^6]+\th_2(2\cdot)\frac{1}{2}\th_2^6
=P_{E_7^{(0)}} \;\; ;\label{efoureven}
\end{eqnarray}
similarly $E_{4,1 \,odd}= P_{E_7^{(1)}}$.  
 
The last relation in (\ref{efoureven}) follows
by noting the lattice decomposition of $P_{E_7^{(0)}}$:
$P_{E_7^{(0)}}=P_{D_6^{(0)}}\cdot P_{A_1^{(0)}}+
P_{D_6^{(S)}}\cdot P_{A_1^{(1)}}$.
Here one uses the following lattice sums for $A_1$,
which 
has the root lattice
$\Lambda_{A_1}^{(0)}=\sqrt{2}{\bf Z}$ and two conjugacy classes:  
\begin{eqnarray}
P_{A_1^{(0)}}&=& \sum_{x\in (0)}^{A_1}
q^{\frac{1}{2}x^2}=\sum_{n\in {\bf Z}} q^{n^2}=
\th_3(2\cdot)\;\;,\nonumber\\
P_{A_1^{(1)}}&=&\sum_{x\in (1)}^{A_1}q^{\frac{1}{2}x^2}=
\sum_{n\in {\bf Z}}  q^{(n-1/2)^2}=
\th_2(2\cdot) \;\;.
\end{eqnarray}

Thus we obtain
\begin{eqnarray}
2\widehat{E_{4,1}}&=&\th_2^6[\th_2(2\cdot)+\th_3(2\cdot)]+
\th_3^6[\th_2(2\cdot)+\th_3(2\cdot)]+
\th_4^6[\th_3(2\cdot)-\th_2(2\cdot)]\nonumber\\
&=&\th_2^6\cdot\widehat{\th_2^2(\tau,z)}+\th_3^6\cdot\widehat{\th_3^2(\tau,z)}+
\th_4^6\cdot\widehat{\th_4^2(\tau,z)}\nonumber\\
&=&2(P_{E_7^{(0)}}+P_{E_7^{(1)}}) 
\;\; ,
\end{eqnarray}
which also holds, as is easily seen, in the dehatted version.
Now we understand that the breaking of $E_8$ to $E_7$ by turning
on a Wilson line, i.e.
the splitting off of an $A_1^{\rm{Wilson}}$, precisely corresponds
to the replacement of $E_4$ by the hatted modular function $\widehat{E_{4,1}}$.

On the other hand, note that the truncation $V\rightarrow 0$ 
\begin{eqnarray}
E_{4,1}(\tau,0)=E_4=(E_{4,1})_{ev}\th_3(2\cdot)+(E_{4,1})_{odd}\th_2(2\cdot)
\end{eqnarray}
reflects the decomposition of $E_8 \supset E_7 \times A_1$
\begin{eqnarray}
P_{E_8}=P_{E_7^{(0)}}\cdot P_{A_1^{(0)}}+P_{E_7^{(1)}}\cdot P_{A_1^{(1)}}
\;\;.
\end{eqnarray}

Let us again demonstate the hatting procedure by considering
the Wilson line breaking of 
$D_2=A_1 \times 
A_1^{\rm{Wilson}}$ to $A_1$.
The lattice decomposition of $D_2$ under $A_1\times A_1$ has the form
\begin{eqnarray}
P_{D_2^{(0)}}&=&{\th_3^2+\th_4^2\over 2}=P_{A_1^{(0)}}\cdot P_{A_1^{(0)}}=
\th_3(2\cdot )^2,\nonumber\\
P_{D_2^{(V)}}&=&{\th_3^2-\th_4^2\over 2}=P_{A_1^{(1)}}\cdot P_{A_1^{(1)}}=
\th_2(2\cdot )^2,\nonumber\\
P_{D_2^{(S,C)}}&=&{\th_2^2\over 2}=P_{A_1^{(0)}}\cdot P_{A_1^{(1)}}=
\th_2(2\cdot )\th_3(2\cdot ).
\end{eqnarray}
Thus the corresponding hatted Jacobi forms become
\begin{eqnarray}
{\widehat{\th_3^2(\tau,z)}+\widehat{\th_4^2(\tau,z)}
\over 2}&=&P_{A_1^{(0)}}=
\th_3(2\cdot ),\nonumber\\
{\widehat{\th_3^2(\tau,z)}-\widehat{\th_4^2(\tau, z)}
\over 2}&=&P_{A_1^{(1)}}=
\th_2(2\cdot ),\nonumber\\
{\widehat{\th_2^2(\tau,z)}\over 2}&=&{1\over 2}(P_{A_1^{(0)}}+ P_{A_1^{(1)}})=
{1\over 2}(\th_2(2\cdot )+\th_3(2\cdot )).
\end{eqnarray}

Finally, 
going back from the conjugacy class picture to the boundary condition
picture   one has 
\begin{eqnarray}
NS^{\pm}_{A_1}=P_{A_1^{(0)}}\pm P_{A_1^{(1)}}=\th_3(2\cdot)\pm \th_2(2\cdot)
=\widehat{\th_{3/4}^2(\tau,z)} \;\;, \\
R^+_{A_1}=P_{A_1^{(0)}}+ P_{A_1^{(1)}}=\th_3(2\cdot)+ \th_2(2\cdot)
=\widehat{\th_{2}^2(\tau,z)} \;\; .
\end{eqnarray}

\newpage

\subsection{Tables}

The first table displays some expansion coefficients of the Jacobi forms
$E_{4,1}$, $E_{6,1}$, $\phi_{10,1}$, $\phi_{12,1}$ and of the Siegel
forms ${\cal C}_5$, ${\cal C}_{30}$. \\

\hspace{0.5cm}

\begin{tabular}{|c||c|c|c|c||c|c|} \hline
$N$ &$e_{4,1}(N)$&$e_{6,1}(N)$&$c_{10,1}(N)$&$c_{12,1}(N)$&$f(N)
$&$f_2^{\prime}(N)$\\ \hline \hline
--4&      - &      - &         - &         - & -    &   1  \\
--1&      - &      - &         - &         - &  1   &  --1  \\
0 &      1 &      1 &         0 &         0 & 10   &  60  \\
3 &     56 &    --88 &         1 &         1 &--64   &32448 \\
4 &    126 &   --330 &        --2 &        10 &108   &131868\\
7 &    576 &  --4224 &       --16 &       --88 &--513  & ***  \\
8 &    756 &  --7524 &        36 &      --132 & 808  & ***  \\
11&   1512 & --30600 &        99 &      1275 &--2752 & ***  \\
12&   2072 & --46552 &      --272 &       736 & 4016 & ***  \\
15&   4032 &--130944 &      --240 &     --8040 &--11775& ***  \\
16&   4158 &--169290 &      1056 &     --2880 & 16524& ***  \\
19&   5544 &--355080 &      --253 &     24035 &  *** & ***  \\
20&   7560 &--464904 &     --1800 &     13080 &  *** & ***  \\ \hline
\end{tabular}

\hspace{0.5cm}

\hspace{0.5cm}

In the following 
table some expansion coefficients of ${E_{4,1}E_6\over \Delta}$,
${E_4E_{6,1}\over \Delta}$ and of $A_n$ (see eq. (\ref{An})) for $n=0,1,2,12$
are listed.\\

\hspace{0.5cm}

\begin{tabular}{|c||c|c||c|c|c|c|} \hline
$N$ & $E_{4,1}E_6/\Delta$ & $E_4E_{6,1}/\Delta$ 
&$2A_0$&$2A_1$&$2A_2$&$2A_{12}$\\ \hline \hline
--4 &  1  &  1 & 2 & 2 & 2 & 2\\
--1 & 56  & --88  & --32 & --44 & --56 & --176\\
 0 &--354 & --66  & --420 & --396 & --372 & --132\\
 3 &--26304 & --27456  & --52760 & --53356 & --53952 & --54912\\
 4 &--88128 & --86400  & --174528 & --174384 & --174240 & --172800\\ \hline
\end{tabular}

\hspace{0.5cm}

%
%

\newpage

\section{Bibliograpy}


\begin{enumerate}






\bibitem{ABK}
I. Antoniadis, C. P. Bachas and C. Kounnas, Nucl. Phys. {\bf B 289} (1987) 87.


\bibitem{AlFoIbaQue1} 
G. Aldazabal, A. Font, L. E. Ib\'a\~nez and
F. Quevedo, Nucl. Phys. {\bf B 461} (1996) 85, hep-th/9510093.



\bibitem{AlFoIbaQue2} G. Aldazabal, A. Font, L. E. Ib\'a\~nez and
F. Quevedo,
Phys. Lett. {\bf B 380} (1996) 33, hep-th/9602097.


\bibitem{gaume}  
L. Alvarez-Gaum\'{e} and E. Witten, 
Nucl. Phys. {\bf B 234} (1984) 269.

\bibitem{ABFPT}
I. Antoniadis, C. Bachas, C. Fabre, H. Partouche and T.R. Taylor, 
hep-th/9608012.

\bibitem{AFGNT}
I. Antoniadis, S.Ferrara, E. Gava, K. S. Narain and T. R. Taylor, Nucl. Phys.
{\bf B 447} (1995) 35.

\bibitem{AFT}
I. Antoniadis, S. Ferrara and T. R. Taylor,
Nucl.Phys. {\bf B 460} (1996) 489, hep-th/9511108.

\bibitem{AGN} I. Antoniadis, E. Gava,  and K. S. Narain,
Phys. Lett. {\bf B 283} (1992) 209,
hep-th/9203071; Nucl. Phys. {\bf B 383} (1992) 109, hep-th/9204030.


\bibitem{AGNTtop}
I. Antoniadis, E. Gava, K. S. Narain and T. R. Taylor, Nucl. Phys. {\bf B 413}
(1994) 162, hep-th/9307158.



\bibitem{AGNTtop1}
I. Antoniadis, E. Gava, K. S. Narain and T. R. Taylor, Nucl. Phys. {\bf B 455}
(1995) 109, hep-th/9507115; Nucl. Phys. {\bf B 476}
(1996) 1033, hep-th/9604077. 


\bibitem{AGNT}
I. Antoniadis, E. Gava, K.S. Narain, T.R. Taylor,
Nucl.Phys. {\bf B 455} (1995) 109,
 hep-th/9507115.


\bibitem{AP}
I. Antoniadis, H. Partouche, 
Nucl. Phys. {\bf B 460} (1996) 470,
hep-th/9509009.

\bibitem{AM}
P. Aspinwall and D. Morrison, hep-th/9404151.

\bibitem{A}
P. Aspinwall, Phys. Lett. {\bf B 357} (1995) 329, hep-th/9507012; 
Phys. Lett. {\bf B 371} (1996) 231, hep-th/9511171.


\bibitem{AG}
P. Aspinwall and M. Gross, Phys. Lett. {\bf B 382} (1996) 81, hep-th/9602118.

\bibitem{AL}
P. Aspinwall and J. Louis, Phys. Lett. {\bf B 369} (1996) 233, hep-th/9510234.

\bibitem{BKKM}
P. Berglund, S. Katz, A. Klemm and P. Mayr, Nucl. Phys. {\bf B 483} (1997) 209,
hep-th/9605154.

\bibitem{BKKM}
P. Berglund, S. Katz, A. Klemm and  P. Mayr, hep-th/9605154.

\bibitem{bele}
M. Berkooz, R. Leigh, J. Polchinski,
J. Schwarz, N. Seiberg and E. Witten,  
Nucl.\ Phys.\ {\bf B 475} (1996) 115, 
hep-th/9605184.

\bibitem{BCOV}
M. Bershadsky, S. Cecotti, H. Ooguri, C. Vafa, Nucl. Phys. {\bf B405}
 (1993) 279; Comm. Math. Phys. {\bf 165} (1994) 311.


\bibitem{BSV1}
M. Bershadsky, V. Sadov and C. Vafa,
Nucl. Phys. {\bf B 463} (1996) 398, hep-th/9510225.

\bibitem{BSV2}
M. Bershadsky, V. Sadov and C. Vafa,
Nucl. Phys. {\bf B 463} (1996) 420, hep-th/9511222.


\bibitem{BORCH} 
R. E. Borcherds. Adv. Math. {\bf 83} (1990) No. 1;  
Invent. Math. {\bf 109} (1992) 405.


\bibitem{CF}
P. Candelas and A. Font, hep-th/9603170.


\bibitem{CdlOFKM}
P. Candelas, X.C. de la Ossa, A. Font, S. Katz, D. Morrison, Nucl. Phys.
 {\bf B416} (1994) 481.


\bibitem{CCL}
G. Lopes Cardoso, G. Curio and D. L\"ust, hep-th/9608154.


\bibitem{CCLMR}
G. Lopes Cardoso, G. Curio, D. L\"ust, T. Mohaupt and S.-J. Rey, 
Nucl.Phys. {\bf B 464} (1996) 18, hep-th/9512129.

\bibitem{CCLM}
G. Lopes Cardoso, G. Curio, D. L\"ust and T. Mohaupt,
Phys. Lett {\bf B 382} (1996) 241, hep-th/9603108.


\bibitem{CdWLMR}
G. Lopes Cardoso, B. de Wit, D. L\"ust, T. Mohaupt and S.-J. Rey,
Nucl. Phys. {\bf B 481} (1996) 353, hep-th/9607184. 

\bibitem{CLM1}
G. Lopes Cardoso, D. L\"ust and T. Mohaupt, Nucl. Phys. {\bf B 455} (1995) 131,
hep-th/9507113.

\bibitem{CLM2} 
G. Lopes Cardoso, D. L\"ust and T. Mohaupt, 
Nucl. Phys. {\bf B 450} (1995) 115,
hep-th/9412209.

\bibitem{CdAF}
L. Castellani, R. D'Auria and S. Ferrara, Phys. Lett. {\bf B 241} (1990) 57;
Class. Quant. Grav. {\bf 7} (1990) 1767.

\bibitem{CdAFLL}
L. Castellani, R. D'Auria, S. Ferrara, W. Lerche and J. Louis, Int. Journ.
Mod. Phys. {\bf A8} (1993) 79, hep-th/9204035.

\bibitem{CFG}
S. Cecotti, S. Ferrara and L. Girardello, Int. Journ. Mod. Phys {\bf A4}
(1989) 2457.

\bibitem{CFIV}
S. Cecotti, P. Fendley, K. Intrilligator and C. Vafa, Nucl. Phys. {\bf B 386}
(1992) 405, hep-th/9204102.

\bibitem{CV}
S. Cecotti and C. Vafa, Comm. Math. Phys. {\bf 158} (1993) 569, hep-th/9211097.

\bibitem{CdAFvP}
A. Ceresole, R. D'Auria, S. Ferrara, A. Van. Proyen, Nucl. Phys. {\bf B 444} 
(1995) 92.

\bibitem{CHSW}
P. Candelas, G.T. Horowitz, A. Strominger and E. Witten,
Nucl. Phys. {\bf B 258} (1985) 46.

\bibitem{CRTvP1}
B. Craps, F. Rose, W. Troost and A. van Proeyen, hep-th/9606073.

\bibitem{CRTvP}
B. Craps, F. Rose, W. Troost and A. van Proeyen, hep-th/9703082.

\bibitem{C}
G. Curio, Phys. Lett {\bf B 366} (1996) 131, hep-th/9509042; 
Phys. Lett {\bf B 368} (1996) 78, hep-th/9509146.

\bibitem{CL}
G. Curio and D. L\"ust, hep-th/9703007.

\bibitem{dWvP}
B. de Wit and A. van Proeyen,
Nucl. Phys. {\bf B 245} (1984) 89.


\bibitem{dWLvP} 
B. de Wit, P.G. Lauwers and A. van Proeyen, 
Nucl. Phys. {\bf B 255} (1985) 569.


\bibitem{dWKLL}
B. de Wit, V. Kaplunovsky, J. Louis, D. L\"ust, 
Nucl. Phys. {\bf B 451} (1995) 53, hep-th/9504006.


\bibitem{DKL}
L. Dixon, V. Kaplunovsky and J. Louis, Nucl. Phys. {\bf B 329} (1990) 27.

\bibitem{DGW}
R. Donagi, A. Grassi and E. Witten, 
Mod.Phys.Lett. {\bf A11} (1996) 2199,
hep-th/9607091.

\bibitem{D}
M. Duff, Nucl. Phys. {\bf B 442} (1995) 47, hep-th/9501030.


\bibitem{Duff} 
M. J. Duff, James T. Liu, R. Minasian,
Nucl. Phys. {\bf B 452} (1995) 261, hep-th/9506126.

\bibitem{DK} 
M.J. Duff and R.R. Khuri,
Nucl.\ Phys.\ {\bf B 411} (1994) 473, hep-th/9305142.


\bibitem{DLR}
M. Duff, J. T. Liu and J. Rahmfeld, 
Nucl.Phys. {\bf B 459} (1996) 125, hep-th/9508094.


\bibitem{DuMinWit} 
M. J. Duff, R. Minasian and E. Witten, 
Nucl.Phys. {\bf B 465} (1996) 413, hep-th/9601036.


\bibitem{EZ}
M. Eichler and D. Zagier, 
{\it The Theory of Jacobi Forms} (Birkh\"auser, Basel, 1985).



\bibitem{EOTY}
T. Eguchi, H. Ooguri, 
A. Taormina and S.--K. Yang, Nucl. Phys. {\bf B 315} (1989) 193.

\bibitem{erler} 
J. Erler, 
J. Math. Phys. {\bf 35} (1994) 1819, hep-th/9304104.


\bibitem{FHSV}
S. Ferrara, J. Harvey, A. Strominger, C. Vafa, 
Phys. Lett. {\bf B 361} (1995) 59, hep-th/9505162.

\bibitem{FKLZ}
S. Ferrara, C. Kounnas, D.L\"ust and F.Zwirner, Nucl. Phys. {\bf B365} (1991) 431.

\bibitem{FvP}
S. Ferrara and A. van Proeyen, Class. Quant. Grav. {\bf 6} (1989) 243.

\bibitem{FL}
S. Foerste and J. Louis, hep-th/9612192.


\bibitem{FILQ}
A. Font, L.E. Ibanez, D. L\"ust and F. Quevedo, Phys. Lett. {\bf B 245}
(1990) 401.

\bibitem{gimon} 
E.G.~Gimon and J.~Polchinski, 
Phys.\ Rev.\ {\bf D 54} (1996) 1667, 
hep-th/9601036.

\bibitem{green-schwarz} 
M.B. Green and J. Schwarz, 
Phys. Lett. {\bf B 149} (1984) 117.



\bibitem{gswest} 
M.B. Green, J. Schwarz and P. C. West, 
Nucl. Phys. {\bf B 254} (1985) 327.

\bibitem{GP}
B. R. Greene and M. R. Plesser, Nucl. Phys. {\bf B 338} (1990) 15.



\bibitem{GN2}
V. A. Gritsenko and  V. V. Nikulin, alg-geom/9603010.


\bibitem{GN1}
V. A. Gritsenko and V. V. Nikulin, alg-geom/9510008.

\bibitem{GHMR}
D.J. Gross, J.A. Harvey, E. Martinec and R. Rohm, Nucl. Phys. {\bf B 256}
(1985) 253.

\bibitem{GSII}
M. Green and J. Schwarz, Phys. Lett. {\bf B 109} (1981) 444, Nucl. Phys.
{\bf B 181} (1981) 502, Nucl. Phys. {\bf B 198} (1982) 252,441.

\bibitem{GSW}
M. Green, J. Schwarz and E. Witten, "Superstring theory", Cambridge
University Press (1987).

\bibitem{HM}
J. Harvey and G. Moore, 
Nucl. Phys. {\bf B 463} (1996) 315, hep-th/9510182.


\bibitem{HS}
J. Harvey and A. Strominger, Nucl.Phys. {\bf B 449} (1995) 535, hep-th/9504047.



\bibitem{HKTY}
S. Hosono, A. Klemm, S. Theisen and S.-T. Yau, Comm. Math. Phys. {\bf 167}
(1995) 301, hep-th/9308122.



\bibitem{HT}
C. Hull, P. Townsend, Nucl. Phys. {\bf B438} (1995) 109, hep-th/9410167.



\bibitem{I}
J. Igusa, Amer. J. Math {\bf 84} (1962) 175, {\bf 86} (1964) 392,
{\bf 88} (1966) 817.


\bibitem{KV}
S. Kachru, C. Vafa, 
Nucl. Phys. {\bf B 450} (1995) 69, hep-th/9505105.

\bibitem{KKLMV}
S. Kachru, A. Klemm, W. Lerche, P. Mayr, C. Vafa, 
Nucl. Phys. {\bf B 459} (1996) 537, hep-th/9508155

\bibitem{KL}
V. Kaplunovsky and J. Louis, Nucl. Phys. {\bf B 444} (1995) 191, 
hep-th/9502077.

\bibitem{KLTh}
V. Kaplunovsky, J. Louis, S. Theisen, 
Phys. Lett. {\bf B 357} (1995) 71, hep-th/9506110.

\bibitem{KLT}
H. Kawai, D. C. Lewellen and S.-H. H. Tye, Phys. Rev. {\bf D 34} (1986) 3794.
Nucl. Phys {\bf B 288} (1987) 1.

\bibitem{KYY}
T. Kawai, Y. Yamada
and S.-K. Yang, Nucl. Phys. {\bf B 414} (1994) 191, hep-th/9306096.

\bibitem{K1}
T. Kawai, Phys. Lett. {\bf B 371} (1996) 59, hep-th/9512046.

\bibitem{K2}
T. Kawai, hep-th/9607078.

\bibitem{KaMoPle} 
S. Katz, D. R. Morrison and R. Plesser,
Nucl.Phys. {\bf B 477} (1996) 105, hep-th/9601108.

\bibitem{KLM}
A. Klemm, W. Lerche, P. Mayr, 
Phys. Lett. {\bf B 357} (1995) 313, hep-th/9506112.

\bibitem{KLMVW}
A. Klemm, W. Lerche, P. Mayr, C. Vafa and N. Warner,
Nucl.Phys. {\bf B 477} (1996) 746, hep-th/9604034.

\bibitem{KM} 
A. Klemm and P. Mayr, hep-th/9601014.

\bibitem{LLS}
W. Lerche, D. L\"ust and A. N. Schellekens, Nucl. Phys. {\bf B 287} (1987) 477.


\bibitem{Le}
W. Lerche, Nucl. Phys. {\bf B 308} (1988) 102.

\bibitem{L}
W. Lerche, hep-th/9507011.

\bibitem{LY}
B. H. Lian and S.-T. Yau, hep-th/9507151 and hep-th/9507153.

\bibitem{LF}
J. Louis and K. F\"orger, hep-th/9611184.

\bibitem{LSTY}
J. Louis, J. Sonnenschein, S. Theisen and S. Yankielowicz, 
Nucl. Phys. {\bf B 480} (1996) 185, hep-th/9606049.

\bibitem{LT}
D. L\"ust and S. Theisen, "Lectures on String Theory", Springer (1989),
Lecture Notes in Physics 346.

\bibitem{MV}
D. Morrison and C. Vafa, Nucl. Phys. {\bf B 473} (1996) 74, hep-th/9602114;
Nucl. Phys. {\bf B 476} (1996) 437, hep-th/9603161.

\bibitem{NEU} C. D. D. Neumann, hep-th/9607029.

\bibitem{OV}
H. Ooguri and C. Vafa, 
Nucl. Phys. {\bf B 463} (1996) 55, hep-th/9511164.

\bibitem{vP}
A. van Proeyen, Lectures given in the 1995 Trieste Summer School,
hep-th/9512139.

\bibitem{P}
J. Polchinski, 
Phys.Rev.Lett. {\bf 75} (1995) 4724, hep-th/9510017.

\bibitem{Sagn} 
A. Sagnotti, Phys. Lett. {\bf B 294} (1992) 196.



\bibitem{SchW}
A. N. Schellekens and N. P. Warner, Nucl. Phys. {\bf B 287} (1987) 317.


\bibitem{schwarzanom} 
J. Schwarz, 
Phys. Lett. {\bf B 371} (1996) 223, hep-th/9512053.

\bibitem{Seib}
N. Seiberg, Phys. Lett. {\bf B 206} (1988) 75.

\bibitem{Sei}
N. Seiberg, Nucl. Phys. {\bf B303} (1988) 286.

\bibitem{SeiWit} N. Seiberg and E. Witten, Nucl. Phys. {\bf B 471} (1996) 
121, hep-th/9603003.

\bibitem{Sen}
A. Sen, Phys. Lett. {\bf B 329} (1994) 217, hep-th/9402032.

\bibitem{S+SS}
J. Schwarz and A. Sen,
Phys.Lett. {\bf B 312} (1993) 10, hep-th/9305185; 
Nucl.Phys. {\bf B 411} (1994) 35, hep-th/9304154 .

\bibitem{Str}
A. Strominger, Comm. Math. Phys. {\bf 133} (1990) 163.

\bibitem{S}
A. Strominger, Nucl. Phys.{\bf B451} (1995) 96, hep-th/9504090.

\bibitem{SW}
N. Seiberg and E. Witten, Nucl. Phys. {\bf B 426} (1994) 19.

\bibitem{V}  
C. Vafa, Nucl. Phys. {\bf B 447} (1995) 261, hep-th/9505023.

\bibitem{V-F}
C. Vafa, Nucl. Phys. {\bf B 469} (1996) 493, hep-th/9602022.

\bibitem{VW}
C. Vafa and E. Witten, hep-th/9507050.

\bibitem{W-anom}
E. Witten, Nucl. Phys. {\bf B 234} (1983) 269.

\bibitem{W}
E. Witten, Nucl. Phys. {\bf B443} (1995) 85, hep-th/9503124.


\bibitem{Wi}
E. Witten, hep-th/9507121.


\bibitem{W2}
E. Witten, 
Nucl. Phys. {\bf B 460} (1996) 335, hep-th/9510135.

\bibitem{Wit2} 
E. Witten, Nucl. Phys. {\bf B 460} (1996) 541, hep-th/9511030.


\end{enumerate}
\end{document}